\documentclass[twocolumn]{aastex701}

\usepackage{tabularx} 
\usepackage{float}
\usepackage{booktabs}
\usepackage{threeparttable} 
\usepackage{multirow} 
\usepackage{graphicx}
\usepackage{amsmath} 
\usepackage{amssymb}
\usepackage{subfigure}
\usepackage{hyperref} 
\usepackage{float}
\bibpunct{(}{)}{;}{a}{}{,}  
\makeatletter
\renewcommand{\@thesubfigure}{\hskip\subfiglabelskip}
\makeatother
\makeatletter

\graphicspath{{./}{Fig/}}

\received{2025 July 5}
\revised{2025 October 1}
\accepted{2025 October 5}

\shorttitle{Further Evidence for the Outer Component of AGNs}
\shortauthors{Wang et al.}

\begin{document}
\title{The UV/optical Continuum Reverberation Mapping of Eight Active Galactic Nuclei with Swift: Further Evidence for the Outer Component}
\author[0009-0002-5955-4932]{Chi-Zhuo Wang}
\affiliation{Department of Astronomy, School of Physics, Peking University, Beijing 100871, People's Republic of China}
\affiliation{Kavli Institute for Astronomy and Astrophysics, Peking University, Beijing 100871, People's Republic of China}
\email{chizhuowang@stu.pku.edu.cn}

\author[0000-0002-7350-6913]{Xue-Bing Wu}
\affiliation{Department of Astronomy, School of Physics, Peking University, Beijing 100871, People's Republic of China}
\affiliation{Kavli Institute for Astronomy and Astrophysics, Peking University, Beijing 100871, People's Republic of China}
\email{wuxb@pku.edu.cn}

\author[0009-0001-5163-5781]{Yuanzhe Jiang}
\affiliation{Department of Astronomy, University of Illinois at Urbana-Champaign, Urbana, IL 61801, USA}
\email{yuanzhe@illinois.edu}

\author[0000-0003-0827-2273]{Qinchun Ma}
\affiliation{Department of Astronomy, School of Physics, Peking University, Beijing 100871, People's Republic of China}
\affiliation{Kavli Institute for Astronomy and Astrophysics, Peking University, Beijing 100871, People's Republic of China}
\email{maqinchun@pku.edu.cn}

\author[0009-0007-7679-0001]{Huapeng Gu}
\affiliation{Department of Astronomy, School of Physics, Peking University, Beijing 100871, People's Republic of China}
\affiliation{Kavli Institute for Astronomy and Astrophysics, Peking University, Beijing 100871, People's Republic of China}
\email{guhuapeng@pku.edu.cn}

\author{Yuhan Wen}
\affiliation{Department of Astronomy, School of Physics, Peking University, Beijing 100871, People's Republic of China}
\affiliation{Kavli Institute for Astronomy and Astrophysics, Peking University, Beijing 100871, People's Republic of China}
\email{wenyuhan@pku.edu.cn}
\correspondingauthor{Chi-Zhuo Wang, Xue-Bing Wu}
\email{chizhuowang@stu.pku.edu.cn, wuxb@pku.edu.cn}

\begin{abstract}
In our previous work, we applied the ICCF-Cut method to the continuum reverberation mapping (CRM) of six active galactic nuclei (AGNs)  based on the published Swift data. Extending this work, we perform a systematic AGN CRM study utilizing the Swift archive. We enlarge our sample with eight additional AGNs at $z$ \textless $ 0.05$ with high-cadence (\textless $ 3$ days) and multiband photometric observations. Time series analysis of these light curves shows two main results: (1) The interband lags are broadly consistent with $\tau \propto \lambda^{4/3}$, while the average interband lags are larger than those predicted by the standard thin accretion disk model. (2) For most targets, there exists a $U$ band lag excess, which is probably due to the diffuse continuum emission from the broad-line region (BLR). We employ the ICCF-Cut method to extract the possible diffuse continuum component from the $U$ band light curves and calculate the diffuse continuum lags ($\tau_{cut}$), which are generally consistent with the lags ($\tau_{jav}$) derived by the JAVELIN Photometric Reverberation Mapping Model. Further analysis with our sample indicates a positive correlation between the diffuse continuum region size and the BLR size ($R_{DCR}-R_{BLR}$ relation), as well as another correlation with the luminosity ($R_{DCR}-L$ relation). These findings provide further evidence for a significant contribution of diffuse continuum emission from the BLR to the AGN continuum lags.
\end{abstract}


\section{Introduction} \label{sec:intro}
Active Galactic Nuclei (AGNs) are among the most luminous and energetic objects in the universe, powered by accretion onto supermassive black holes (SMBHs; \citealt{1969Natur.223..690L}). The SMBH is surrounded by a complex structure, including the accretion disk, jet, broad-line region (BLR), and dusty torus \citep{1993ARA&A..31..473A,1995PASP..107..803U}. Understanding the physical processes and structures in these regions is crucial for unraveling the nature of AGNs and their role in galaxy evolution \citep{2012ARA&A..50..455F,2013ARA&A..51..511K}. However, the small angular sizes of most AGNs make it challenging to resolve their structure spatially with current facilities. 

Continuum reverberation mapping (CRM) has proven to be a powerful technique for investigating the structure and dynamics of AGNs \citep{1991ApJ...366...64C,2021iSci...24j2557C}. According to the standard thin disk model proposed by \cite{1973A&A....24..337S}, the accretion disk is optically thick and geometrically thin with a temperature profile $T(R)  \propto R^{-3/4}$. Given the time delay (lag) $\tau \approx R/c$ and Wien’s law $\lambda\propto 1/T$, this model predicts a relation between lag and wavelength, $\tau \approx \lambda^{4/3}$, where the variations from the smaller, hotter inner disk are expected to precede those from the larger, cooler outer disk regions \citep{2007MNRAS.380..669C}. By monitoring the variability of AGN continuum across different wavelengths, CRM allows us to measure the lags between variations in the ultraviolet (UV), optical, and X-ray bands. These lags provide insights into the size and temperature structure of the accretion disk and the contribution of other components, such as the BLR.

The past decade has witnessed significant advancements in CRM studies, driven by high-cadence, multiband monitoring campaigns utilizing both ground and space telescopes \citep{2014ApJ...788...48S,2014MNRAS.444.1469M,2018MNRAS.480.2881M,2015ApJ...806..129E,2017ApJ...840...41E,2019ApJ...870..123E,2024ApJ...973..152E,2016ApJ...821...56F,2018ApJ...854..107F,2018ApJ...857...53C,2020ApJ...896....1C,2023ApJ...958..195C,2021MNRAS.504.4337V,2022MNRAS.512L..33V,2021ApJ...922..151K,2023ApJ...947...62K,2024ApJ...964..167L,2024ApJ...974..271L,2024MNRAS.527.5569G}. The key results of these CRM campaigns can be summarized as follows: (1) The UV/optical variations are strongly correlated, while the correlation between X-ray and UV variations is much weaker than the correlations among the UV/optical bands. (2) The wavelength-dependent interband lags largely follow the expected $\tau \approx \lambda^{4/3}$ relation predicted by the reprocessing from a standard disk, but the disk sizes derived from CRM are larger than expected by a factor of 2–4. (3) The $u/U$ band lags around the Balmer jump (3646 Å) are significantly larger than expected based on an extrapolation of the other UV/optical lags.

These observational results challenge various aspects of the current standard model of AGN central engines, prompting a surge in theoretical investigations. For instance, various models have been proposed, including the inhomogeneous accretion disk model \citep{2011ApJ...727L..24D,2017ApJ...835...65S,2018ApJ...855..117C,2020ApJ...891..178S,2022MNRAS.513.1046N}, the secondary reprocessing model \citep{2017MNRAS.470.3591G,2023MNRAS.521..251H,2024MNRAS.530.4850H}, the disk wind model \citep{2019MNRAS.483.2275L,2019MNRAS.482.2788S,2025ApJ...978...54C}, the modifying disk and corona model \citep{2018ApJ...857...86S,2019ApJ...879L..24K,2021ApJ...907...20K,2021MNRAS.503.4163K,2023MNRAS.526..138K,2025arXiv250100806J}, the non-blackbody disk model \citep{2018ApJ...854...93H}, the obscuration effect \citep{lewin2025}, and the underappreciated non-disk components model \citep{2019NatAs...3..251C}. One leading explanation suggests that the diffuse continuum emission from the BLR significantly contributes to the observed continuum lags \citep{2001ApJ...553..695K,2019MNRAS.489.5284K,2018MNRAS.481..533L,2020MNRAS.494.1611N,2022MNRAS.509.2637N}. This diffuse emission is dominated by the free–free and free–bound processes in the BLR gas. Since the BLR is located at larger distances than the accretion disk, the observed continuum light curves integrate signals from both the accretion disk and the outer component, leading to systematically larger observed lags than those predicted by the standard disk reprocessing models. In parallel to the BLR diffuse continuum interpretation, some studies have demonstrated that the disk reprocessing of X-rays can quantitatively reproduce many of the key observational results. For example, \citet{2021ApJ...907...20K,2021MNRAS.503.4163K,2023MNRAS.526..138K} have shown that the state-of-the-art simulations of the X-ray-illuminated disk model, which incorporate more elaborate mechanisms and parameters, can quantitatively reproduce the observed lag–wavelength relations without invoking additional BLR components. \citet{2022ApJ...935...93P} and \citet{2022A&A...661A.135D} demonstrated that the variability power spectra and the energy spectral distribution are well reproduced by disk reprocessing. \citet{2022ApJ...941...57P} also suggested that a low UV/X-ray correlation is well expected in the case of an X-ray illuminated accretion disk, when the dynamic variability of the X-ray source is taken into account. In addition, the frequency-resolved analysis has been applied to further probe the origin of the observed variability \citep{2022ApJ...925...29C,2023ApJ...954...33L,2024ApJ...974..271L,2025ApJ...983..132P}. These studies reveal that high-frequency lags are consistent with disk reprocessing, while low-frequency lags may be influenced by the extended reprocessors, likely associated with the BLR. Nevertheless, despite the success of modifying disk reprocessing models, the contribution of diffuse continuum emission from the BLR remains an important factor, particularly in the $u/U$ band, where the Balmer continuum and the Balmer jump (3646 Å) strongly enhance the excess lags. This motivates our focus on disentangling the diffuse continuum component in the present study.

\begin{table*}[t!]
\begin{center}
\caption{Physical Properties of the AGN Sample}
\label{tab:Properties}
\setlength\tabcolsep{3.8mm}{
\begin{tabular}{ccccccc}
\toprule
\toprule
Object & Redshift & $\log_{10}(M_{BH}/M_\odot)$ & $\log L_{5100}$/erg s$^{-1}$ & $\dot{m}_{Edd}$ & $R_{BRL}$/lt-day  & References\\ 
(1) & (2) & (3) & (4) & (5) & (6) & (7) \\
\midrule
Fairall 9      & 0.047 & $8.299^{+0.078}_{-0.116}$ & $43.98\pm0.04$& $0.020^{+0.004}_{-0.004}$   & $17.4_{-4.30}^{+3.20}$ &1,2,3,4\\
3C 120         & 0.033 & $7.745^{+0.083}_{-0.040}$ & $44.00\pm0.10$& $0.420^{+0.100}_{-0.090}$   & $20.2_{-4.20}^{+5.00}$&1,2,5\\
MCG+08-11-011 & 0.021 & $7.288^{+0.046}_{-0.054}$ & $43.33\pm0.11$& $0.054^{+0.040}_{-0.025}$  & $15.7_{-0.50}^{+0.50}$ &1,2,6\\
Mrk 110        & 0.035 & $7.292^{+0.101}_{-0.097}$ & $43.66\pm0.12$& $0.433_{-0.074}^{+0.120}$    & $25.6_{-7.20}^{+8.90}$&1,2,3,7\\
Mrk 279        & 0.030 & $7.435^{+0.099}_{-0.133}$ & $43.71\pm0.07$& $0.210_{-0.032}^{+0.056}$& $16.7_{-3.90}^{+3.90}$ &1,2,3\\
Mrk 335        & 0.026 & $7.230^{+0.042}_{-0.044}$ & $43.76\pm0.07$& $0.100_{-0.030}^{+0.080}$   & $14.0_{-3.40}^{+4.60}$&1,2,8\\
Mrk 817        & 0.031 & $7.586^{+0.064}_{-0.072}$ & $43.74\pm0.09$& $0.200_{-0.025}^{+0.051} $   & $19.9_{-6.70}^{+9.90}$&1,2,3,9\\
NGC 6814       & 0.005 & $7.230^{+0.042}_{-0.044}$ & $42.12\pm0.28$& $0.026_{-0.001}^{+0.001}$& $6.60_{-0.90}^{+0.90}$ &1,2,10\\ 
\bottomrule
\end{tabular}}
\end{center}
\textbf{Note.} $R_{BLR}$ is represented by H$\beta$ lag. Reference: (1) AGN Black Hole Mass Database \citep{2015PASP..127...67B}, (2) \cite{2019ApJ...886...42D}, (3) \cite{2009MNRAS.392.1124V}, (4) \cite{2020MNRAS.498.5399H}, (5) \cite{2020MNRAS.497.2910H}, (6) \cite{2018ApJ...854..107F}, (7) \cite{2011A&A...527A.127M}, (8) \cite{2020MNRAS.499.1266T}, (9) \cite{2021ApJ...922..151K}, (10) \cite{ 2024MNRAS.527.5569G}.
\end{table*}

To quantify the contribution of the diffuse continuum to the observed continuum lags, \cite{2018MNRAS.481..533L} and \cite{2019MNRAS.489.5284K} simulated the ionization state of the BLR under different AGN model assumptions and found that the diffuse continuum can significantly affect the total observed lags. Their simulations demonstrated that the diffuse continuum emission originating from the BLR clouds contributes $\sim 40 \%$ of the total continuum flux near the Balmer jump (3646 Å). \cite{2020MNRAS.494.1611N,2022MNRAS.509.2637N} further confirmed these results, and found that the observations align with the modeled lags by assuming a typical Balmer continuum lag of $0.5\tau_{H \beta}$. In addition, \cite{2022MNRAS.509.2637N} presented a new lag-luminosity relationship, showing that the continuum lags scale with $L_{5100}^{1/2}$. It is similar in the slope to the well-known
radius-luminosity relation of the broad H$\beta$ line, but scaled down by a factor of about 6. This relationship has been confirmed in several recent studies using various AGN samples \citep{2022ApJ...940...20G,2023ApJ...948L..23W,2025ApJ...985...30M,2025A&A...695A..10H}. Collectively, these findings suggest a significant contribution to the UV/optical continuum may come from the BLR itself, which appears to be a common occurrence in AGNs.

Although numerous studies indicated that the diffuse continuum emission from BLR contributes to the UV/optical continuum, especially in the $u/U$ band, directly quantifying and isolating this component remains challenging. It is still unclear whether this outer component can reverberate to the disk emissions and result in larger observed lags than those predicted by the accretion disk models. In our previous work \citep{2024ApJ...966..149J}, we applied the ICCF-Cut method \citep{2023ApJ...949...22M,2024ApJ...966....5M} to CRM of 6 AGNs and successfully decomposed the light curves of the potential diffuse continuum from the Swift $U$ band data. We found that the extracted outer component was highly correlated with the central disk emission and resulted in a larger lag than the original continuum lags. In this study, we will further expand our sample to the Swift database and revisit the lag-wavelength relation. We employ the ICCF-Cut method to isolate the diffuse continuum component and check the reliability of the lag measurements using the JAVELIN Pmap Model \citep{2011ApJ...735...80Z,2013ApJ...765..106Z,2016ApJ...819..122Z}. Based on the enlarged sample, we further investigate the relationships between the diffuse continuum region size and the BLR size, as well as the luminosity. These investigations provide further evidence that the outer component, which significantly contributes to the UV/optical continuum, originates from the diffuse continuum in the BLR.

This paper is organized as follows: Section \ref{sec:data} introduces the sample selection and data reduction. In Section \ref{sec:analysis}, we present the time series analysis in Section \ref{subsec:time-series-analysis}, describe the ICCF-Cut method for CRM in Section \ref{subsec:ICCF-Cut}, and the JAVELIN Pmap Model method in Section \ref{subsec:Javelin}. The specific sample analysis and results are presented in Section \ref{subsec:sample}. Finally, we discuss these results in Section \ref{sec:discussion} and give our conclusions in Section \ref{sec:conclusion}.

\begin{figure*}
    \centering
    \includegraphics[width=0.7\linewidth]{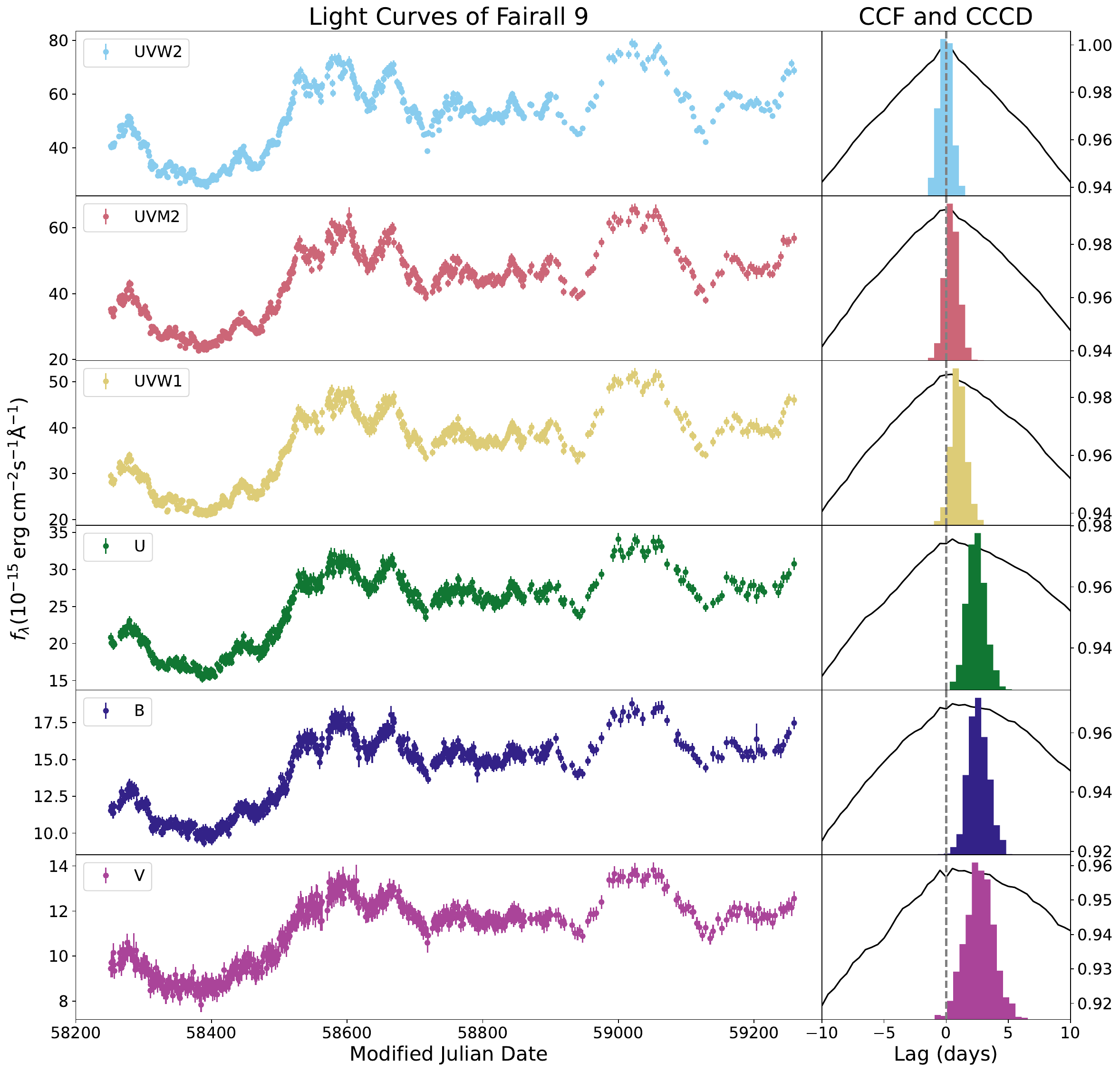}
    \caption{Left: Swift UVOT light curves of Fairall 9. Data are ordered by wavelength from the shortest (top) to the longest (bottom). Right: CCFs (in black) and FR/RSS centroid distributions (in color) for each band relative to the $UVW2$ band. A positive value means the band lags behind $UVW2$ band.}
    \label{fig:Fairall9_lc}
\end{figure*}
\section{Sample and Data Reduction} \label{sec:data}
\subsection{Sample Selection}\label{subsec:sample selection}
The Neil Gehrels Swift Observatory, launched in 2004, is a multiband observatory dedicated to studying gamma-ray burst science \citep{2004ApJ...611.1005G}. It carries three instruments that enable high-cadence monitoring across the X-ray, ultraviolet (UV), and optical bands. The Ultraviolet and Optical Telescope (UVOT; \citealt{2005SSRv..120...95R}) is one of these three instruments, which captures data in six filters: $UVW2$, $UVM2$, $UVW1$, $U$, $B$, and $V$ bands. Over the past decade, it has been a powerful instrument in studying AGN CRM, leading to a surge in related observation proposals and a significant accumulation of data in the Swift archive. 

We select a sample of AGNs from the Swift archive, focusing on those targets with redshifts smaller than 0.05. The selected AGNs were required to have high-quality, multiband photometric observations with a cadence of less than 3 days. In addition, these AGNs should exhibit clearly detectable broad emission lines during the observation period, which is crucial for conducting reliable reverberation mapping analysis. Ultimately, we identified 14 AGNs that meet our criteria, and six of which have been analyzed by \cite{2024ApJ...966..149J} using the published Swift data. Eight additional AGNs in this analysis are Fairall 9, 3C 120, MCG+08-11-011, Mrk 110, Mrk 279, Mrk 335, Mrk 817, and NGC 6814. Table \ref{tab:Properties} lists the physical properties of these targets, including redshift, black hole mass ($M_{BH}$), optical luminosity at 5100 Å ($L_{5100}$), Eddington ratio ($\dot{m}_{Edd}$), and H$\beta$ BLR size. These parameters are collected from the reference papers and the AGN Black Hole Mass Database \footnote{\url{http://www.astro.gsu.edu/AGNmass/}} \citep{2015PASP..127...67B}. For individual targets with multiple H$\beta$ lag measurements, we prioritize those overlapping with the Swift observation period. If concurrent measurements are not available, we employ the average H$\beta$ lag to represent the BLR size.

\subsection{UVOT Data Reduction}\label{subsec:data-raducation}
The Swift UVOT analysis largely follows the general procedure detailed in previous works \citep{2015ApJ...806..129E,2017ApJ...840...41E,2019ApJ...870..123E,2020MNRAS.498.5399H,2020ApJ...896....1C}. The data were processed with HEASOFT version 6.32. For each observation, fluxes are measured using the \texttt{uvotsource} tool. Source extractions are measured using a circular region with a radius of $5^{\prime \prime}$, while the background is measured in an annulus from $10^{\prime \prime}$ to $20^{\prime \prime}$, from which any stars that fall within the background annulus are excluded. Considering that the host galaxy contributes a fraction of the observed flux within the UVOT apertures, we will remove this contamination in Section \ref{subsubsec:host}. The resulting light curves exhibit occasional, anomalously low points (“dropouts”), which may be due to the localized low-sensitivity regions or tracking problems. The general filtering strategy is mapping the data onto the detector plane, delineating boxes around bad data, and using these boxes as a mask to filter out data. However, the detector masks are sometimes too aggressive, eliminating many points that are consistent with the light curves within their measurement errors \citep{2021ApJ...922..151K,2023ApJ...947...62K}. Therefore, we use a more conservative method. Dropouts are identified by comparing the flux deviation relative to the neighbor points to a threshold based on their respective errors. It is very similar to the dropout identification method given by \cite{2017ApJ...840...41E}, but we did not further screen out all data in these detector mask regions, ensuring that only points with significant deviations are removed.

\begin{figure*}
    \centering
    \begin{minipage}[b]{0.45\textwidth}
		\includegraphics[width=\linewidth]{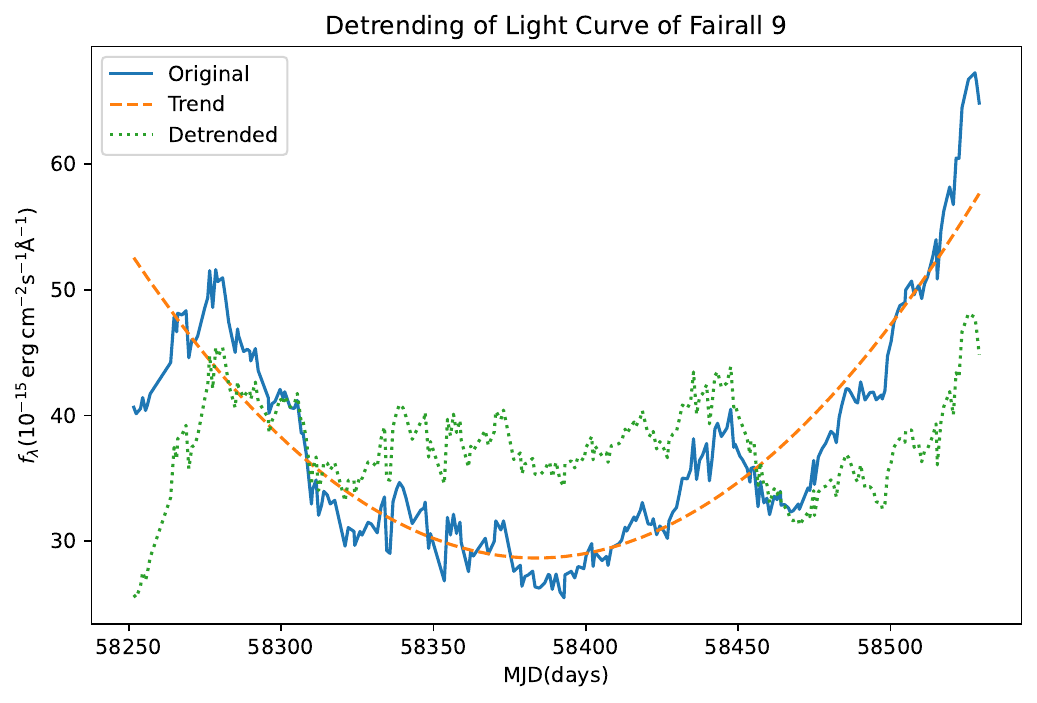}
	\end{minipage}%
	\begin{minipage}[b]{0.45\textwidth}
		\includegraphics[width=\linewidth]{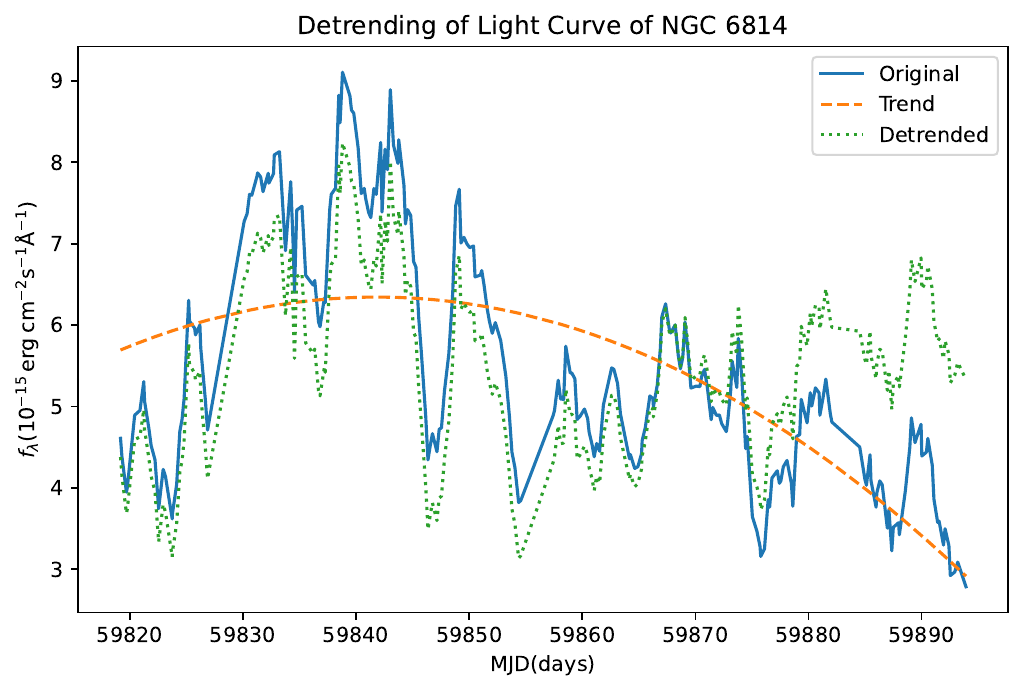}
	\end{minipage}%
    \caption{Left panel: $UVW2$ band light curves for Fairall 9. Right panel: $UVW2$ band light curves for NGC 6814. Both panels show the original light curve (blue solid), the long-term trend fit by a second-order (orange dashed), and the detrended light curve (green dotted).}
    \label{fig:detrendlc}
\end{figure*}

\subsection{Light Curves}\label{subsec:lightcurves}
The resulting light curves for Fairall 9 are plotted in the left panels of Figures \ref{fig:Fairall9_lc}-\ref{fig:Fairall9_lc2}. The light curves for the remaining seven targets are shown in the appendix, see Figure Set \ref{FigSetA}. They are in order of descending frequency with the $UVW2$ band at the top and the $V$ band at the bottom. But for 3C 120, only five band light curves are shown in Figure Set \ref{FigSetA} because the $B$ band data are unavailable in the Swift archive. Since the Swift monitoring for some targets spans many years, previous studies have presented partial light curves for these targets, including Fairall 9 \citep{2020MNRAS.498.5399H,2024ApJ...973..152E}, 3C 120 \citep{2018ApJ...867..128M}, Mrk 110 \citep{2022MNRAS.512L..33V}, Mrk 335 \citep{2023ApJ...947...62K}, Mrk 817 \citep{2023ApJ...958..195C}, and NGC 6814 \citep{2024MNRAS.527.5569G}. In this work, we not only reproduce these light curves but also provide several sets of unpublished light curves based on new data. All these data of final light curves are available in Zenodo \footnote{\url{https://zenodo.org/records/14930797}}. Following previous studies, we divide the light curves of Fairall 9 and Mrk 817 into two segments, respectively. The first segments correspond to the published light curves in \cite{2020MNRAS.498.5399H,2023ApJ...958..195C}, while the second segments are new observations. Given the large seasonal gaps in the observations, we also divide the light curves of Mrk 279 into three segments. Furthermore, the results of \cite{2020MNRAS.498.5399H} for Fairall 9 and \cite{2024MNRAS.527.5569G} for NGC 6814 indicate that the light curves of these targets exhibit both short-term and long-term variations. The variability on short timescales is likely related to the reprocessing of X-rays in the accretion disk, while the variability on long timescales may arise from the changes in the accretion flow. Therefore, we perform a quadratic fitting to the light curves and subtract the long-term trend represented by the fitted curve. Taking the $UVW2$ band light curves as an example, we exhibit the initial and detrended light curves of Fairall 9 and NGC 6814 in Figure \ref{fig:detrendlc}. Note that the mean flux of the detrended light curves will be shifted to equivalence with the original light curves. Following \cite{2020MNRAS.498.5399H}, we only detrend the first segment light curves for Fairall 9. To facilitate the description of different types of light curves for the same target, we employ the numerals 1, 2, and 3 following the target name to denote the first, second, and third segments of the light curves, respectively. The suffix '-D' indicates that the light curves have subtracted the long-term trend. For example, Fairall 9 (1-D) refers to the first segment detrended light curves of Fairall 9.

\begin{figure*} % 
    \centering

    \begin{minipage}[t]{0.48\textwidth}
        \centering
        \includegraphics[width=\linewidth]{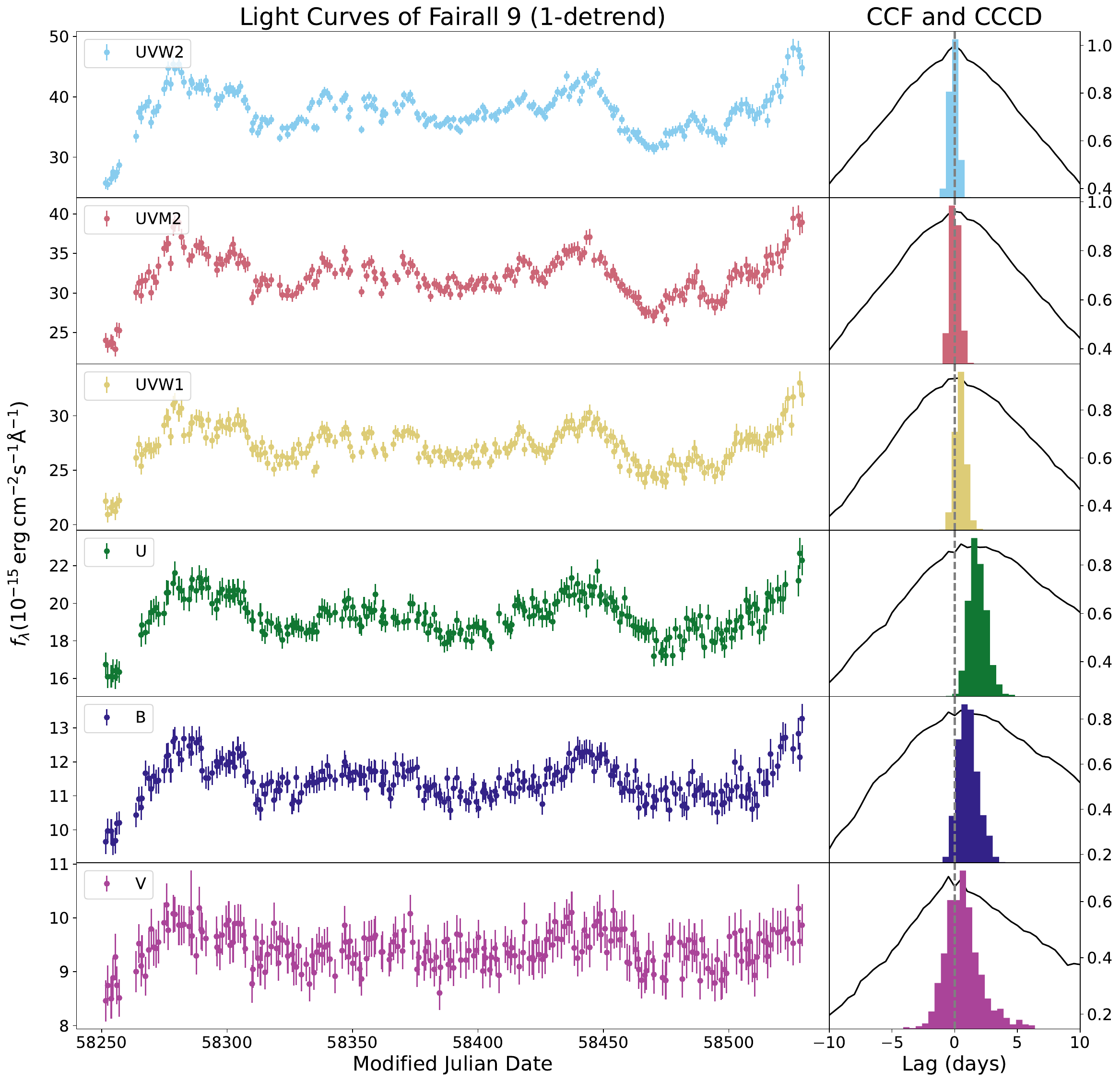}
        \caption{The same as Figure \ref{fig:Fairall9_lc}, but for the first part detrended light curves of Fairall 9.}
        \label{fig:Fairall9_lc1_detrend}
    \end{minipage}
    \hfill 
    \begin{minipage}[t]{0.48\textwidth}
        \centering
        \includegraphics[width=\linewidth]{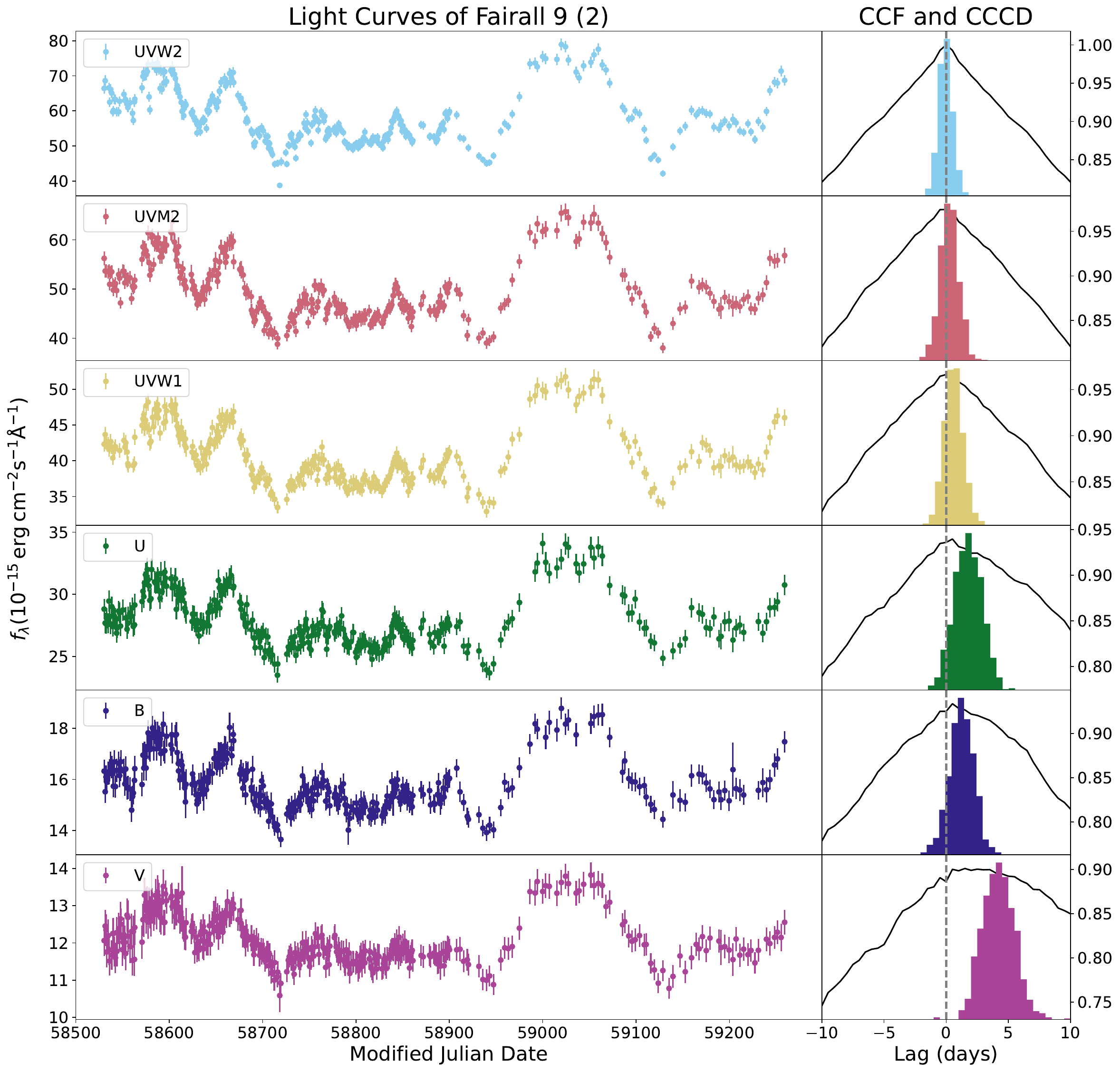}
        \caption{The same as Figure \ref{fig:Fairall9_lc}, but for the second part light curves of Fairall 9.}
        \label{fig:Fairall9_lc2}
    \end{minipage}
\end{figure*}

To quantify the variability amplitude of light curves, we calculate the fractional variability ($F_{var}$; \citealt{2003MNRAS.345.1271V}) 
\begin{equation}\label{eq:Fvar}
	{F_{var}=\sqrt{\frac{S^2-\overline{\sigma^2_{err}}}{\overline{X}^2}}},
\end{equation}
where $S^{2}$, $\overline{X}$, and $\overline{\sigma_{err}^2}$ are the total variance, mean value, and mean square error of the light curves, respectively. The $F_{var}$ results for each band are shown in Table \ref{tab:lc}. We find that $F_{var}$ generally decreases with increasing wavelength. This trend is primarily intrinsic to the AGN variability process, as shorter wavelengths originate from smaller, more rapidly responding regions of the accretion disk, while longer wavelengths trace larger, more slowly varying regions \citep{2016ApJ...821...56F,2018MNRAS.480.2881M,2019ApJ...870..123E,2018MNRAS.478.2557G}. Additionally, contamination from the host galaxy's starlight may contribute to the reduced variability at redder wavelengths, particularly in the optical bands \citep{2018MNRAS.478.2557G,2019ApJ...870..123E,2020MNRAS.498.5399H,2022A&A...659A..13F,2024MNRAS.527.5569G}. To isolate the variable AGN component from the invariable host galaxy emission, we will perform a flux–flux analysis in Section \ref{subsubsec:host}. In addition, we compute the observation epochs and cadence of the $UVW2$ band light curve for each target, listed in columns 2 and 3 of Table \ref{tab:lc}. Most targets exhibit high-quality, high-cadence, multiband-monitored light curves, which are ideal for CRM and specific analyses.

\begin{table*}[t]
\begin{center}
\caption{Multiband Light Curves Properties}
\label{tab:lc}
\setlength\tabcolsep{4.4mm}{
\begin{tabular}{cccccccccc}
\toprule
\toprule
Object& Segment& Epoch& Cadence& \multicolumn{6}{c}{$F_{var}$ ($\%$)}\\
\cline{5-10}& &  &(days)& $F_{w2}$&$F_{m2}$ &$F_{w1}$ & $F_{u}$&$F_{b}$ &$F_{v}$\\ 
(1) & (2) &  (3) &(4) & (5) & (6)& (7)& (8)& (9)&(10)\\
\midrule
Fairall 9      & All&  595
&1.07& 26.0 & 24.9& 22.3& 19.6& 17.3&12.9\\
Fairall 9      & 1-D&  244
&1.00& 9.58 & 8.25& 6.36& 4.87& 4.15&3.60\\
Fairall 9      & 2         &  351
&1.19& 12.8 & 11.5& 9.41& 6.87& 5.78&4.34\\
3C 120         & All&  103&2.20& 16.8 & 15.0& 12.2& 11.9& Null&7.63\\
MCG+08-11-011& All&  223
&0.34& 14.9 & 12.4& 9.92& 10.1& 8.72&5.57\\
Mrk 110        & All&  53&2.16& 10.3 & 10.8& 7.42& 7.31& 6.19&3.76\\
Mrk 279        & 1         &  112
&2.04& 47.1 & 42.6& 34.6& 32.7& 18.8&9.89\\
Mrk 279        & 2         &  165
&0.80& 23.1 & 21.0& 18.1& 17.1& 13.0&8.34\\
Mrk 279        & 3         &  72
&1.98& 19.0 & 16.8& 13.3& 13.5& 10.2&8.58\\
Mrk 335        & All&  204&0.40& 8.67 & 7.83& 6.26& 5.54& 4.20&2.66\\
Mrk 817        & All&  568&1.39& 21.0 & 15.7& 13.7& 11.6& 9.25&7.24\\
Mrk 817        & 1         &  317
&1.06& 17.0 & 15.7& 13.7& 11.6& 9.25&7.24\\
Mrk 817        & 2         &  251
&1.93& 15.3 & 18.9& 12.1& 9.88& 9.38&6.70\\
NGC 6814       & All&  252&0.27
& 26.4
 & 23.2& 18.8& 16.5& 8.60&4.47\\ 
NGC 6814       & All-D&  252
&0.27& 19.6 & 17.0& 12.6& 10.9& 5.19&2.26\\ 
\bottomrule
\end{tabular}}

\end{center}
\textbf{Note.} Column 1: Object name. Column 2: Numbers 1, 2, and 3 represent the first, second, and third segments of the light curves, respectively. “All” represents the whole light curves. The suffix “-D” represents the detrended light curves. Column 3: Number of good data points in $UVW2$ band. Column 4: Sampling interval in $UVW2$ band. Column $5\sim10$: The fractional variability amplitude in $UVW2$, $UVM2$, $UVW1$, $U$, $B$, and $V$ band.
\end{table*}

\section{Analysis and Results}\label{sec:analysis}
\subsection{Time series analysis}\label{subsec:time-series-analysis}

\begin{table*}[t]
\begin{center}
\caption{Continuum Lags and Lag-Wavelength Fitting Results}
\label{tab:lags}
\setlength\tabcolsep{2.3mm}{
\begin{tabular}{cccccccccc}
\toprule
\toprule
Object& Segment& \multicolumn{6}{c}{Interband Lags}& $\tau_0$&$R_{u}$ \\
\cline{3-8}
& & $\tau_{w2}$/days 
& $\tau_{m2}$/days& $\tau_{w1}$/days& $\tau_{u}$/days& $\tau_{b}$/days& $\tau_{v}$/days& (days)&\\ 
(1) & (2) & (3) & (4) & (5) & (6)& (7)& (8)&(9)&(10)\\
\midrule
Fairall 9      & All& $0.00_{-0.49}^{+0.49}$& $0.50_{-0.50}^{+0.51}$& $1.01_{-0.50}^{+0.73}$& $2.48_{-0.74}^{+0.74}$& $2.71_{-0.93}^{+0.77}$& $2.75_{-1.20}^{+1.01}$&1.15 
&1.82 
\\
Fairall 9      & 1-D& $0.00_{-0.27}^{+0.26}$& $0.02_{-0.28}^{+0.46}$& $0.48_{-0.47}^{+0.48}$& $1.76_{-0.53}^{+0.72}$& $1.02_{-0.73}^{+0.91}$& $0.50_{-1.17}^{+1.46}$&0.39 
&3.83 
\\
Fairall 9      & 2         & $0.00_{-0.49}^{+0.50}$& $0.24_{-0.72}^{+0.72}$& $0.52_{-0.57}^{+0.77}$& $1.75_{-1.03}^{+1.14}$& $1.23_{-0.97}^{+0.97}$& $4.20_{-1.27}^{+1.30}$&1.17 
&1.26 
\\
3C 120         & All& $0.00_{-0.72}^{+0.72}$& $0.72_{-1.03}^{+0.97}$& $1.23_{-1.18}^{+0.99}$& $2.95_{-1.17}^{+1.02}$& Null& $1.97_{-1.77}^{+2.03}$&0.94 
&2.66 
\\
MCG+08-11-011& All& $0.00_{-0.25}^{+0.25}$& $0.19_{-0.44}^{+0.31}$& $0.74_{-0.26}^{+0.26}$& $1.28_{-4.30}^{+3.20}$& $1.25_{-0.26}^{+0.26}$& $1.54_{-0.29}^{+0.46}$&0.61 
&1.77 
\\
Mrk 110        & All& $0.00_{-0.73}^{+0.73}$& $-0.22_{-0.73}^{+0.74}$& $-0.27_{-0.92}^{+0.75}$& $1.70_{-0.96}^{+1.00}$& $1.25_{-0.93}^{+0.97}$& $1.71_{-1.67}^{+1.47}$&0.64 
&2.25 
\\
Mrk 279        & 1         & $0.00_{-0.93}^{+0.77}$& $0.73_{-0.95}^{+0.92}$& $1.19_{-1.10}^{+0.98}$& $1.71_{-0.99}^{+0.99}$& $1.95_{-1.17}^{+1.20}$& $2.60_{-1.85}^{+1.72}$&0.98 
&1.48 
\\
Mrk 279        & 2         & $0.00_{-0.47}^{+0.47}$& $-0.72_{-0.49}^{+0.48}$& $-0.25_{-0.69}^{+0.50}$& $0.76_{-0.71}^{+0.71}$& $0.51_{-0.73}^{+0.91}$& $1.01_{-1.40}^{+1.42}$&0.31 
& 
2.04 
\\
Mrk 279        & 3         & $-0.01_{-0.72}^{+0.71}$& $-0.22_{-0.75}^{+0.91}$& $0.21_{-0.98}^{+0.97}$& $0.56_{-0.80}^{+0.73}$& $0.26_{-1.17}^{+1.17}$& $0.65_{-1.43}^{+1.39}$&0.21 
&2.21 
\\
Mrk 335        & All& $0.00_{-0.45}^{+0.26}$& $-0.29_{-0.48}^{+0.48}$& $0.26_{-0.48}^{+0.49}$& $1.72_{-0.50}^{+49}$& $1.76_{-0.53}^{+0.68}$& $1.80_{-0.77}^{+0.89}$&0.74 
& 
1.95 
\\
Mrk 817        & All& $0.00_{-0.48}^{+0.50}$& 
$0.25_{-0.69}^{+0.71}$& $0.92_{-0.89}^{+0.76}$& $3.62_{-1.13}^{+1.34}$& $1.46_{-0.94}^{+0.98}$& $2.96_{-1.96}^{+1.60}$&1.12 
&2.73 
\\
Mrk 817        & 1         & $0.00_{-0.45}^{+0.45}$& $0.24_{-0.45}^{+0.45}$& $0.74_{-0.51}^{+0.68}$& $2.23_{-0.73}^{+0.89}$& $1.91_{-0.94}^{+0.98}$& $1.94_{-1.18}^{+1.02}$&0.85 
& 
2.20 
\\
Mrk 817        & 2         & $0.01_{-0.50}^{+0.26}$& 
$0.20_{-0.70}^{+0.74}$& $0.76_{-0.58}^{+0.71}$& $2.10_{-0.66}^{+0.81}$& $1.49_{-0.78}^{+0.77}$& $3.43_{-0.96}^{+0.96}$&1.09 
&1.62\\
NGC 6814       & All& $0.00_{-0.01}^{+0.01}$& $0.23_{-0.22}^{+0.01}$& $0.22_{-0.22}^{+0.02}$& $0.26_{-0.01}^{+0.22}$& $0.25_{-0.23}^{+0.23}$& $0.03_{-0.28}^{+0.45}$&0.07 
&3.06\\ 
NGC 6814       & All-D& $0.00_{-0.01}^{+0.01}$& $0.23_{-0.22}^{+0.01}$& $0.23_{-0.22}^{+0.01}$& $0.26_{-0.01}^{+0.02}$& $0.25_{-0.01}^{+0.02}$& $0.24_{-0.24}^{+0.25}$&0.11 
&1.89\\ 
\bottomrule
\end{tabular}}

\end{center}
\textbf{Note.} Column 1: Object name. Column 2: Numbers 1, 2, and 3 represent the first, second, and third segments of the light curves, respectively. “All” represents the whole light curves. The suffix “-D” represents the detrended light curves. Column 3: Number of good data points in the $UVW2$ band. Column $3\sim8$: The ICCF centroid lags and the $68\%$ confidence intervals from the FR/RSS. A positive value means the comparison band lags behind the reference band ($UVW2$). Column 9: Derived fit parameter $\tau_0$ for the lag-wavelength model $\tau=\tau_0[(\lambda/\lambda_0)^{4/3}-1]$. Column 10: Excess ratio of the observed lag relative to the fitted prediction lag at $U$ band.
\end{table*}

\subsubsection{Lag Measurement and Uncertainty Estimation}\label{subsubsec:lags}
The light curves show significant variability and strong correlation across all wavelengths, with prominent features such as peaks and troughs appearing consistently. Therefore, we measure the interband lags using the interpolated cross-correlation function (ICCF) combined with flux randomization and random subset sampling (FR/RSS), as implemented by \cite{2004ApJ...613..682P}.
All these interband lags are measured relative to the $UVW2$ band because it is the shortest UV wavelength band with the highest $F_{var}$, and is closest to the thermal peak of the accretion disk. The cross-correlation function (CCF) is calculated from $-30$ to $30$ days but only shown from $-10$ to $10$ days for clarity. The lags are estimated from the centroid value with CCF values higher than $80\%$ of the peak value. Uncertainties are estimated using the FR/RSS method with $10^3$ light-curve realizations. For each realization, we measure the CCF values and their centroid lags. Finally, it yields a distribution of centroid lags, called the centroid cross-correlation distribution (CCCD). The final centroid lag is the median of the CCCD, and its 1$\sigma$ uncertainty is from the $16\%$ and $84\%$ quantiles. The interband lags for each target are shown in columns 3–8 of Table \ref{tab:lags}.

\subsubsection{Lag-Wavelength Fits}\label{subsubsec:lags-fit}
The lamp-post reprocessing model predicts a lag-wavelength dependence of  $\tau \propto \lambda^{4/3}$ \citep{2007MNRAS.380..669C}. The interband lags obtained from Section \ref{subsubsec:lags} allow us to test this prediction. Therefore, we fit this relation for each target using the following form:
\begin{equation}\label{eq:lag-wave-fit}
	\tau=\tau_0[(\lambda/\lambda_0)^{\alpha}-1],
\end{equation}
where $\lambda_0$ = 1928 Å is the central wavelength of the reference $UVW2$ band, $\alpha=4/3$ is the power-law index for the standard thin disk, and $\tau_0$ is the fitted lag between wavelength zero and $\lambda_0$. The $UVW2$ autocorrelation function lag is identically zero, so the fit is forced to pass through this point. The fitted lags $\tau_0$ are listed in column 9 of Table \ref{tab:lags}, and the best-fitting trends are shown by the black solid lines in Figure \ref{fig:lag_sample}. 

\cite{2016ApJ...821...56F} provide a method to estimate the expected value of $\tau_0$ and describe the lag-wavelength relation for the thin disk model. The equation is given as follows:
\begin{equation}\label{eq:lag-wave-thin-disk}
\tau_{0} = \frac{1}{c} \left( X \frac{k \lambda_0}{h c} \right)^{4/3} \left[ \left( \frac{G M}{8 \pi \sigma} \right) \left( \frac{L_{\text{Edd}}}{\eta c^2} \right) (3 + \kappa) \dot{m}_{\text{Edd}} \right]^{1/3},
\end{equation}
where $X$ is a multiplicative scaling factor for converting the annulus temperature $T$ to wavelength $\lambda$ at a characteristic radius $R$, $\kappa$ is the local ratio of external to internal heating, assumed to be constant with radius, $\eta$ is the radiative efficiency, $L_{Edd}$ is the Eddington luminosity, and $\dot{m}_{Edd}$ is the Eddington ratio. Here, we assume $X = 2.49$, $\kappa=0$ (i.e., negligible external heating compared to internal heating) and $\eta=0.1$, following previous works \citep{2017ApJ...840...41E,2024ApJ...966..149J}. The other parameters for each target are listed in Table \ref{tab:Properties}. The lag-wavelength relations predicted by the thin disk model are shown by the black dashed lines in Figure \ref{fig:lag_sample}.

\subsubsection{Diffuse Continuum Emission from the BLR}\label{subsubsec:DCE_BLR}
\begin{figure*}[t]
	\centering 
	\begin{minipage}[b]{0.32\textwidth}
		\includegraphics[width=\linewidth]{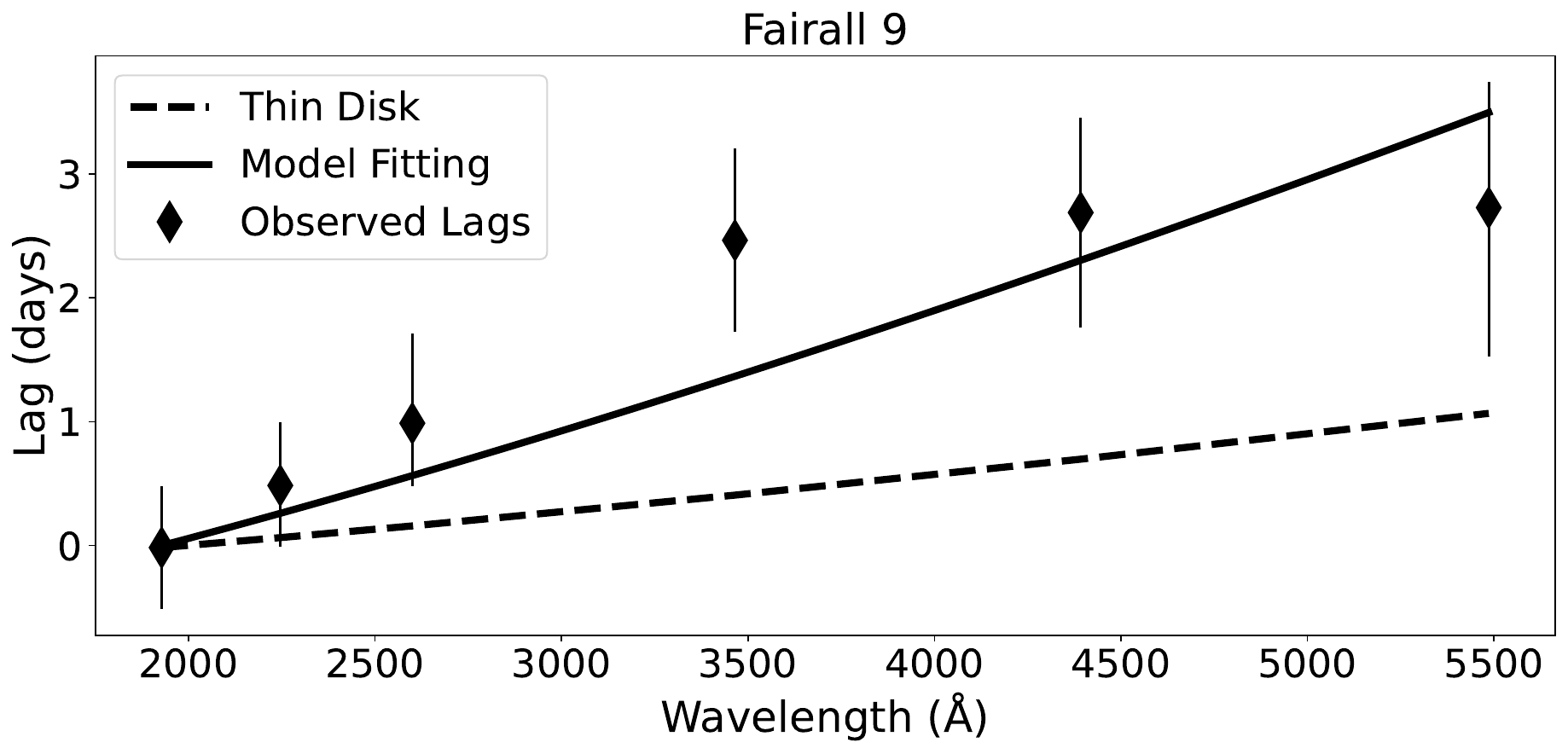}
	\end{minipage}%
	\begin{minipage}[b]{0.32\textwidth}
		\includegraphics[width=\linewidth]{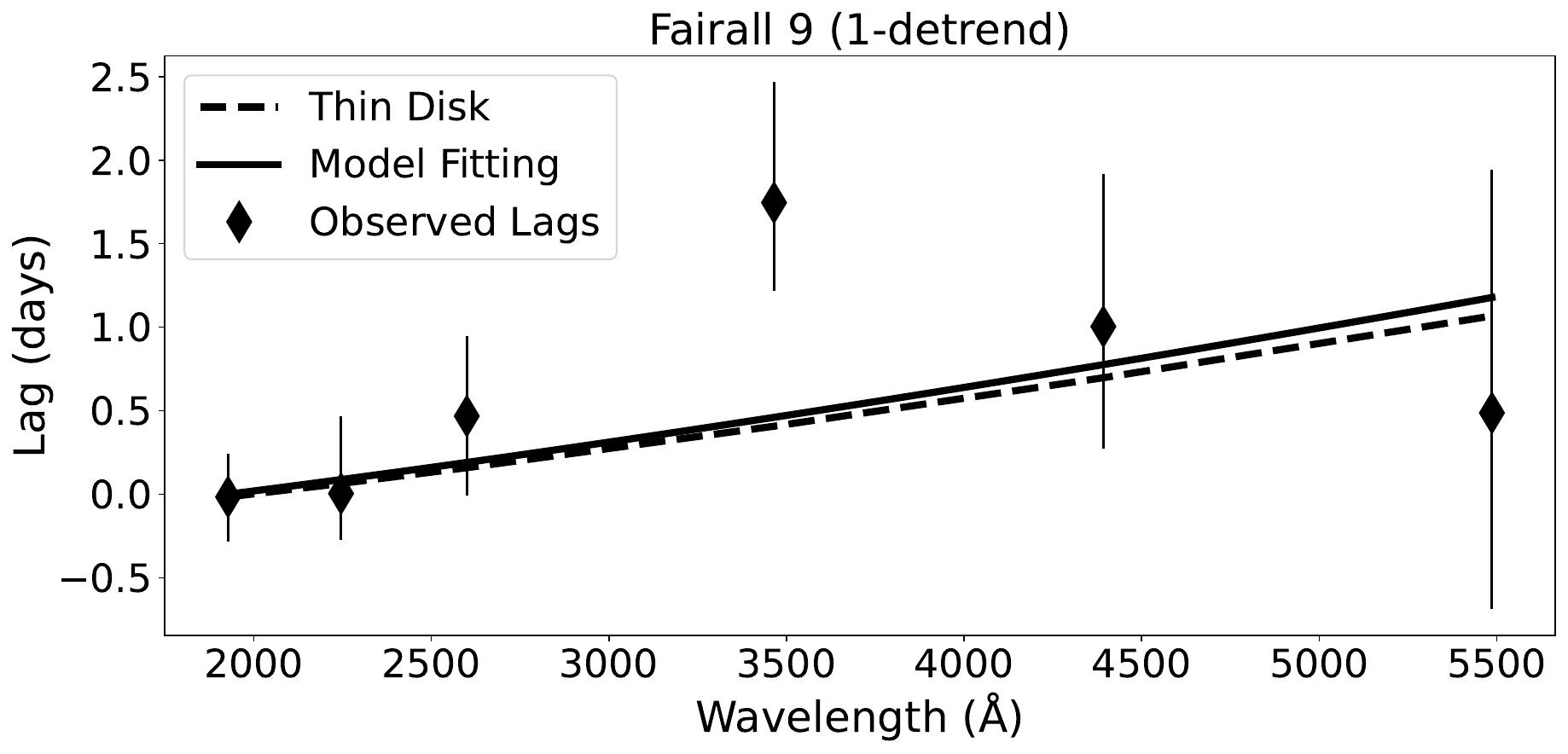}
	\end{minipage}%
	\begin{minipage}[b]{0.32\textwidth}
		\includegraphics[width=\linewidth]{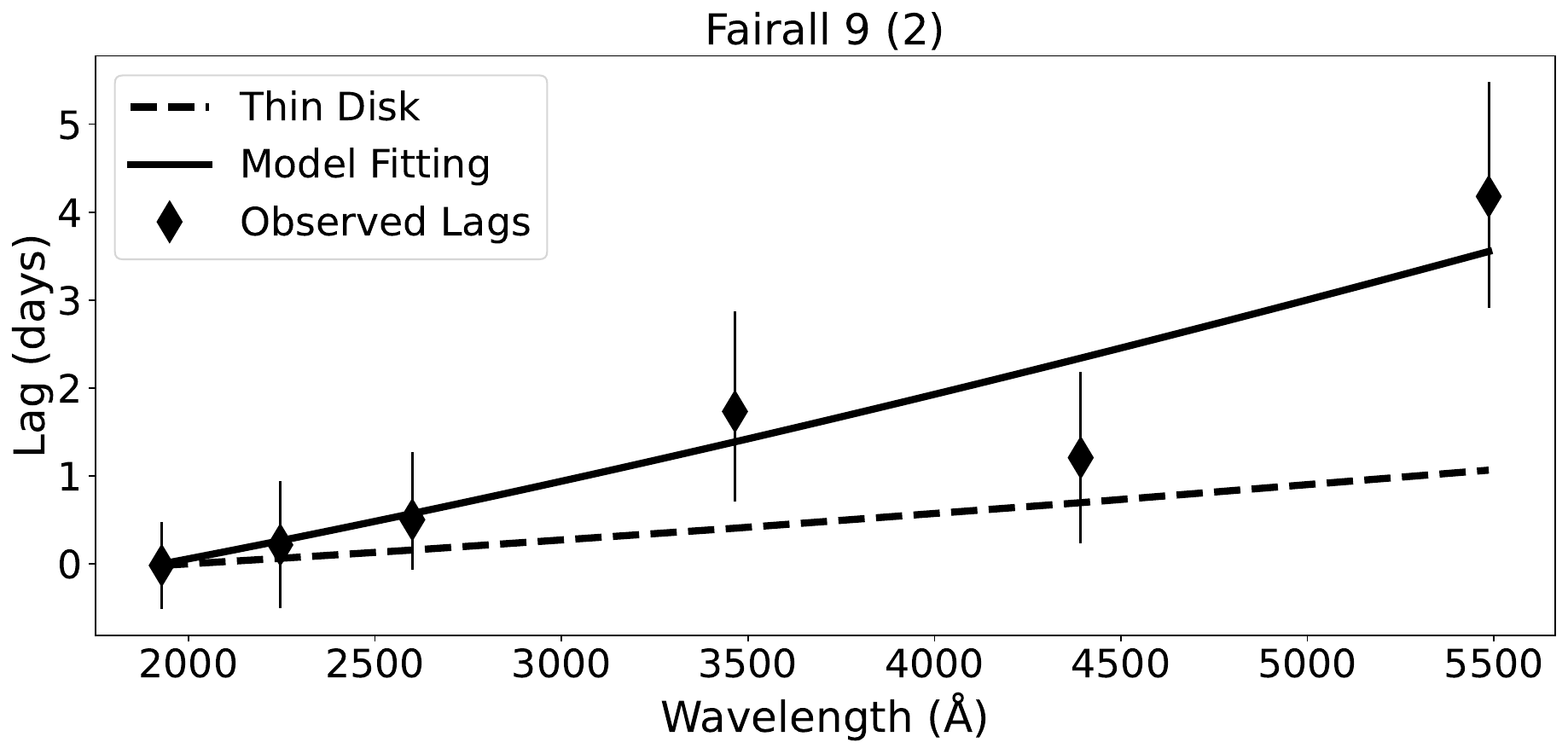}
	\end{minipage}

	\begin{minipage}[b]{0.32\textwidth}
		\includegraphics[width=\linewidth]{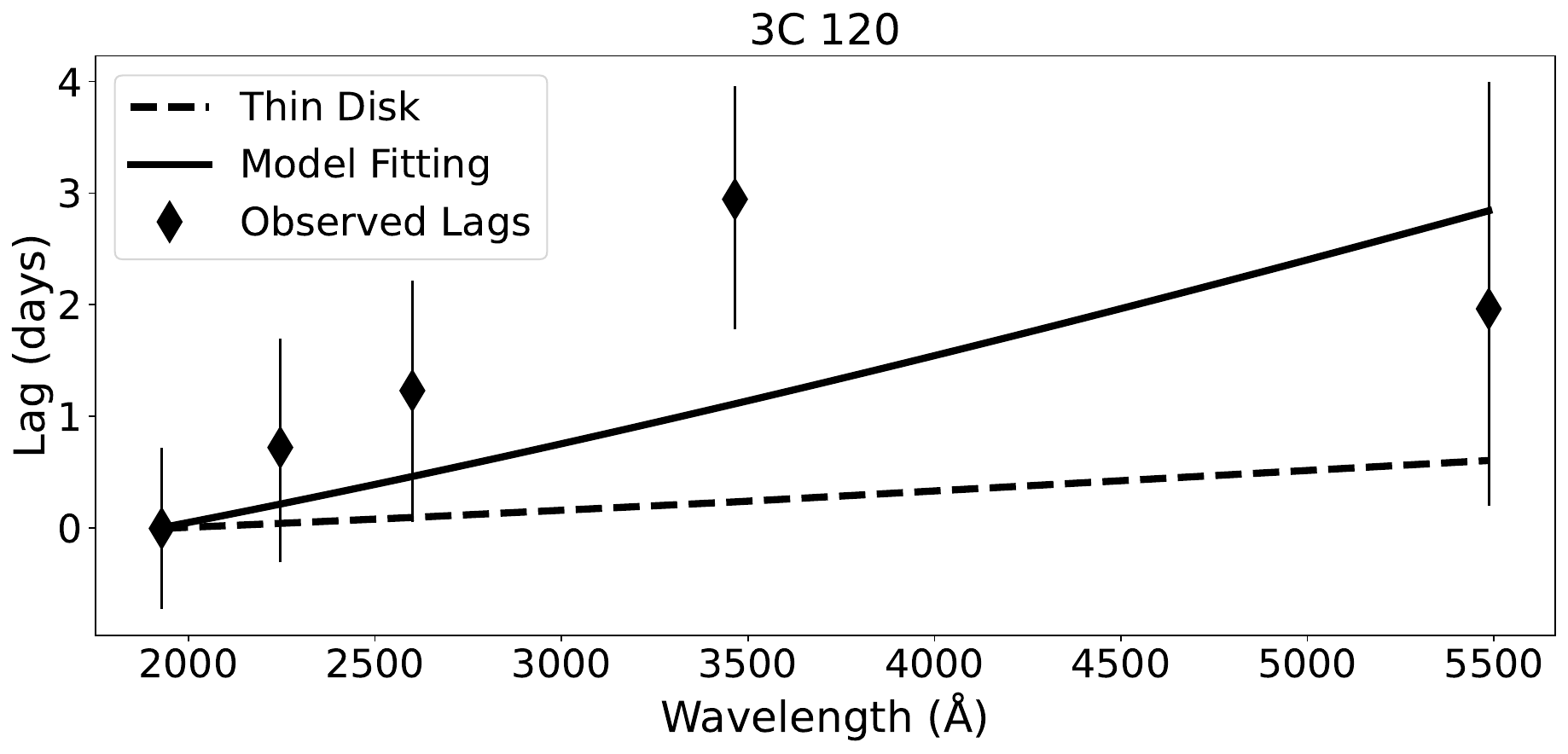}
	\end{minipage}%
	\begin{minipage}[b]{0.32\textwidth}
		\includegraphics[width=\linewidth]{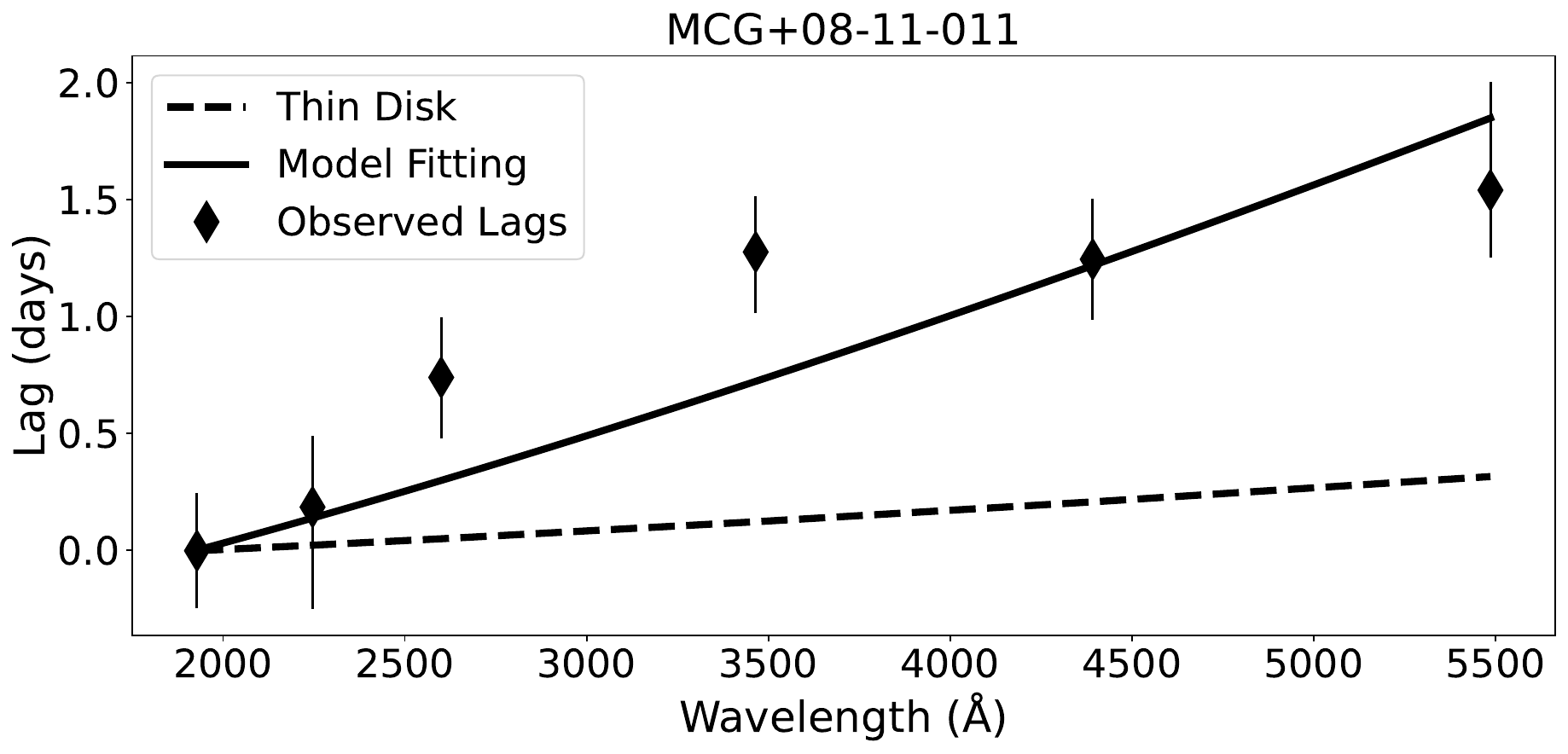}
	\end{minipage}%
	\begin{minipage}[b]{0.32\textwidth}
		\includegraphics[width=\linewidth]{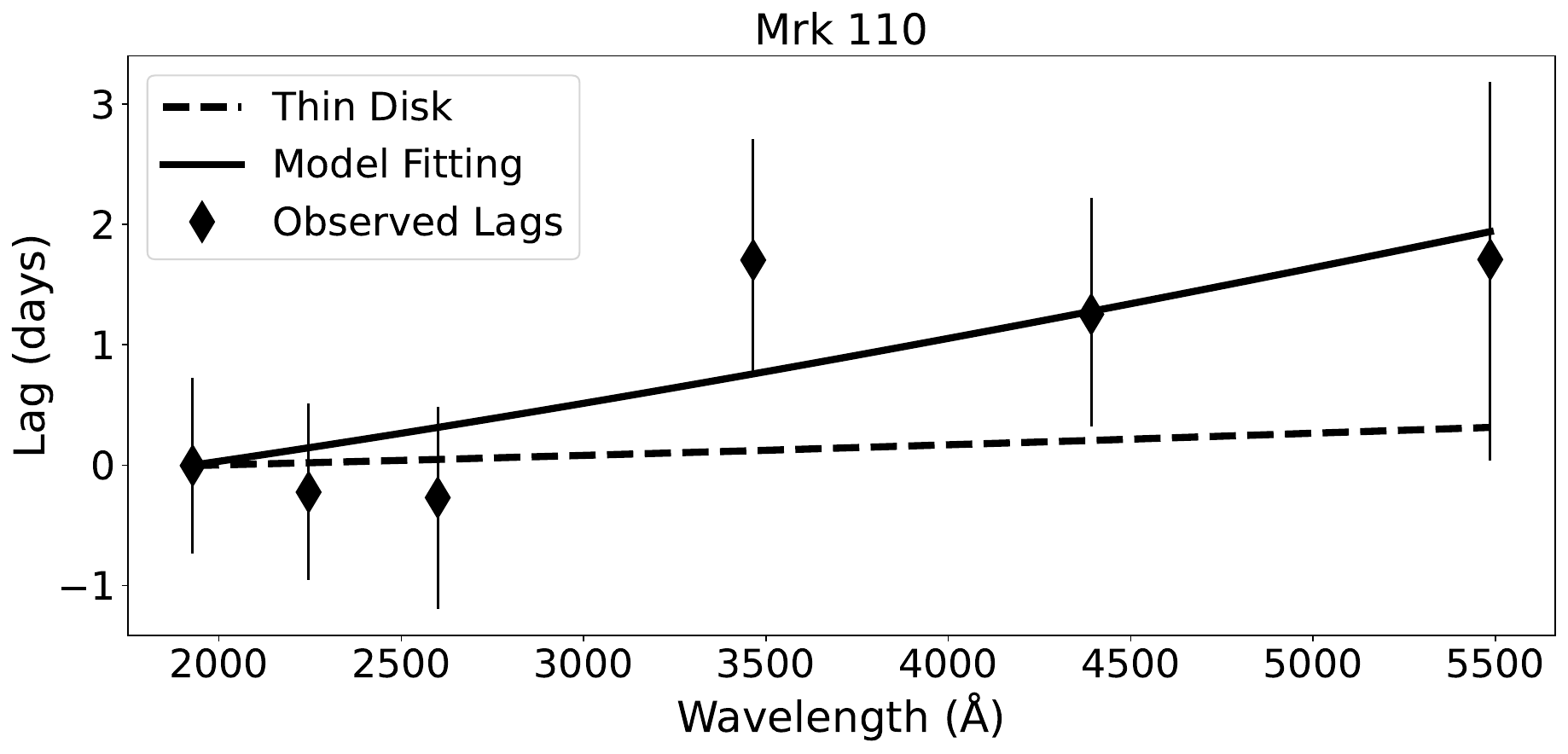}
	\end{minipage}

	\begin{minipage}[b]{0.32\textwidth}
		\includegraphics[width=\linewidth]{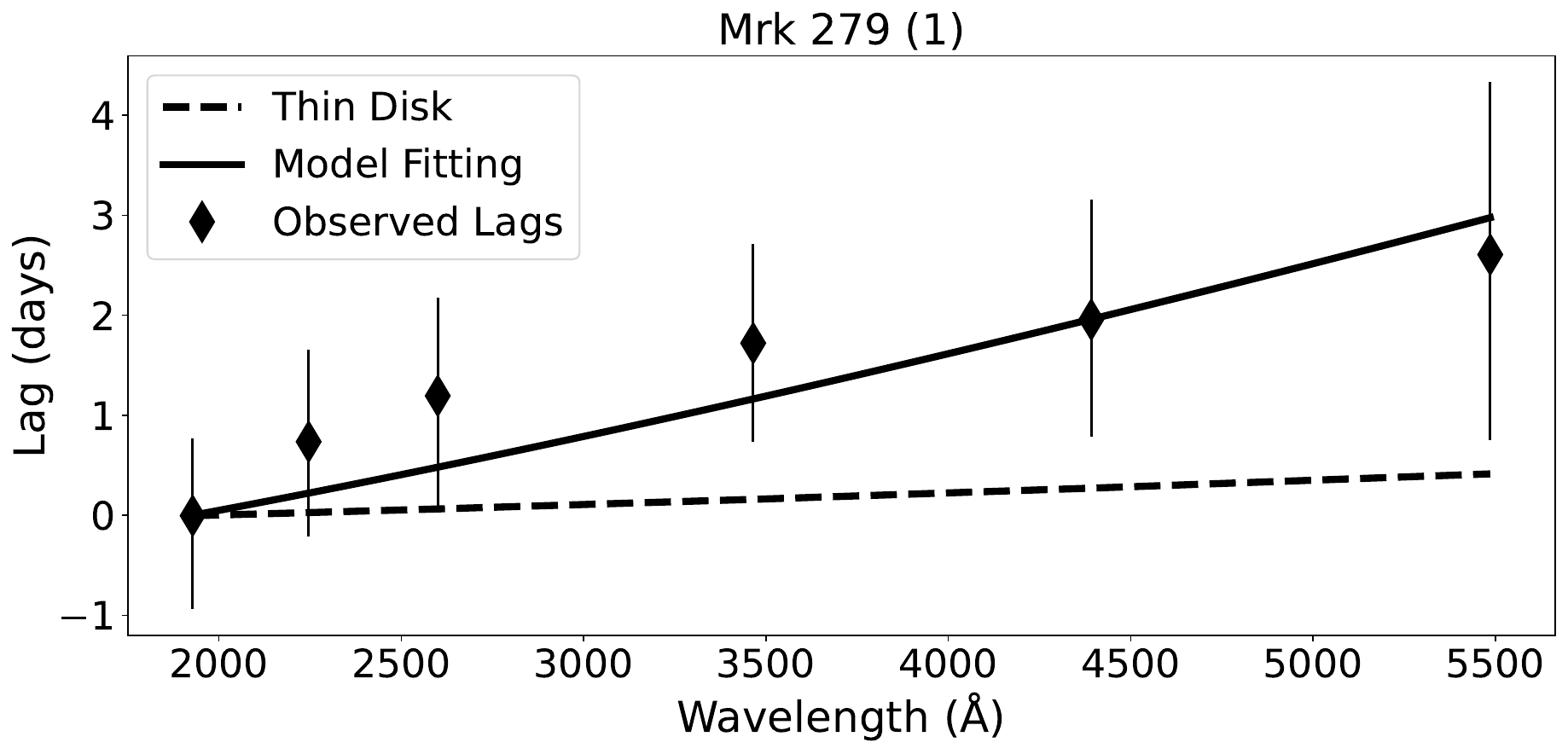}
	\end{minipage}%
	\begin{minipage}[b]{0.32\textwidth}
		\includegraphics[width=\linewidth]{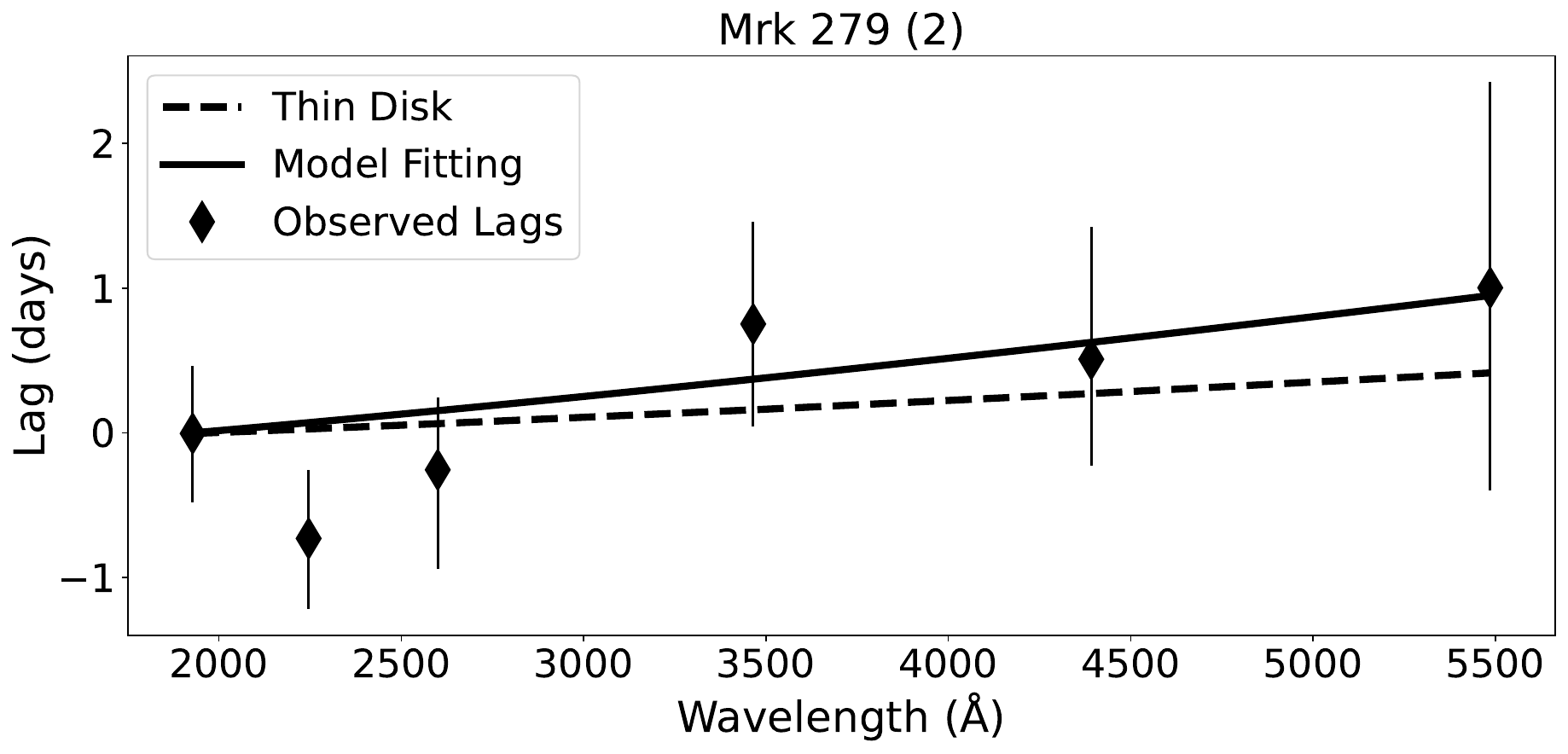}
	\end{minipage}%
	\begin{minipage}[b]{0.32\textwidth}
		\includegraphics[width=\linewidth]{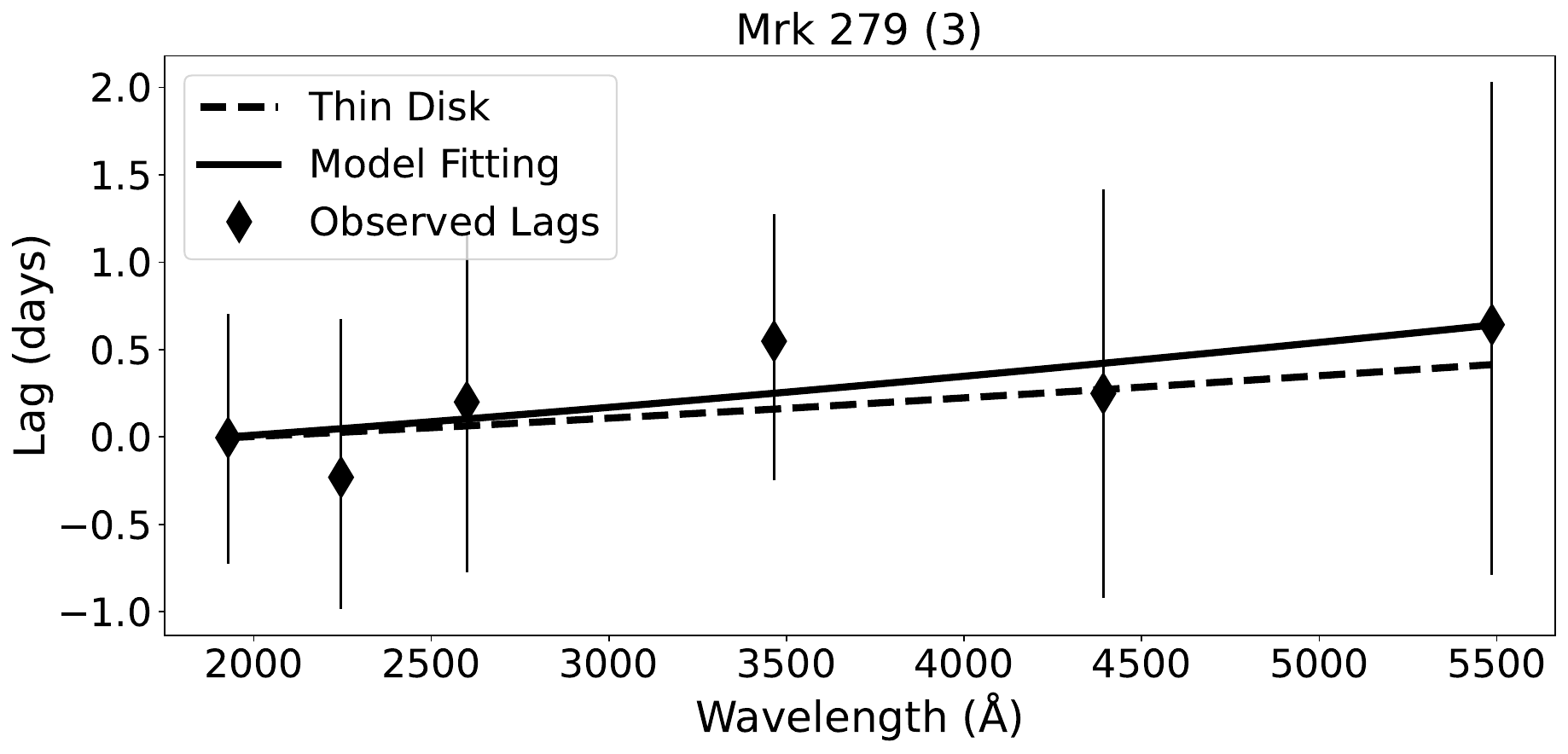}
	\end{minipage}

	\begin{minipage}[b]{0.32\textwidth}
		\includegraphics[width=\linewidth]{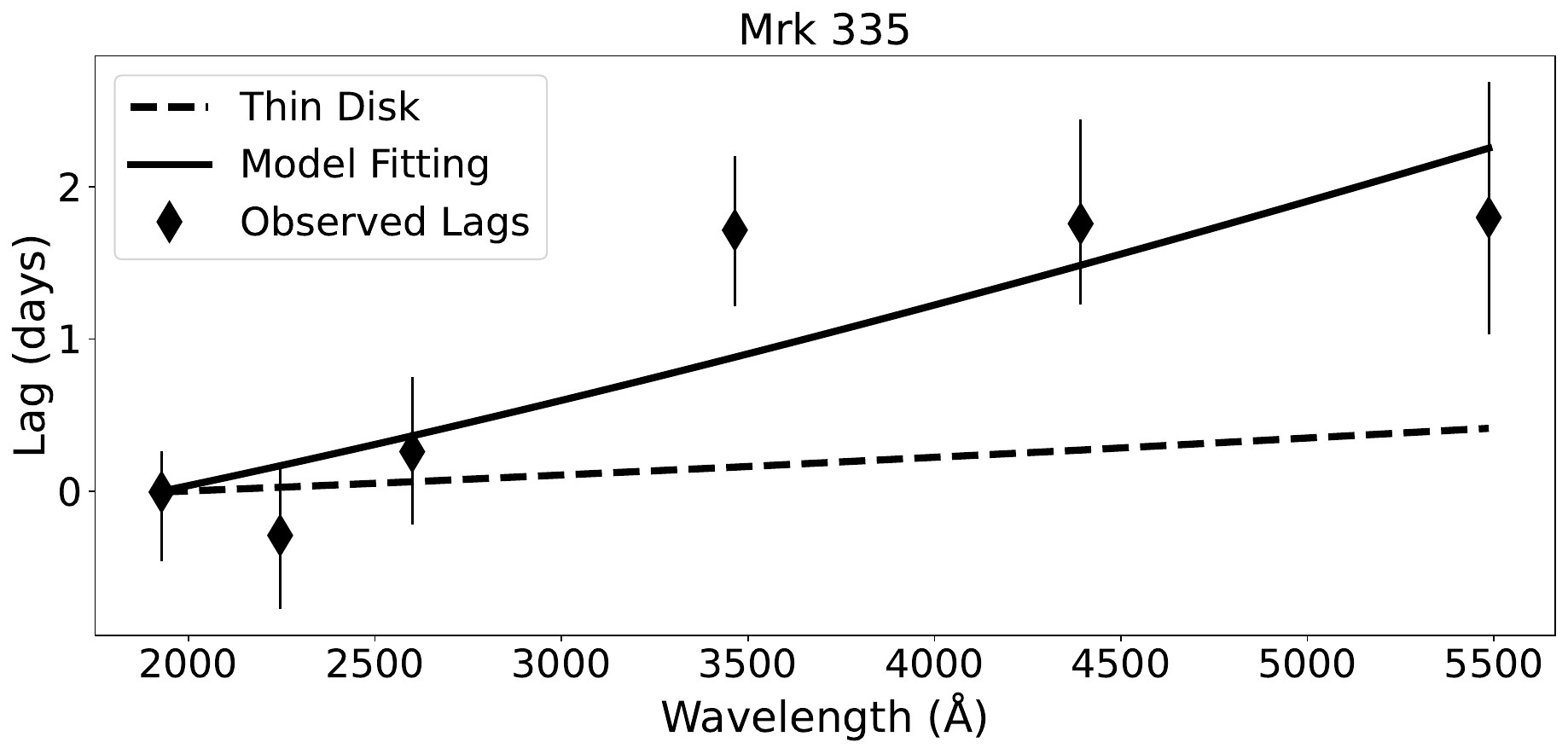}
	\end{minipage}%
        \begin{minipage}[b]{0.32\textwidth}
		\includegraphics[width=\linewidth]{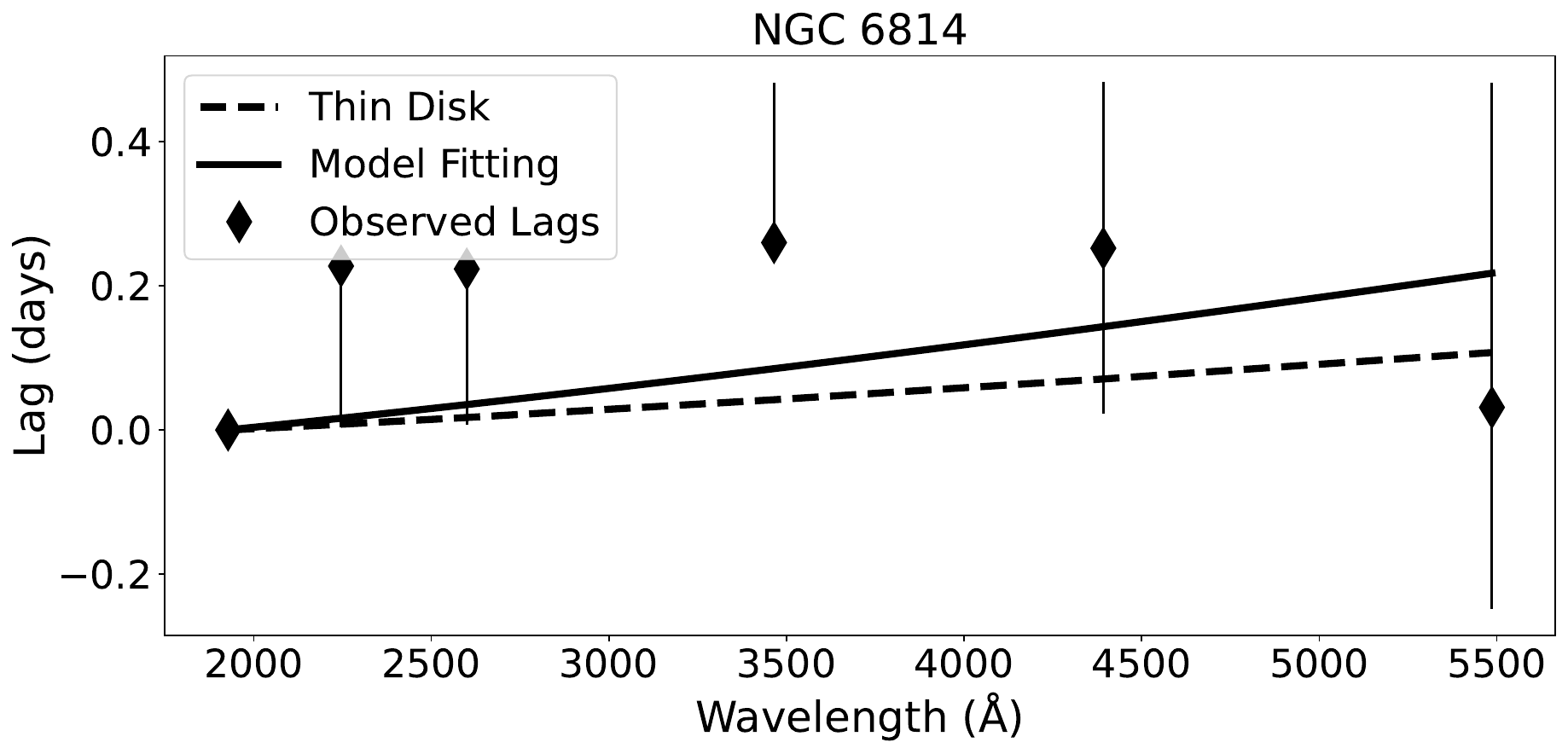}
	\end{minipage}%
	\begin{minipage}[b]{0.32\textwidth}
		\includegraphics[width=\linewidth]{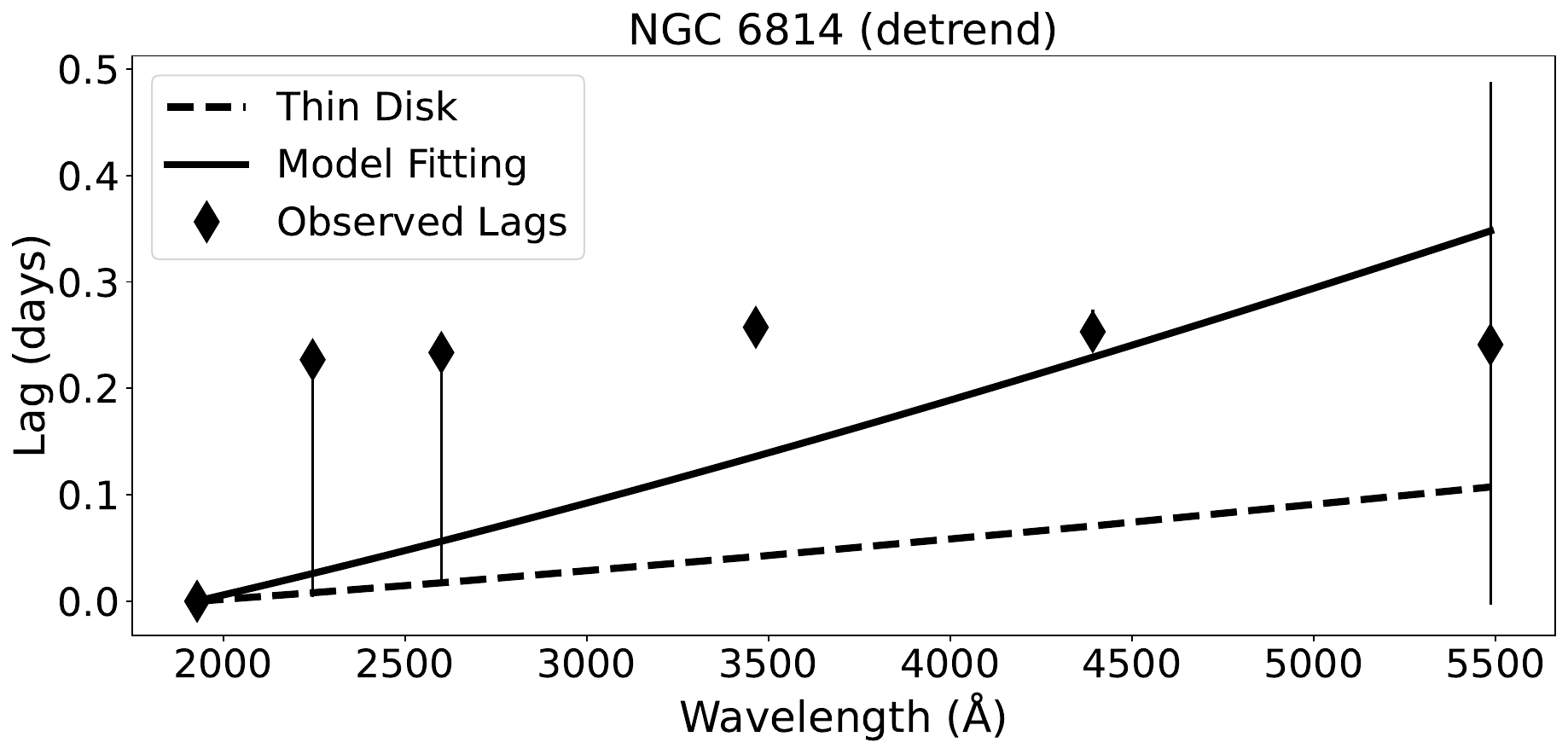}
	\end{minipage}

        \begin{minipage}[b]{0.32\textwidth}
		\includegraphics[width=\linewidth]{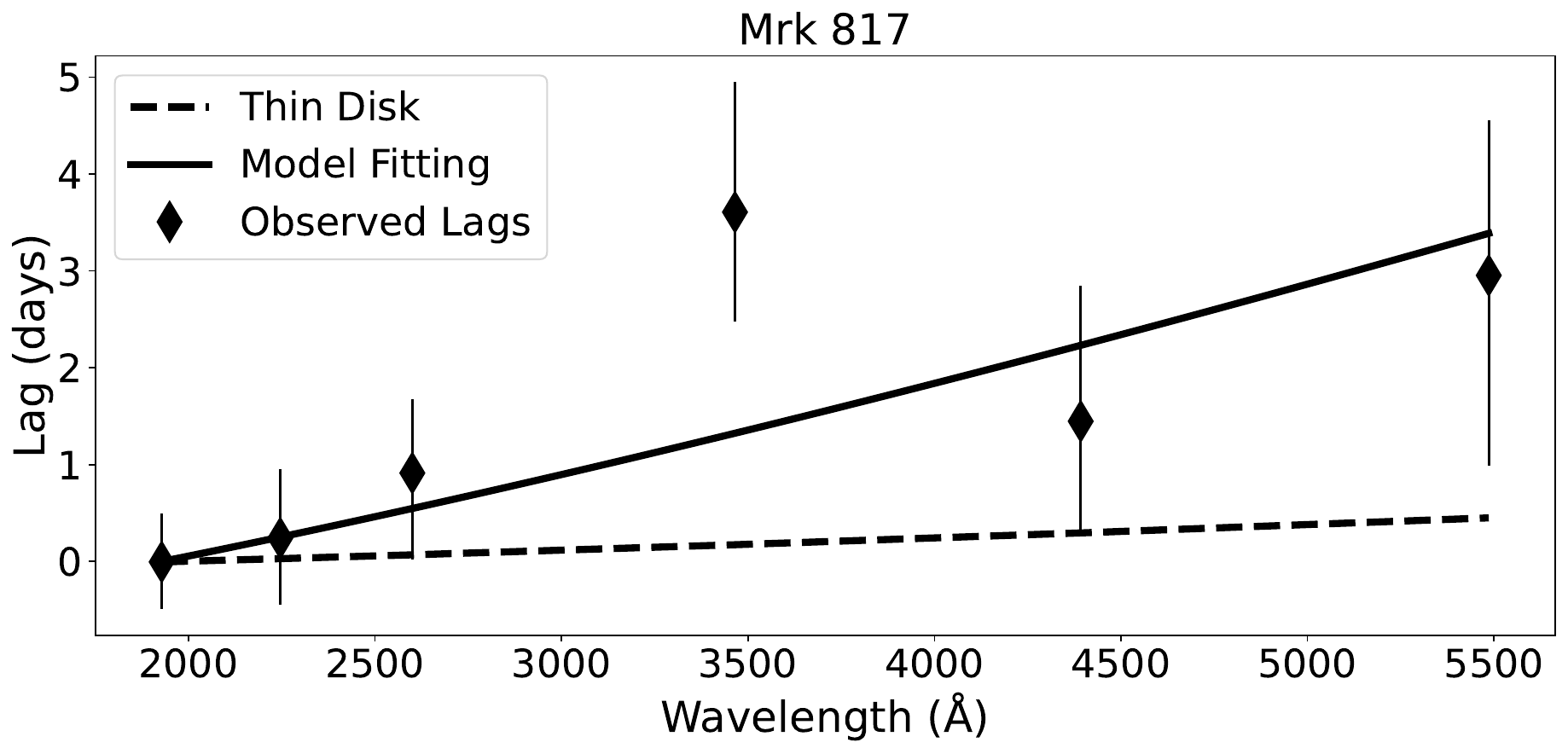}
	\end{minipage}%
	\begin{minipage}[b]{0.32\textwidth}
		\includegraphics[width=\linewidth]{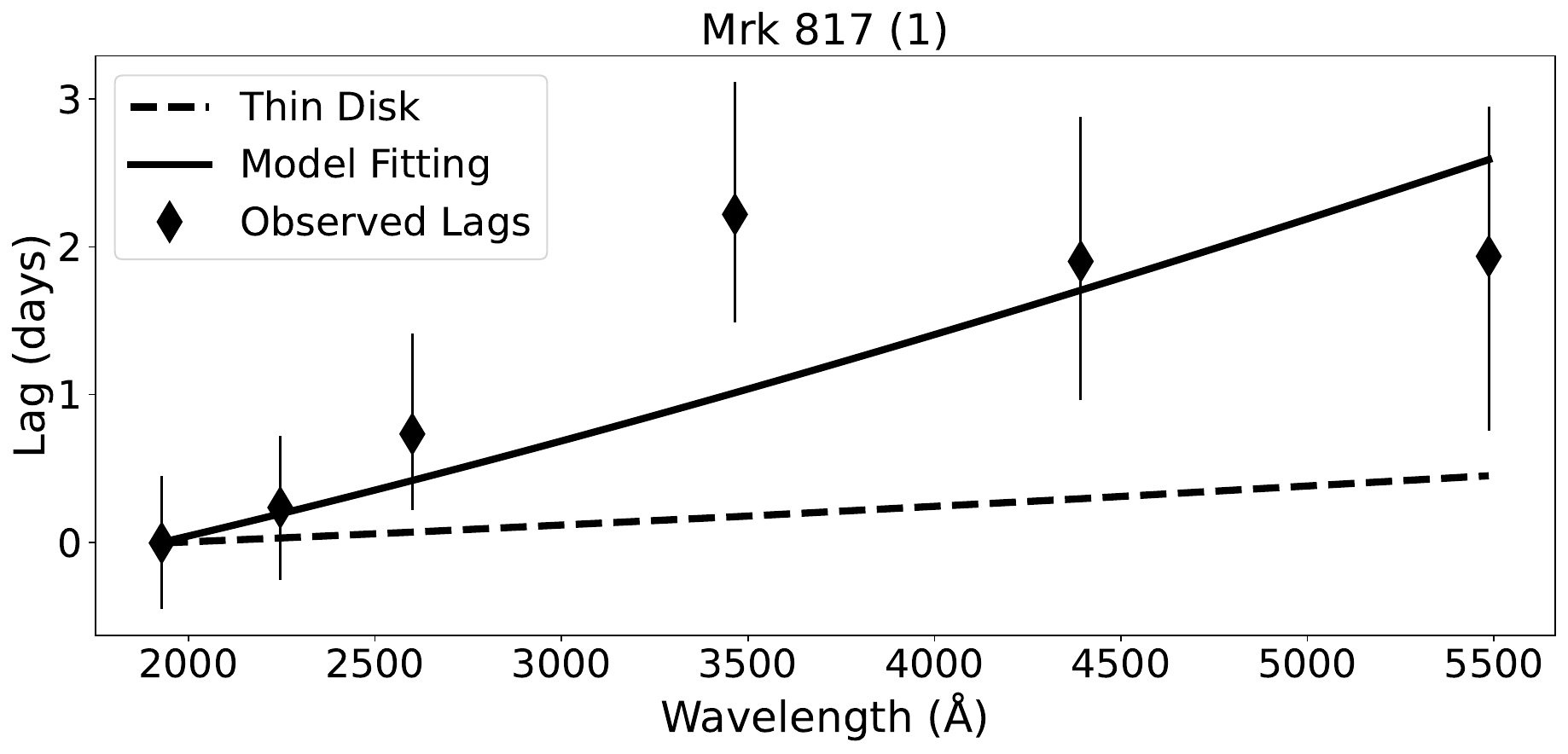}
	\end{minipage}%
        \begin{minipage}[b]{0.32\textwidth}
		\includegraphics[width=\linewidth]{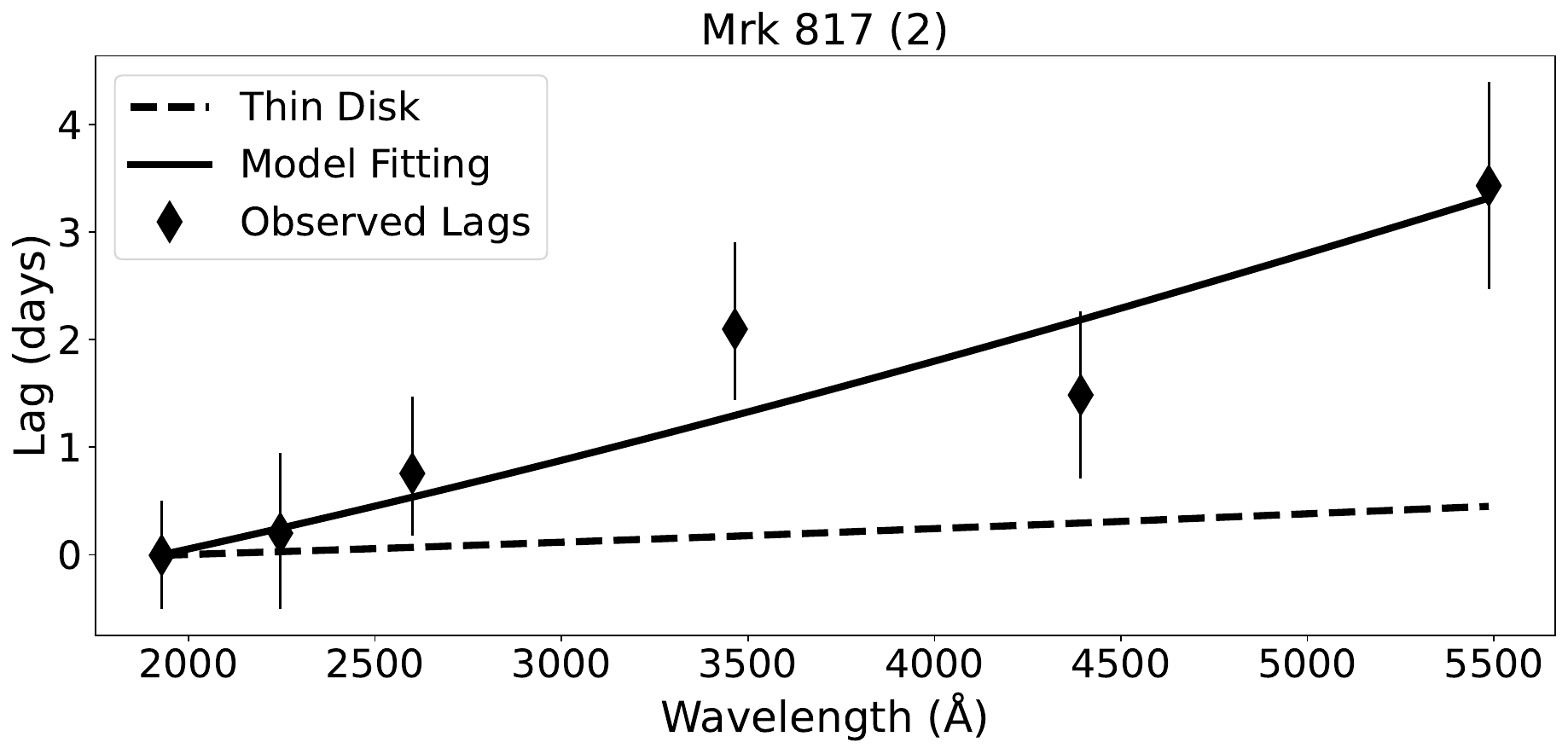}
	\end{minipage}
	\caption{Plots of measured median ICCF centroid lags (black circles) as a function of wavelength for each target. All lags are measured with respect to the $UVW2$ band. The best fits of $\tau_0$ with $\alpha=4/3, \lambda=1928$ Å to the function $\tau=\tau_0[(\lambda/\lambda_0)^{\alpha}-1]$ are shown by the black solid lines. The predictions from the standard thin disk model (Equation \ref{eq:lag-wave-thin-disk}) are shown by the black dashed lines.}
	\label{fig:lag_sample}
\end{figure*}

In Figure \ref{fig:lag_sample}, the relationship of $\tau$ increasing with $\lambda$ holds for most targets. However, the best-fitting trends (solid lines) exceed those predicted by the thin disk model (dashed lines). The larger observed lags indicate that the disk sizes derived from CRM are larger than the standard thin disk model predictions. This ‘\textit{too-big disk}’ problem has been noted in previous campaigns \citep{2014ApJ...788...48S,2015ApJ...806..129E,2016ApJ...821...56F,2017ApJ...836..186J,2018MNRAS.480.2881M,2020ApJ...896....1C,2023ApJ...958..195C}. In addition, we also notice the significant excess lags in the $U$ band compared to both the $\tau \propto \lambda^{4/3}$ and the accretion disk fit, as well as the surrounding $UVW1$ and $B$ band lags. To quantify the magnitude of this effect, we calculate the $U$ band excess ratio $R_u$, defined as the ratio of the observed lag to the best-fitting prediction at the $U$ band. The results are listed in column 10 of Table \ref{tab:lags}. It shows that on average the excess is a factor of $2.2$ larger than that expected from the best-fitting lag-wavelength relation, with values ranging from $1.3$ to $3.8$. The possible explanation is that the continuum emission from the BLR contributes significantly to the measured fluxes in the UV/optical continuum windows. This emission has a significant discontinuity at the Balmer jump (3646 Å), which leads to an increase in the lags, particularly around that wavelength \citep{2001ApJ...553..695K}. Although the above analysis demonstrates that both the BLR continuum and the disk component contribute to the UV/optical interband lags of these AGN, detailed BLR modeling is required to elucidate the precise contributions of each component \citep{2018MNRAS.481..533L,2019MNRAS.489.5284K,2020MNRAS.494.1611N,2022MNRAS.509.2637N}. We will apply the ICCF-Cut method to isolate the diffuse continuum component from the $U$ band light curves in the next section.

\subsection{ICCF-Cut Method for CRM}\label{subsec:ICCF-Cut}
The ICCF-Cut method, proposed by \cite{2023ApJ...949...22M}, provides a novel technique to measure emission-line lags using broadband photometric light curves. This method is based on the idea of removing the continuum emission from the “line” band, which is a band containing both continuum and significant emission-line contributions. We define the band with negligible emission-line contamination as the continuum band. Utilizing the single-epoch spectra, we estimate the continuum flux level in the “line” band and subtract the continuum component. The resulting “cut” light curve roughly contains only line emission. Finally, we employ the ICCF to calculate the lag between the emission line and the continuum band. More details can be found in \cite{2023ApJ...949...22M,2024ApJ...966....5M} for a comprehensive understanding of this method. Also, the code for the ICCF-Cut method is available at \href{https://github.com/PhotoRM/ICCF-Cut}{https://github.com/PhotoRM/ICCF-Cut.} 

\cite{2024ApJ...966..149J} applied the ICCF-Cut method in CRM for the first time. Different from the original design for isolating emission-line components from broadband photometric light curves, the method was modified to extract an outer diffuse continuum component embedded in a specific band light curve. Consequently, we will replace the light curve in the “continuum band” with one that has a pure disk component and decompose the embedded light curve using the diffuse continuum model. This allows us to examine whether the extracted light curves for a possible diffuse continuum correlate well with the accretion disk light curves and exhibit larger lags than the previously observed continuum lags.

\subsubsection{Disk Component}\label{subsubsec:disk}
To isolate the outer component, we should first remove the disk component contribution from the light curve. Following our previous work \citep{2024ApJ...966..149J}, we utilize the BLR model described by \cite{2020MNRAS.494.1611N,2022MNRAS.509.2637N} and decompose the continuum spectrum through CLOUDY simulations \citep{2017RMxAA..53..385F}. Specifically, the BLR cloud structure is determined by the radiation pressure of the central source, which is called a radiation-pressure-confined (RPC) cloud \citep{2002ApJ...572..753D,2014MNRAS.438..604B,2016ApJ...819..130S}. The SED used here is the AD1 described in \cite{2020MNRAS.494.1611N}, which combines a standard thin disk SED with an X-ray power-law continuum. Given the RPC-AD1 cloud model, we can derive the luminosity of the simulated disk and diffuse continuum through CLOUDY. This enables us to estimate the fraction of the diffuse continuum $p_{dc}$ at different wavelengths. In principle, $p_{dc}$ may vary with time, but current observations and modeling techniques cannot provide reliable time-resolved constraints on this parameter. Therefore, following previous works \citep{2020MNRAS.494.1611N,2022MNRAS.509.2637N}, we adopt a constant value of $p_{dc}$ predicted by the CLOUDY simulations.

The most important parameter of the model, which determines the luminosity and lag, is the distance-dependent covering factor, $c_f$(r). Following \cite{2022MNRAS.509.2637N}, we adopt the average $c_f$ = 0.2 for all targets in our sample because the lag spectra predicted by it are similar to the observational lags in general. In addition, \cite{2023ApJ...948L..23W} found 
that the BLR size is about 8.1 larger than the continuum emission size at 5100 Å, using a sample of 21 AGNs. This result is consistent with the model prediction assuming $c_f$ = 0.2, which further reinforces the appropriateness of this covering factor for our analysis. 

\begin{figure}
    \centering
    \includegraphics[width=1\linewidth]{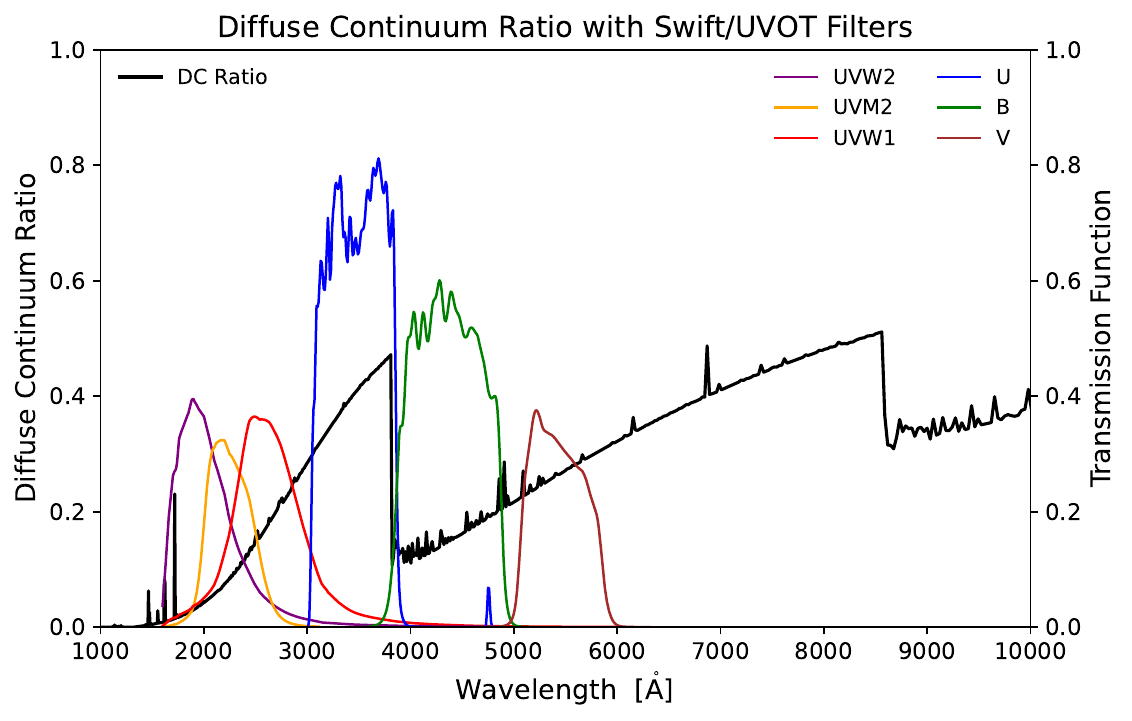}
    \caption{The diffuse continuum ratio spectrum of Fairall 9. The colored curves show the transmission functions of the various Swift/UVOT filters}. The black line shows the wavelength dependencies of the diffuse continuum ratio predicted by the CLOUDY simulation for a standard RPC cloud model.
    \label{fig:Fairall9_ratio}
\end{figure}

Figure \ref{fig:Fairall9_ratio} shows an example of a simulated diffuse continuum ratio spectrum for Fairall 9, and other targets have similar diffuse continuum ratio spectra. These spectra display strong Balmer and Paschen continua, rising towards longer wavelengths, with a significant drop in emission redwards of the Balmer (3648 Å) and Paschen (8204 Å) jumps. The diffuse continuum ratio in the $UVW2$ band is far smaller than that in the $U$ band. Therefore, we assume that the $UVW2$ band can be seen as a pure disk component band compared to other bands at longer wavelengths. Since observational evidence shows that the Balmer continuum contribution is especially prominent in the $u/U$ band, we will focus on deriving an outer component in the Swift $U$ band. Specifically, the $UVW2$ band light curve is the driving light curve, and the $U$ band light curve contains a disk component driven by the $UVW2$ band light curve and an outer component from the diffuse continuum. Thus, these two observational light curves can be written as:
\begin{gather}
    L_{W2}(t) = L_{W2,disk}(t), \label{eq:lcW2} \\
    L_U(t) = L_{U,disk}(t) + L_{U,dc}(t), \label{eq:lcU}
\end{gather}
where $L_{W2}(t)$, $L_{W2,disk}(t)$ represent the total and disk light curves at $UVW2$ band. $L_U(t)$, $L_{U,disk}(t)$, and $L_{U,dc}(t)$ represent the total, disk, and diffuse continuum light curves at $U$ band, respectively. According to the integral of the diffuse continuum ratio spectrum, we can calculate the disk and diffuse continuum proportion in the $U$ band, $p_{dc} = 1- p_{disk}=1-L_{U,disk}/L_U$. The $p_{dc}$ for each target is listed in Table \ref{Table:ICCF-Cut}. To remove the disk component in the $U$ band, we define a simple scaling parameter $\alpha$ to convert the disk flux from the $UVW2$ band to the $U$ band. According to the thin disk model described by Equation (\ref{eq:lag-wave-thin-disk}), there is a small inter-continuum lag $\tau_{disk}$ between the continuum in the $UVW2$ and $U$ bands. So we should shift the $UVW2$ light curves by a lag $\tau_{disk}$ to transfer the disk continuum. Then the scaling parameter $\alpha$ can be derived as 
\begin{equation}\label{eq:alpha}
\begin{aligned}
    \alpha &= \frac{L_{U,disk}(t)}{L_{W2,disk}(t - \tau_{disk})} \\
    &= \frac{L_{U,disk}(t)}{L_{U}(t)} \cdot \frac{L_{U}(t)}{L_{W2}(t - \tau_{disk})} \\
    &= (1 - p_{dc}) \times \text{Median}\left[ \frac{L_{U}(t)}{L_{W2}(t - \tau_{disk})} \right],
\end{aligned}
\end{equation}
where the first term is measured from the diffuse continuum ratio spectrum, and the second term is directly decided by the $UVW2$ and $U$ band light curves. Finally, according to the Equations (\ref{eq:lcW2}), (\ref{eq:lcU}) and (\ref{eq:alpha}), we can derive the diffuse continuum light curve in the $U$ band:
\begin{equation}\label{eq:lcdc}
\begin{aligned}
    L_{U,dc}(t)  &= L_U(t) - \alpha L_{W2,disk}(t - \tau_{disk}) \\
    &\approx L_U(t) - \alpha L_{W2}(t - \tau_{disk}).
\end{aligned}
\end{equation}

\begin{figure}
    \centering
    \includegraphics[width=1\linewidth]{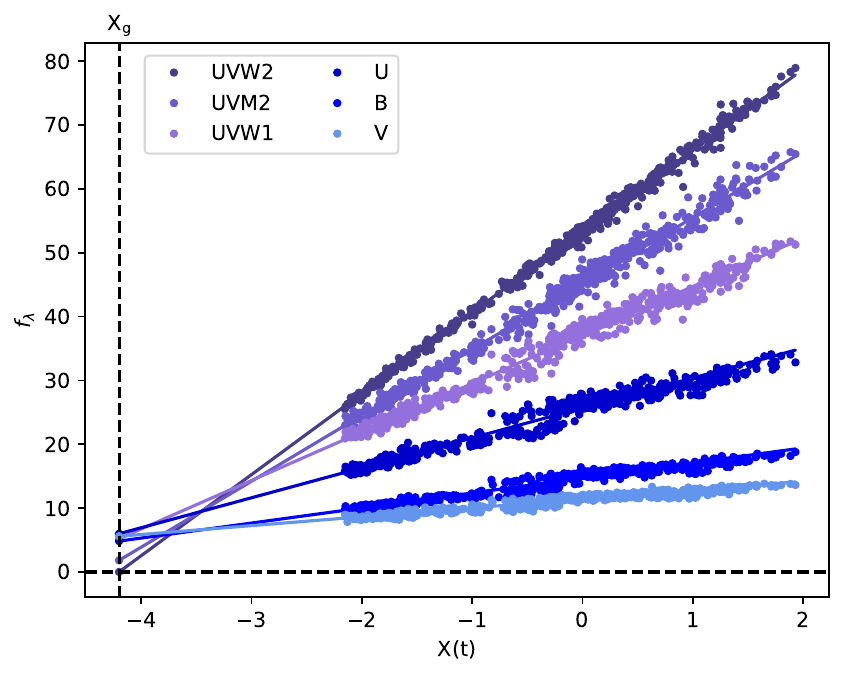}
    \caption{The flux–flux diagram of Swift UVOT light curves of Fairall 9. Best-fitting relations are shown as solid lines. $X_g$ indicates the value of $X(t)$ where $f_\lambda=0$ for the $UVW2$ band. Flux–flux relations for other bands evaluated at $X_g$ give the underlying galaxy contribution. Here, we only care about the $U$ band.}
    \label{fig:Fairall9-flux-flux}
\end{figure}

\subsubsection{Host Galaxy Component}\label{subsubsec:host}
For most AGNs, the host galaxy can contribute a constant flux to the total flux across multiple wavelength ranges. Although it has little influence on the intrinsic variation of the light curves and the lag measurements in the CRM, we cannot ignore the host-galaxy contamination in our scheme. As mentioned above, the diffuse continuum ratio given by CLOUDY is based on the pure AGN contribution without the host-galaxy contamination. When we estimate the scaling factor $\alpha$, we need to use the ratio of two fluxes after the host galaxy component has been excluded. Therefore, we perform the flux-flux analysis by breaking the flux into constant and variable components representing the host galaxy and the AGN, respectively \citep{2007MNRAS.380..669C,2020ApJ...896....1C,2017ApJ...835...65S,2018MNRAS.480.2881M,2023ApJ...953..137M,2023ApJ...958..195C}. We fit the light curves using the following linear model:
\begin{equation}\label{eq:flux-flux}
    f_{\lambda}(\lambda, t) = A_{\lambda}(\lambda) + R_{\lambda}(\lambda)X(t),
\end{equation}
where $A_{\lambda}$ is the average spectrum, $R_{\lambda}$ is the rms spectrum, and $X(t)$ is a dimensionless light curve normalized to a mean of zero and a standard deviation of unity. To estimate the host-galaxy contribution to the different bands, we extrapolate the fits to where the uncertainty envelope of the shortest wavelength $UVW2$ band crosses $f_{\lambda}= 0$, which we define as $X(t) = X_g$. This serves as a reference point for the other bands, and determining $f_{\lambda}= 0$ at $X(t) = X_g$ provides a lower limit on the constant component in each band. In Figure \ref{fig:Fairall9-flux-flux}, we show the flux–flux relations for Fairall 9 and see a good linear response for all the bands. Other targets also have the similar flux–flux relations so that we can remove the host-galaxy contamination. Compared to the image-subtraction method, previous work found that the flux–flux analysis may overestimate the host flux \citep{2024Univ...10..282C}. More precisely, the flux derived from the flux–flux analysis corresponds to the constant component in the light curves, which may include not only the host galaxy emission but also other non-variable (or slowly varying) components\citep{2023ApJ...958..195C}. From this perspective, adopting the flux obtained from the flux–flux analysis in the ICCF-Cut procedure is reasonable, since our goal is to isolate the variable AGN component in the $U$ band light curves. This impact on the lag measurement will be discussed in Section \ref{subsec:Inconsistent-lags}.

\subsubsection{ICCF and CCCD}
Following the strategy mentioned in Section \ref{subsubsec:disk}, we can derive the diffuse continuum light curve in the $U$ band for all targets in our sample. For convenience, we refer to the $UVW2$ band light curve for the disk component as the “driving” light curve, and the derived light curve for the diffuse continuum component as the “cut” light curve. Similar to the previous time series analysis in Section \ref{subsec:time-series-analysis}, we use the ICCF method to measure the lags $\tau_{cut}$ between the driving and cut light curves. \cite{2020MNRAS.494.1611N} predicted that the diffuse continuum lag is about $\tau_{dc}\simeq0.5\tau_{H\beta}$. So we set the lag search range from $-3\tau_{dc}$ to $3\tau_{dc}$ with a grid of 0.5 days. If the $3\tau_{dc}$ is less than 15 days, we will increase the lag search range to $\pm15$ days to avoid missing longer lags. To estimate the lag errors, we use the Monte Carlo simulations to perform the FR/RSS, wherein the CCCD is built from cross-correlating $10^3$ realizations of both light curves.

\subsection{JAVELIN Pmap Model}\label{subsec:Javelin}
We also employ a parallel technique, JAVELIN, to estimate the lag as a comparison. It assumes that the driving light curve is well-modeled by a Damped Random Walk (DRW), and the other light curves are related to it via a transfer function \citep{2011ApJ...735...80Z,2013ApJ...765..106Z,2016ApJ...819..122Z}. We can calculate the maximum likelihood to determine the lags and relevant parameters by using Markov Chain Monte Carlo (MCMC) methods. Here, we use the JAVELIN Pmap Model \citep{2016ApJ...819..122Z}, which can separate the continuum component and the strong emission-line component blended in the “line” band. The Pmap Model assumes that the light curves in the continuum band and “line” band are $f_c=c(t)+u_c$ and $f_l=\alpha\cdot c(t)+l(t)+u_l$, where $\alpha$ is the ratio between the continuum variabilities in two bands. $u_c$ and  $u_l$ are constants representing any contaminating flux from the narrow emission lines and the host galaxy. $c(t)$ and $l(t)$ represent the variability of the continuum and emission line in the corresponding band. The continuum variability and line variability can be related by a top-hat transfer function centered on the time lag $\tau$ with width $\omega$ and amplitude $A$: 
\begin{gather}
l(t) = \int \Psi(t - t') c(t) \, dt', \\
\Psi(t) = \frac{A}{\omega} \quad \text{for} \quad \tau - \frac{\omega}{2} \leq t \leq \tau + \frac{\omega}{2}.
\end{gather}

Similar to the model used in the ICCF-Cut method, we take the $UVW2$ band as the continuum band and the $U$ band as the “line” band containing the diffuse continuum. Then, $c(t)$ represents the pure disk variability in the $UVW2$ band, $l(t)$ represents the diffuse continuum variability in the $U$ band, and $\alpha$ transfers the disk variability between these two bands. The main difference is that the Pmap Model transfers the continuum emission in different bands without considering interband lag. To remain consistent with the ICCF-Cut method, we add the standard thin disk model lag to the disk continuum in the $UVW2$ band. For the JAVELIN run, we take the same lag search range from  $-3\tau_{dc}$ to $3\tau_{dc}$. If $3\tau_{dc} < 15$ or a significantly larger lag appears, this range is extended to $\pm15$ days or $\pm4\sim5\tau_{dc}$. We use the MCMC parameters $n_{chain}=n_{walkers}=n_{burn} = 200$ to provide enough sampling of the posterior over the lag search range. Finally, the estimations of lag $\tau_{jav}$ and other parameters are determined by the median and 68\% confidence intervals of posterior distributions. 

\begin{figure*}[t]
    \centering
    \includegraphics[width=0.9\linewidth]{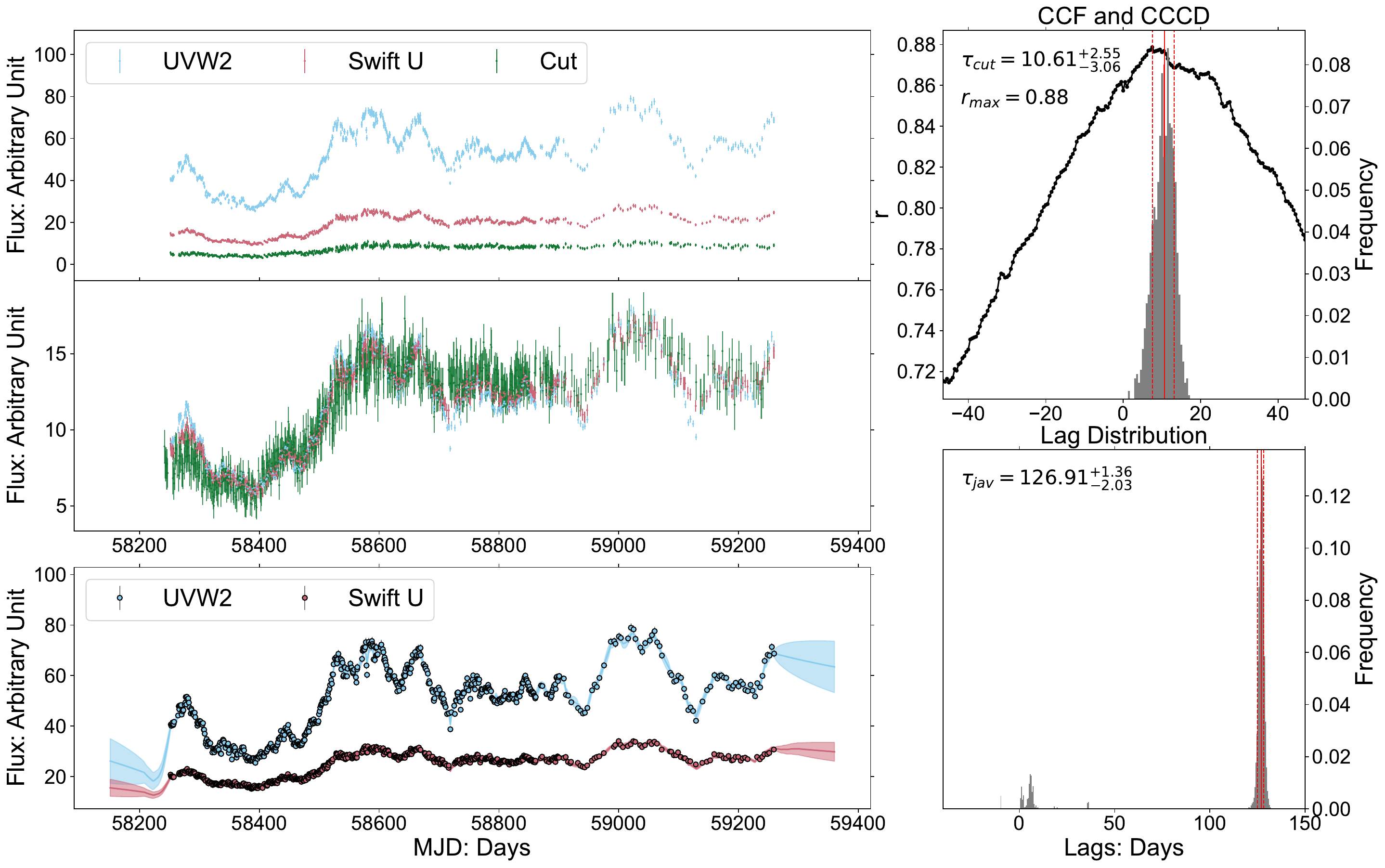}
    \caption{The ICCF-Cut and JAVELIN Pmap Model results for Fairall 9. The top left panel shows the ICCF-Cut light curves, including the observed $UVW2$ (blue), the observed $U$ (red), and the cut (green; i.e., the observed $U$ band light curve minus the predicted disk emission light curve in that band) light curves. All these light curves have subtracted the host-galaxy flux. The middle left panel displays these light curves scaled to the predicted disk flux in the $U$ band and shifted by their respective lags. The bottom left panel shows the observed light curves and the JAVELIN fittings in the $UVW2$ and $U$ bands. The top right panel presents the CCF and CCCD between the $UVW2$ and cut light curves. The bottom right panel shows the posterior distribution of the lags given by the JAVELIN Pmap Model. The masked data in the JAVELIN results are denoted by the light gray. The maximum of the CCF and the lag estimation of the ICCF-Cut method and the JAVELIN Pmap Model, $r_{max}$, $\tau_{cut}$, and $\tau_{jav}$ respectively, can be found in the corresponding panels. All lags presented here are in the rest frame.}
    \label{fig:Fairall9_result}
\end{figure*}

In summary, the Pmap Model provides a parallel method for estimating the diffuse continuum lag without the complicated modeling of the BLR. However, due to many fitting parameters involved, the results of JAVELIN are more reliant on the quality of the light curves, and the distribution of parameters may not converge to give a robust estimation for light curves with short durations and/or large errors. In addition, this method assumes a specific model of AGN variability and a simple transfer function, which may be overly model-dependent and insufficient to fully describe the real situation \citep{2011ApJ...743L..12M,2014ApJ...795....2E,2018MNRAS.480.2881M}. Considering that the ICCF FR/RSS technique has been more extensively tested, we will adopt the ICCF-Cut results as our final results and use the JAVELIN Pmap Model results to evaluate the reliability of the ICCF-Cut results.

\begin{figure*}[t]
    \centering
    \includegraphics[width=0.9\linewidth]{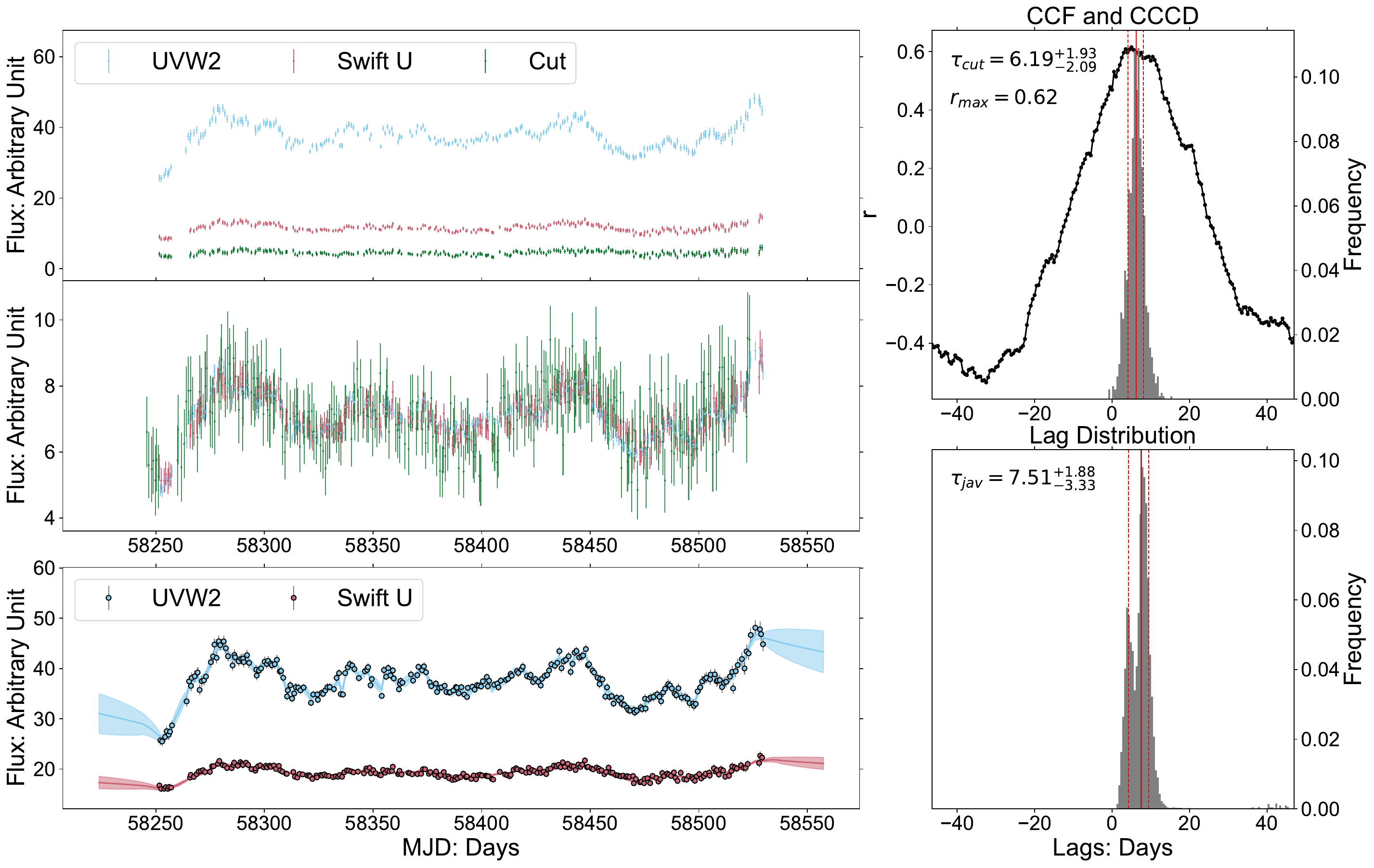}
    \caption{The same as Figure \ref{fig:Fairall9_result}, but for the first part detrended light curves of Fairall 9.}
    \label{fig:Fairall9_1_detrend_result}
\end{figure*}

\subsection{Lag Results}\label{subsec:sample}
\begin{figure*}[t]
    \centering
    \includegraphics[width=0.9\linewidth]{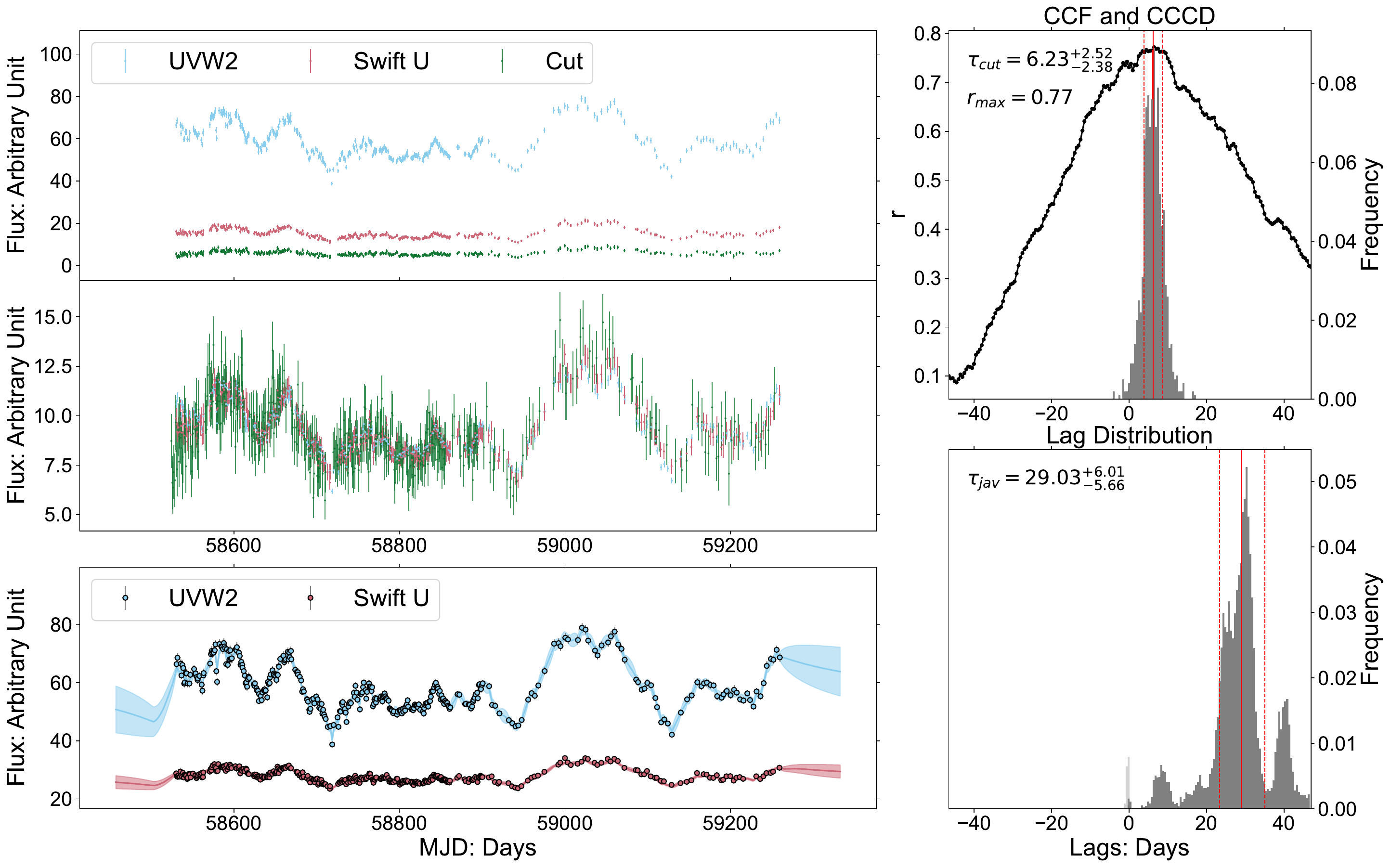}
    \caption{The same as Figure \ref{fig:Fairall9_result}, but for the second part light curves of Fairall 9.}
    \label{fig:Fairall9_2_result}
\end{figure*}

We apply the ICCF-Cut method and the JAVELIN Pmap Model to each target in our sample, with results for Fairall 9 shown in Figures \ref{fig:Fairall9_result}-\ref{fig:Fairall9_2_result} and results for the remaining seven targets presented in Figure Set \ref{FigSetB}. The diffuse continuum lag measurement results are shown in Table \ref{Table:ICCF-Cut}. For the ICCF-Cut results, we find a good correlation and larger lag for each target in our sample. The maximum correlation coefficient $r_{max}$ for each target exceeds 0.6, with 7 out of 8 targets exhibiting $r_{max}$ greater than 0.8. It indicates that the diffuse continuum light curves are highly intrinsically correlated with the driving light curves. By visual examination of the light curves shown in the top left panels of Figures \ref{fig:Fairall9_result}-\ref{fig:Fairall9_2_result} and Figure Set \ref{FigSetB}, it is apparent that the driving, target, and cut light curves show strong consistency after scaling and shifting by the corresponding lags. In addition, we find that the “cut” lags $\tau_{cut}$ are significantly larger than the original lags $\tau_{u}$. The ratios of $\tau_{cut}$ relative to $\tau_{u}$ are listed in column 8 of Table \ref{Table:ICCF-Cut}. The results indicate that the lags increase by approximately a factor of three after cutting. For the JAVELIN Pmap Model results, the lag distributions for most targets show a clear high-significance peak, consistent with the CCCDs given by the ICCF-Cut method. However, the lag distributions occasionally contain lower-significance peaks with unphysical negative lags. Therefore, we will mask them when estimating the final JAVELIN Pmap Model lags. The comparative validation of the lag measurements obtained via the ICCF-Cut method and the JAVELIN Pmap Model indicates that they are generally consistent within the error bars for most targets. However, for Fairall 9 and Mrk 817, where multiple lag measurements are derived from different segments of the light curves, some instances show larger lags estimated by the JAVELIN Pmap Model. The reasons for these inconsistent lags are elucidated in Section \ref{subsec:Inconsistent-lags}. Here we present a detailed description of the diffuse continuum lag measurements for each target in our sample.

\hspace*{1em}\textbf{Fairall 9}: The monitoring of this target is the largest campaign in our sample, with 595 successful visits over more than 1000 days. Firstly, we measure the lag based on the total light curves (see Figure \ref{fig:Fairall9_result}) and find that $\tau_{jav}=126.91_{-3.06}^{+1.36}$ days is too large to be a physical diffuse continuum lag for this target. A similar large $\tau_{jav}$ is found in the lag measurement based on the second segment light curves (see Figure \ref{fig:Fairall9_2_result}). Because the previous CRM studies for the first segment light curves have found a long-term trend with an opposite lag to the reverberation signal \citep{2020MNRAS.498.5399H,2023ApJ...953...43Y}, we perform quadratic detrending operations before measuring the diffuse continuum lag (see Figure \ref{fig:Fairall9_1_detrend_result}). The final result based on the detrended light curves shows consistent lags as $\tau_{cut}=6.19_{-2.09}^{+1.93}$ days and $\tau_{jav}=7.51_{-3.33}^{+1.88}$ days. This result is similar to the predicted Balmer continuum lag $\tau_{dc}=8.70^{+1.60}_{-2.15}$ days. This lag is also generally consistent with the “bowl” model proposed by \cite{2024ApJ...973..152E}, which suggests that the disk thickens at a distance of $\sim10$ light-days and a temperature of $\sim8000$ K, coinciding with the inner edge of the BLR. So we choose it as the final result for this target. As for the large difference $\tau_{jav}$ for the other two measurements, it may be caused by the additional variability features in the long-term trend, which will be discussed in Section \ref{subsec:Inconsistent-lags}.

\hspace*{1em}\textbf{3C 120}: The result for this target is shown in Figure B1. The CCCD given by the ICCF-Cut method exhibits a distinct peak centered at $\tau_{cut}=7.43_{-2.66}^{+2.76}$ days, with a high correlation coefficient $r=0.86$. The results given by the JAVELIN Pmap Model also show a similar lag distribution with $\tau_{jav}=7.37_{-1.37}^{+1.84}$ days.
In addition, the predicted Balmer continuum lag is $10.1^{+2.5}_{-2.1}$ days, generally consistent with our measured lags within error bars. In conclusion, the high correlation coefficients and consistent lags strongly suggest that the measurements for this target are reliable.

\hspace*{1em}\textbf{MCG+08-11-011}: For this target, the result given by the ICCF-Cut method shows a high significant peak centered at $\tau_{cut}=2.75^{+0.70}_{-0.51}$ days. However, the lag distribution given by the JAVELIN Pmap Model shows a primary peak accompanied by other less-significant peaks, which causes a skewed lag measurement with a large uncertainty. The presence of multiple peaks may be due to the short monitoring duration of about 90 days and similar local features in light curves. Nevertheless, the lag measurement may be reasonable because the final lag $\tau_{jav}=4.29_{-1.08}^{+3.50}$ days is consistent with $\tau_{cut}$ within error bars.

\hspace*{1em}\textbf{Mrk 110}: In Figure B3, the CCCD given by the ICCF-Cut shows a clear peak while the measurement given by the JAVELIN Pmap Model shows a broad lag distribution with multiple peaks. This causes an ambiguous lag measurement, even though $\tau_{cut}=4.09_{-2.04}^{+1.91}$ and $\tau_{jav}=11.09_{-5.40}^{+9.36}$ could reach consistency within the error bars. We note that this target only had 53 visits during a monitoring duration of about 120 days, resulting in a limited cadence of 2.16 days. This may explain why this target dose not have good and consistent results.

\begin{table*}
\begin{center}
\caption{ICCF-Cut and JAVELIN Results}
\label{Table:ICCF-Cut}
\setlength\tabcolsep{4.3mm}{
\begin{tabular}{ccccccccc}
\toprule
\toprule
Object& Segment& $p_{dc}$& $r_{max}$& $\tau_{cut}$/days&$\tau_{jav}$/days& $\tau_{dc} $/days&$R_{cut}$& Flag\\
(1) & (2) &  (3) &(4) & (5)& (6)& (7)& (8)&(9)\\
\midrule
Fairall 9      & All&  0.39 &0.88& $10.61_{-3.06}^{+2.55}$& $126.91_{-2.03}^{+1.36}$& $8.70_{-2.15}^{+1.60}$& 4.28&0\\
Fairall 9      & 1-D&  0.39 &0.62& $6.19_{-2.09}^{+1.93}$& $7.51_{-3.33}^{+1.88}$& $8.70_{-2.15}^{+1.60}$& 3.51&1\\
Fairall 9      & 2  &  0.39 &0.77& $6.23_{-2.38}^{+2.52}$& $29.03_{-5.66}^{+6.01}$& $8.70_{-2.15}^{+1.60}$& 3.56&0\\
3C 120         & All&  0.38 &0.86& $7.43_{-2.66}^{+2.76}$& $7.37_{-1.37}^{+1.84}$& $10.1_{-2.10}^{+2.50}$& 2.52&1\\
MCG+08-11-011  & All&  0.37 &0.81& $2.75_{-0.51}^{+0.70}$& $4.29_{-1.08}^{+3.50}$& $7.85_{-0.25}^{+0.25}$& 2.15&1\\
Mrk 110        & All&  0.38 &0.74& $4.09_{-2.04}^{+1.91}$& $11.09_{-5.40}^{+9.36}$& $12.8_{-3.60}^{+4.45}$& 2.40&1\\
Mrk 279        & 1  &  0.38 &0.95& $4.85_{-1.70}^{+1.86}$& $5.31_{-0.87}^{+0.82}$& $8.35_{-1.95}^{+1.95}$& 2.81 &1\\
Mrk 279        & 2  &  0.38 &0.91& $2.02_{-2.45}^{+2.42}$& $4.34_{-0.52}^{+1.06}$& $8.35_{-1.95}^{+1.95}$& 2.67 &0\\
Mrk 279        & 3  &  0.38 &0.75& $2.02_{-1.91}^{+1.53}$& $5.07_{-1.59}^{+9.10}$& $8.35_{-1.95}^{+1.95}$& 3.66 &0\\
Mrk 335        & All&  0.38 &0.77& $3.69_{-1.23}^{+1.25}$& $9.94_{-7.21}^{+2.63}$& $7.00_{-1.70}^{+2.30}$& 2.15 &1\\
Mrk 817        & All&  0.38 &0.88& $7.58_{-1.46}^{+1.30}$& $8.15_{-1.58}^{+1.95}$& $9.95_{-3.35}^{+4.95}$& 2.10 &1\\
Mrk 817        & 1  &  0.38&0.84& $10.98_{-3.17}^{+1.45}$& $32.31_{-2.53}^{+1.56}$& $9.95_{-3.35}^{+4.95}$& 4.94&0\\
Mrk 817        & 2  &  0.38&0.81& $4.62_{-1.55}^{+1.73}$& $8.91_{-1.49}^{+1.54}$& $9.95_{-3.35}^{+4.95}$& 2.20 &0\\
NGC 6814       & All&  0.36&0.84& $0.99_{-1.19}^{+0.25}$& $1.05_{-0.13}^{+0.14}$& $3.30_{-0.45}^{+0.45}$& 3.81&0\\ 
NGC 6814       & All-D&0.36&0.75& $1.21_{-0.24}^{+0.04}$& $0.94_{-0.20}^{+0.19}$& $3.30_{-0.45}^{+0.45}$& 4.70 &1\\ 
\bottomrule
\end{tabular}}
\end{center}
\textbf{Note.} Column 1: Object name. Column 2: Numbers 1, 2, and 3 represent the first, second, and third segments of the light curves, respectively. “All” represents the whole light curves. The suffix “-D” represents the detrended light curves. Column 3: Diffuse continuum ratios $p_{dc}$ in the $U$ band predicted by CLOUDY. Column 4: The maximum cross-correlation coefficient $r_{max}$. Column 5: The lags $\tau_{cut}$ measured by the ICCF-Cut method. Column 6: The lags $\tau_{jav}$ measured by the JAVELIN Pmap Model. All lags presented here are in the rest frame. The $r_{max}$, $\tau_{cut}$, and $\tau_{jav}$ are measured between the driving ($UVW2$ band) light curves and cut light curves (the ones where we subtract the predicted disk components from the Swift $U$ bands). Column 7: The predicted Balmer continuum lags $\tau_{dc}=0.5\tau_{H\beta}$. The observed H$\beta$ lags are shown in Column 6 of Table \ref{tab:Properties}. Column 8: Excess ratio of the ICCF-Cut lags $\tau_{cut}$ relative to the original lags in the $U$ band $\tau_{u}$.  Column 9: Sample flags. If the flag equals 1, it means that this target has consistent ICCF-Cut and JAVELIN Pmap Model lags within the error limit. For those targets with multiple measurements based on the different segments of light curves, we only select the best results flagged as 1. All the measurements flagged as 1 are categorized as the final result and analyzed in Section \ref{sec:discussion}.

\end{table*}

\hspace*{1em}\textbf{Mrk 279}: For this target, there are large seasonal gaps in the light curves, so we divide the total light curves into three segments, and the corresponding measurement results are shown in Figures B4, B5, and B6. For the first segment light curves, the measurement shows highly consistent lags with $\tau_{cut}=4.85_{-1.70}^{+1.86}$ and $\tau_{jav}=5.31_{-0.87}^{+0.82}$. But the measurements based on the other segment light curves show $\tau_{cut}$ are slightly larger than $\tau_{jav}$. Especially for the third segment light curves, the maximum correlation coefficient $r_{max}$ decreases to 0.75, which is much smaller than that for the first ($r_{max}=0.95$) and second ($r_{max}=0.91$) segment light curves. Due to the limited cadence and visits of the third segment light curves, the lag distribution given by the JAVELIN Pmap model shows additional weaker peaks. We notice that $\tau_{jav}$ derived from three segment light curves are similar, approximately 4 to 5 days. These lags are also comparable to $\tau_{cut}$ derived from the first segment light curves. Considering the highest maximum correlation coefficient, and consistent lags, the measurement based on the first segment light curves is selected as the final result for this target.

\hspace*{1em}\textbf{Mrk 335}: The result for this target is shown in Figure B7. We can see a significant CCCD peak given by the ICCF-Cut Method, centered on $\tau_{cut}=3.69^{+1.25}_{-1.23}$ days. The lag distribution given by the JAVELIN Pmap Model shows a primary peak accompanied by a secondary peak of comparable strength. The presence of multiple peaks can skew the lag measurements and produce large uncertainties. Although the secondary peak aligns with the CCCD peak, the final lag $\tau_{jav}=9.94_{-7.21}^{+2.63}$ days is still larger than $\tau_{cut}$ due to the influence of the primary peak centered at a larger lag. The predicted Balmer continuum lag, $7.00_{-1.70}^{+2.30}$ days, is interposed between $\tau_{cut}$ and $\tau_{jav}$. In summary, given that the lag measurements could reach consistency within the error bars, the result is still reasonable for this target.

\hspace*{1em}\textbf{Mrk 817}: The Swift monitoring of this target has 568 visits for approximately 1000 days. Because the light curves have a small seasonal gap around MJD 59600, we divide it into two segments. For the first segment, the Swift daily monitoring of this target has been investigated by the AGN Space Telescope and Optical Reverberation Mapping 2 (AGN STORM 2) Project \citep{2021ApJ...922..151K,2023ApJ...958..195C,2024ApJ...961..219N,2024ApJ...974..271L,2024ApJ...976...59N}. The second segment is from an extended monitoring which has not been reported in previous work. The lag measurements for the total, first segment, and second segment light curves are shown in Figures B8, B9, and B10. Most lag measurements are broadly consistent, ranging from 7 to 11 days, which are comparable to the predicted Balmer continuum lag $9.95_{-3.35}^{+4.95}$ days. We also notice two exceptions, $\tau_{jav}=32.31_{-2.53}^{+1.56}$ days for the first segment light curves and $\tau_{cut}=4.62_{-1.55}^{+1.73}$ days for second segment light curves. These inconsistent lags may be caused by the quasi-periodic variability in their light curves, which will be discussed in Section \ref{subsec:Inconsistent-lags}. In summary, the total light curves show the highest maximum correlation coefficient $r_{max}=0.88$ and the most consistent lag measurement with $\tau_{cut}=7.58_{-1.46}^{+1.30}$ days and $\tau_{jav}=8.15_{-1.58}^{+1.95}$ days, which is a reliable result.

\hspace*{1em}\textbf{NGC 6814}:
Similar to Fairall 9, this target has a long-term variation that is inconsistent with a standard disk reprocessing scenario \citep{2024MNRAS.527.5569G}. Therefore, we perform lag measurements for both the original and detrended light curves. The results are shown in Figures B11 and B12. Although both lag measurements show consistent lags, we still notice that the broad CCCD given by the ICCF-Cut Method is well constrained within a limited lag range after subtracting the long-term variation, and the additional weaker peaks in JAVELIN lag distribution also disappear. Therefore, the detrended light curves give a better lag measurement and associated uncertainty as the final result for this target.

\begin{figure*}
    \centering
    \begin{minipage}[b]{0.45\textwidth}
		\includegraphics[width=\linewidth]{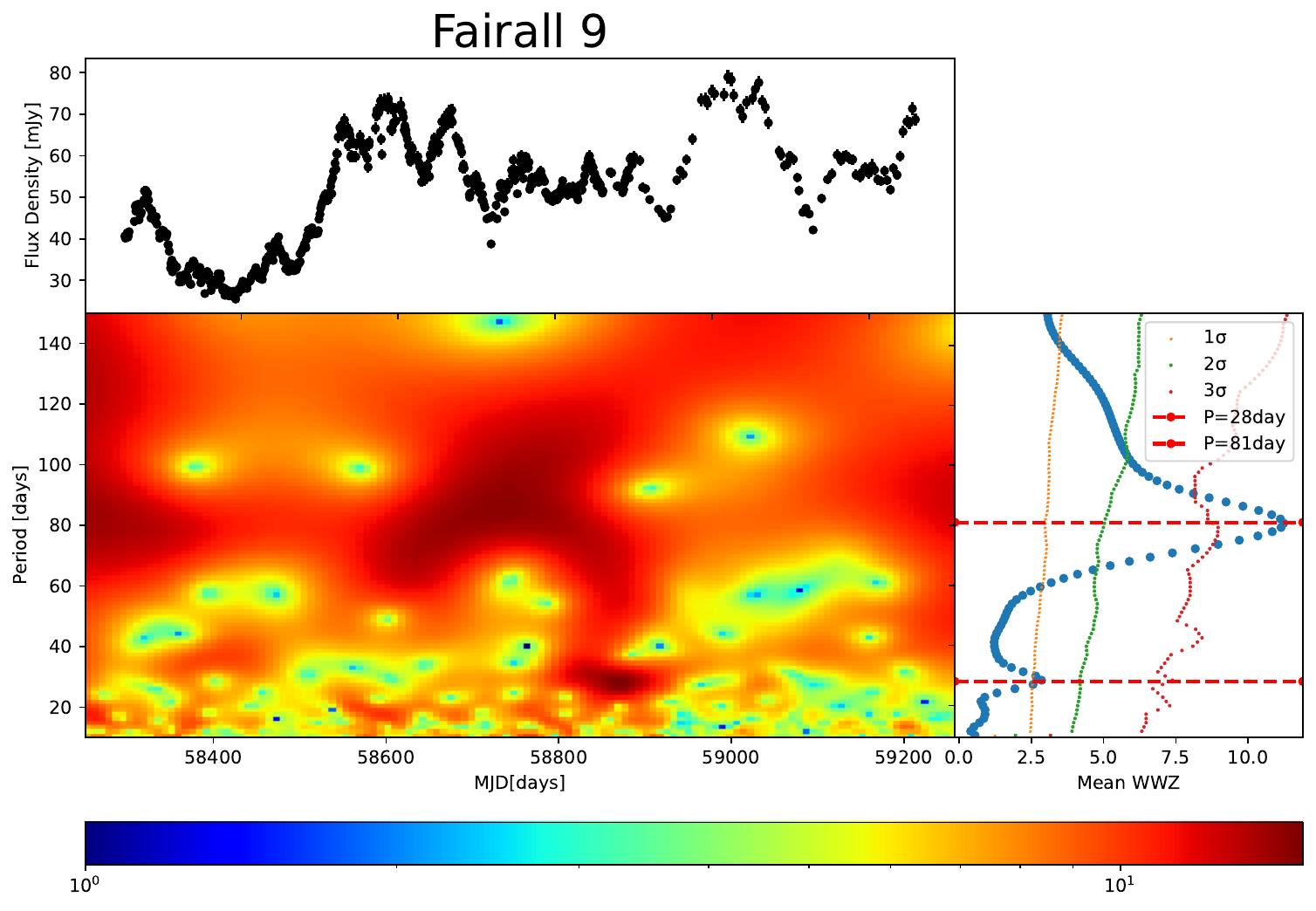}
	\end{minipage}%
	\begin{minipage}[b]{0.45\textwidth}
		\includegraphics[width=\linewidth]{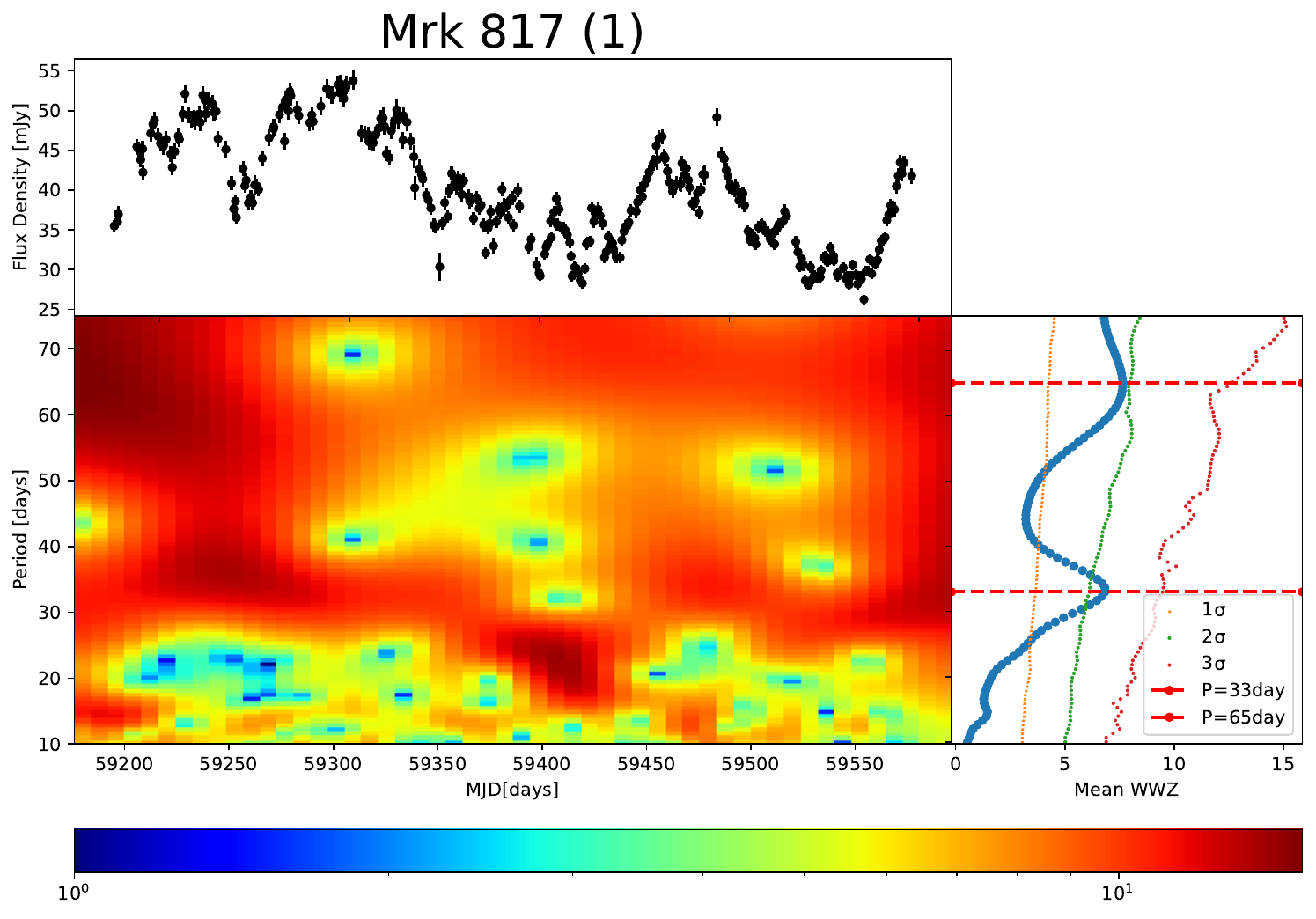}
	\end{minipage}%
    \caption{Left panel: Search for quasi-periodic variability in the whole light curves of Fairall 9. The top subpanel shows the $UW2$ band light curve, and the left bottom subpanel shows the 2D color contour of the WWZ power spectrum for the light curve. In the right bottom subpanel, the blue line represents the time-averaged WWZ power. The orange, green, and red lines represent the 1$\sigma$, 2$\sigma$, and 3$\sigma$ confidence levels, respectively. Right panel: the same as the left panel, but for the first segment light curves of Mrk 817.}
    \label{fig:WWZ}
\end{figure*}

\section{Discussion} \label{sec:discussion}

\subsection{Inconsistent Lags Given by Different Methods}\label{subsec:Inconsistent-lags}
As mentioned in Section \ref{subsec:sample}, the lag measurements given by different methods are consistent within error bars for most targets. However, $\tau_{jav}$ is always slightly larger than $\tau_{cut}$. The inconsistency in lag measurements can be categorized into two types: one arising from the underestimation of $\tau_{cut}$, and the other due to the overestimation of $\tau_{jav}$ caused by the Javelin Pmap Model.

For the underestimation of $\tau_{cut}$, the lag measurements given by the ICCF-Cut method are influenced by several critical factors: the accurate estimation and subtraction of the host-galaxy contribution, the selection of an appropriate diffuse continuum ratio $p_{dc}$ and disk lag $\tau_{disk}$ to derive the diffuse continuum light curves. The host-galaxy contribution is estimated through flux-flux analysis, which always gives a larger host-galaxy flux than that given by the image-subtraction method \citep{2024Univ...10..282C}. According to Equation \ref{eq:alpha}, an overestimated host-galaxy contribution introduces a larger constant subtraction in the $U$ band light curve $L_U(t)$, consequently diminishing the scaling factor $\alpha$. According to Equation \ref{eq:lcdc}, a lower $\alpha$ implies that a portion of the disk flux is still retained in the diffuse light curves after cutting, finally leading to an underestimation of $\tau_{cut}$. Using NGC 5548 as an example, our previous work provides direct evidence that a more accurate image-subtraction method yields a larger lag measurement than that derived from the flux-flux analysis \citep{2024ApJ...966..149J}.

The diffuse continuum ratio is predicted by CLOUDY and a specific BLR model. This model includes a critical parameter, $c_f$, which determines the diffuse continuum ratio. As noted above, we assume an average value of $c_f=0.2$ for all targets, primarily because the predictions under this assumption align well with observational results. However, this assumption may not be valid for all targets. For example, \cite{2022MNRAS.509.2637N} used $c_f=0.1$ for NGC 4593 and Mrk 509. A larger $c_f$ results in a higher diffuse continuum ratio. As demonstrated in our previous simulations (see Figure 9 in \citealt{2024ApJ...966..149J}), a systematic overestimate of $p_{dc}$ would lead to an underestimation of $\tau_{cut}$, while an underestimate of $p_{dc}$ would have the opposite effect. In principle, the diffuse continuum ratio is also expected to vary with time, which would introduce additional systematic uncertainty into the lag measurements. However, a comprehensive investigation of its time-dependent behavior is very challenging and beyond the scope of the present work. In addition, this model only focuses on the disk continuum and the diffuse continuum, while in reality, some line emissions also contribute to the total flux. For example, the CIII] emission line at 1909 Å contributes to the Swift $UVW2$ band, and the FeII pseudo-continuum (a blend of multiple UV/optical transitions spanning 2000–4000 Å) affects the Swift $U$ band \citep{1983ApJ...275..445N,1985ApJ...288...94W,1992ApJS...80..109B,2001AJ....122..549V}. Although these line emissions are relatively weak, ignoring them may artificially inflate the measured continuum levels, biasing estimates of key parameters such as the diffuse continuum ratio.

The disk lag is predicted by a standard thin disk model, which may be too simplistic to sufficiently explain the observations of all targets. \cite{2021ApJ...907...20K,2021MNRAS.503.4163K} constructed a standard Novikov–Thorne accretion disk model that considers more elaborate mechanisms and parameters, such as relativistic effects and disk reflection. The model fits well the overall time-lag spectrum, except the $U$ band excess lags caused by the diffuse BLR emission \citep{2023MNRAS.526..138K}. It suggests that the disk lag predicted by this model will be larger than that predicted by the standard thin disk model. In addition, the predicted disk lag also depends on the choice of $\kappa$ in Equation \ref{eq:lag-wave-thin-disk}. While we adopt $\kappa=0$ in this work, some studies assume $\kappa=1$, corresponding to comparable external and internal heating \citep{2018ApJ...857...53C,2020ApJ...896....1C,2023ApJ...958..195C}. However, as discussed in \citet{2016ApJ...821...56F}, Equation \ref{eq:lag-wave-thin-disk} is relatively insensitive to this parameter. Specifically, increasing $\kappa$ from 0 to 1 increases the predicted disk size by a factor of $(3/4)^{1/3} \approx 0.91$, corresponding to a change of less than 10\%. Therefore, this effect is minor for our analysis. Referring to the simulations by \cite{2024ApJ...966..149J}, an underestimation of the disk lag for some targets would result in an overestimation of $\tau_{cut}$. Although this does not account for the underestimation of $\tau_{cut}$ that we are primarily concerned with, it illustrates that the selection of the disk model can affect lag measurements. Further investigation into the influence of different disk models is beyond the scope of the current paper but is needed in future works.

Compared to the underestimation of $\tau_{cut}$, the overestimation of $\tau_{jav}$ bears greater responsibility for the inconsistencies in lag measurements for most targets. The fundamental reason is that JAVELIN is a model-dependent method, where the validity of its assumptions and the quality of the light curves can significantly affect the lag measurements. According to the lag distributions derived from the JAVELIN Pmap Model, the underestimation of $\tau_{cut}$ is from two cases. One characterized by a significant peak centered at a large lag, as illustrated for Fairall 9 (all) in Figure \ref{fig:Fairall9_result} and Mrk 817 (1) in Figure B9. The other is characterized by multiple peaks, as illustrated for Mrk 110 in Figure B3 and Mrk 335 in Figure B7. We also note that there
are some weak secondary peaks for MCG+08-11-011 in Figure B2 and Mrk 279 (3) in Figure B6. These targets have short observation baselines and/or limited visits. Specifically, the observation epoch is 53 for Mrk 110 and 72 for Mrk 279 (3). The monitoring duration is 88.7 days for MCG+08-11-011 and 96.7 days for Mrk 335. Compared to the other lag measurements based on longer observation baselines and sufficient sampling, they are more likely to result in lower correlations and poorer lag distributions. In summary, the presence of multiple peaks, also known as aliasing, is a potential outcome of lag detection with sparse sampling data, limited baseline length, and quasi-periodic variability, etc. \citep{2017ApJ...851...21G,2019ApJ...887...38G,2020ApJ...901...55H,2024ApJS..275...13W}. 

For Fairall 9 (all) and Mrk 817 (1), the observed strong individual peaks at large lags may be attributed to the similar local features in their light curves. If the light curves exhibit quasi-periodic variability with a period $T$, shifting the light curves by $T$ could align these features well, causing the quasi-periodic variability period $T$ to be mistakenly identified as the time delay. To further investigate the impact of quasi-periodic variability on lag measurements, we utilize the weighted wavelet Z-transform (WWZ) method to search for the quasi-periodic signals in the whole light curves of Fairall 9 and the first segment light curves of Mrk 817. The WWZ method is a common tool used in periodicity analysis \citep{1996AJ....112.1709F}. Compared to other periodicity analysis tools, such as the Lomb-Scargle Periodogram \citep{1976Ap&SS..39..447L,1982ApJ...263..835S} and the Jurkevich method \citep{1971Ap&SS..13..154J}, the WWZ method has a strong localization ability in both time domain and frequency domain. To estimate the significance of quasi-periodic signals, we follow the MCMC recipe given in \cite{2024A&A...689A..35C}. First, we model the power spectral density and the probability distribution function of the observed light curves. Next, based on these models, we generate $10^4$ light curves using the Emmanoulopoulos algorithm \citep{2013MNRAS.433..907E}, and then employ the WWZ method to analyze these artificial light curves. Finally, we will study the distribution of these results and obtain 1$\sigma$, 2$\sigma$, and 3$\sigma$ significance level lines. 

\begin{figure*}[t]
\centering
\includegraphics[width=0.9\textwidth]{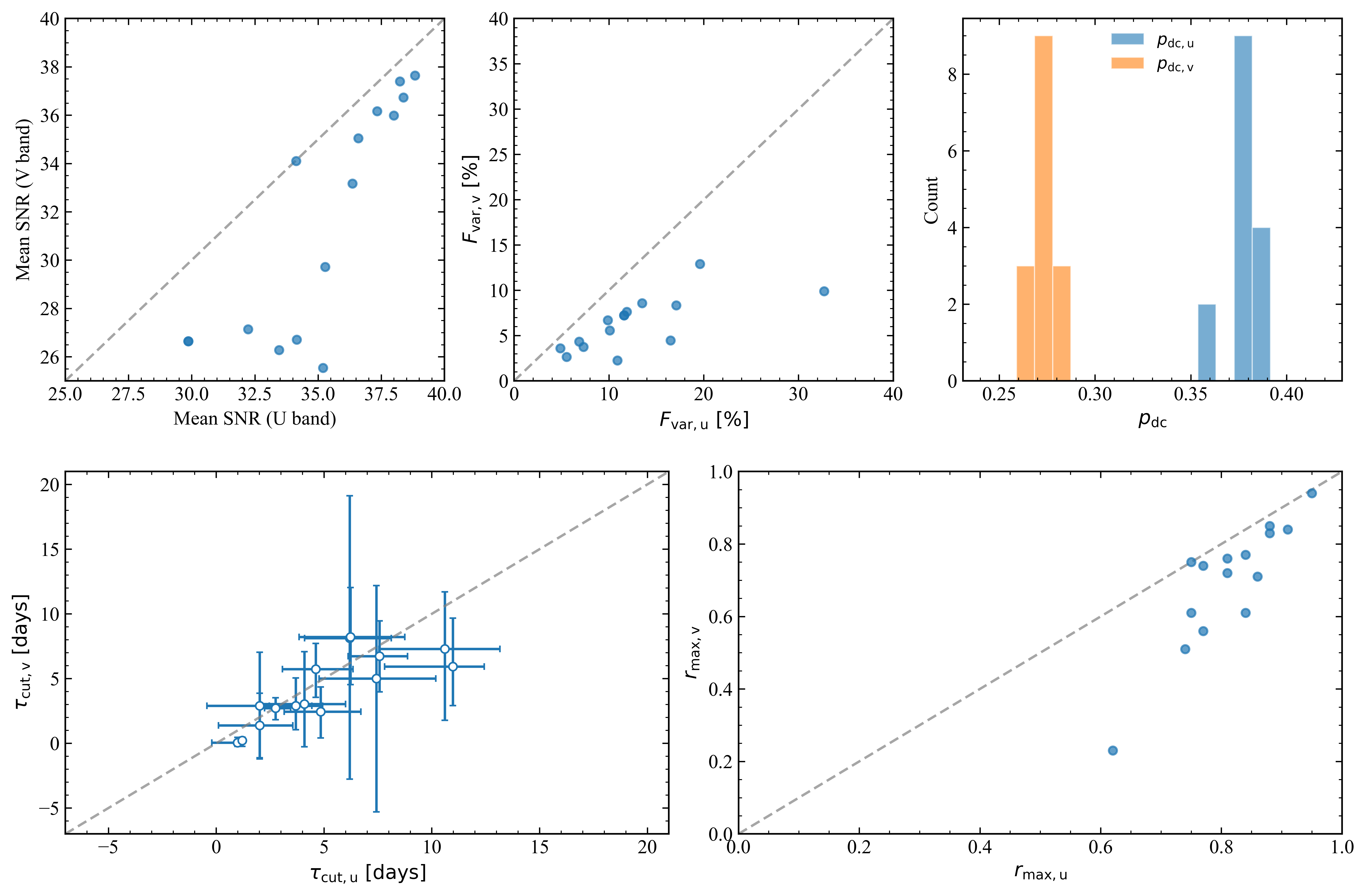}
\caption{Comparison of the $U$ and $V$ band results. 
Top left panel: mean SNRs.
Top middle panel: fractional variability amplitudes.
Top right panel: distributions of diffuse continuum ratios ($U$: orange, $V$: blue). 
Bottom left panel: lags measured by the ICCF-Cut method. 
Bottom right panel: maximum correlation coefficients measured by the ICCF-Cut method. 
The dashed line indicates the one-to-one relation.
}
\label{FigUVCompare}
\end{figure*}

Given that the UV/optical variations appear qualitatively similar for the same target, we perform periodicity analysis exclusively on the driving light curves in the $UVW2$ band. According to the lag measurements and the duration of each light curve, the period search range is [10,150] days for Fairall 9, and
[10,75] days for Mrk 817. In this way, we can reduce the calculation time and exclude the spurious signals produced by the red noise. The detailed results calculated by WWZ are shown in Figure \ref{fig:WWZ}. For Fairall 9, the periodogram exhibits a distinct peak centered around 81 days, whose significance is over 3$\sigma$. Although this period is not precisely consistent with the observed lag $\tau_{jav}=126.91_{-3.06}^{+1.36}$ days, its high significance could provide a partial explanation for the overestimated lag. We also notice a weak peak at around 28 days, while the significance of this peak only exceeds 1$\sigma$. According to the periodogram, this periodic signal appears to emerge from MJD 58800 to 58900, which falls within the second segment light curve for Fairall 9. This segment light curve also yields a large lag measurement of $\tau_{jav}=29.03_{-6.01}^{+5.66}$ days, which is consistent with this periodic signal of $T=28$ days. For Mrk 817, we find two peaks at 33 days and 65 days in the WWZ periodogram. The significance of the peak at 33 days is over 2$\sigma$, while the peak at 33 days approaches but remains below 2$\sigma$. The peak at 65 days may be explained as a harmonic of the fundamental 33-day period. Notably, the 33-day periodic signal aligns well with the JAVELIN-derived overestimated lag of $\tau_{jav}=32^{+1.56}_{-2.53}$ days. In conclusion, the quasi-periodic features in the light curves provide a potential explanation for the inflated lag measurements given by the JAVELIN Pmap Model.

\subsection{Extensions to Other Bands}

In the previous sections, we focused on the Swift $U$ band as a representative window to probe the diffuse continuum contribution, primarily because the $U$ band is located around the Balmer jump where the diffuse continuum emission from the BLR is expected to be the strongest. However, some previous works suggest that the diffuse continuum spreads over the whole UV/optical spectrum and contributes to the observed lags \citep{2018MNRAS.481..533L,2019MNRAS.489.5284K,2022MNRAS.509.2637N}. As shown in Figure \ref{fig:Fairall9_ratio}, the $V$ band is also predicted to feature a non-negligible diffuse continuum contribution. To further test the robustness of our conclusions, we apply the ICCF-Cut method to the Swift $V$ band light curves, following the same procedure as described in Section \ref{subsec:ICCF-Cut} for the $U$ band. The resulting diffuse continuum lags are presented in Appendix (see Figure \ref{FigVbandICCF}).

\begin{figure*}[t]
    \centering
        \begin{minipage}[b]{0.45\textwidth}
		\includegraphics[width=\linewidth]{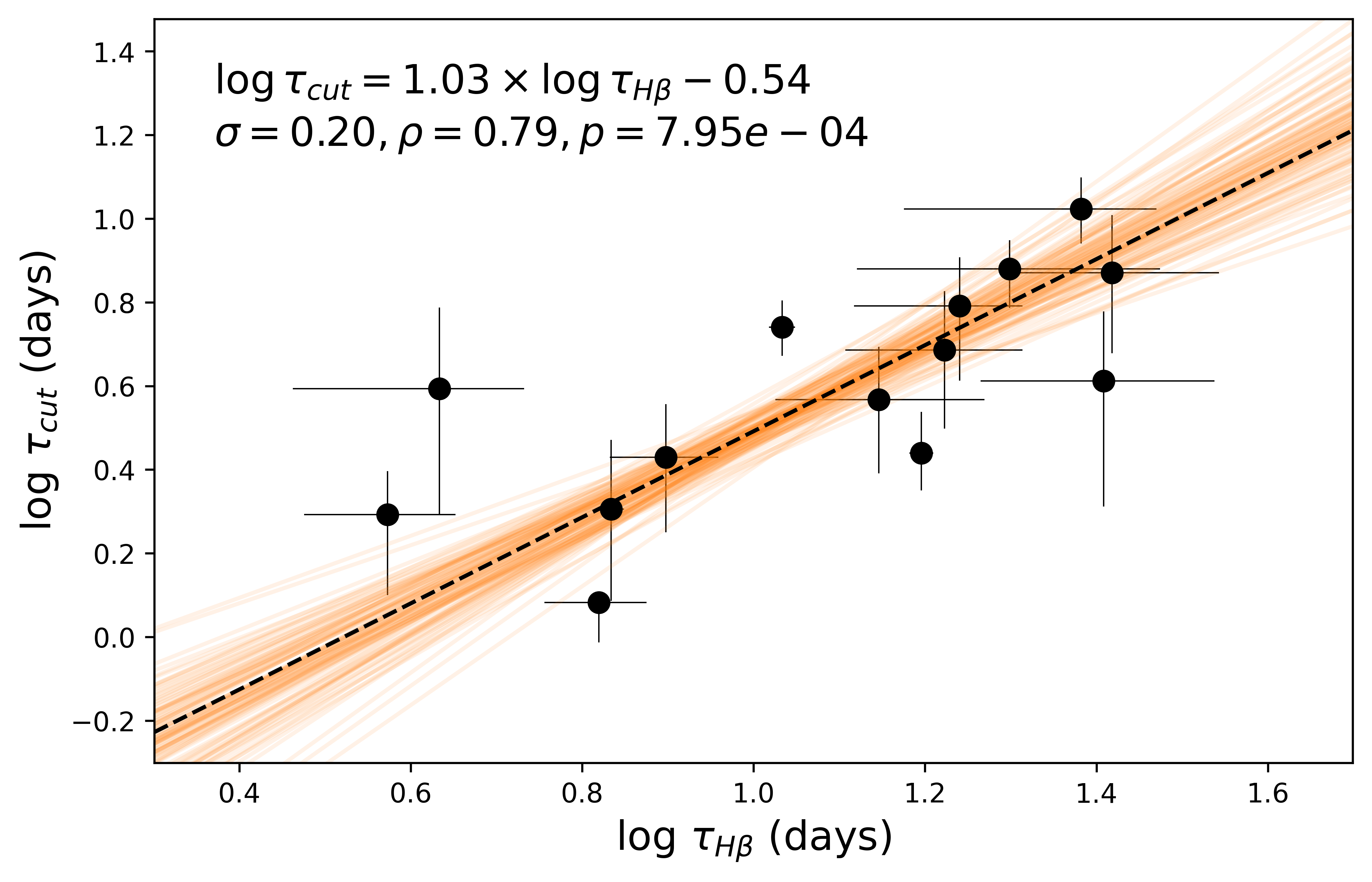}
	\end{minipage}%
	\begin{minipage}[b]{0.45\textwidth}
		\includegraphics[width=\linewidth]{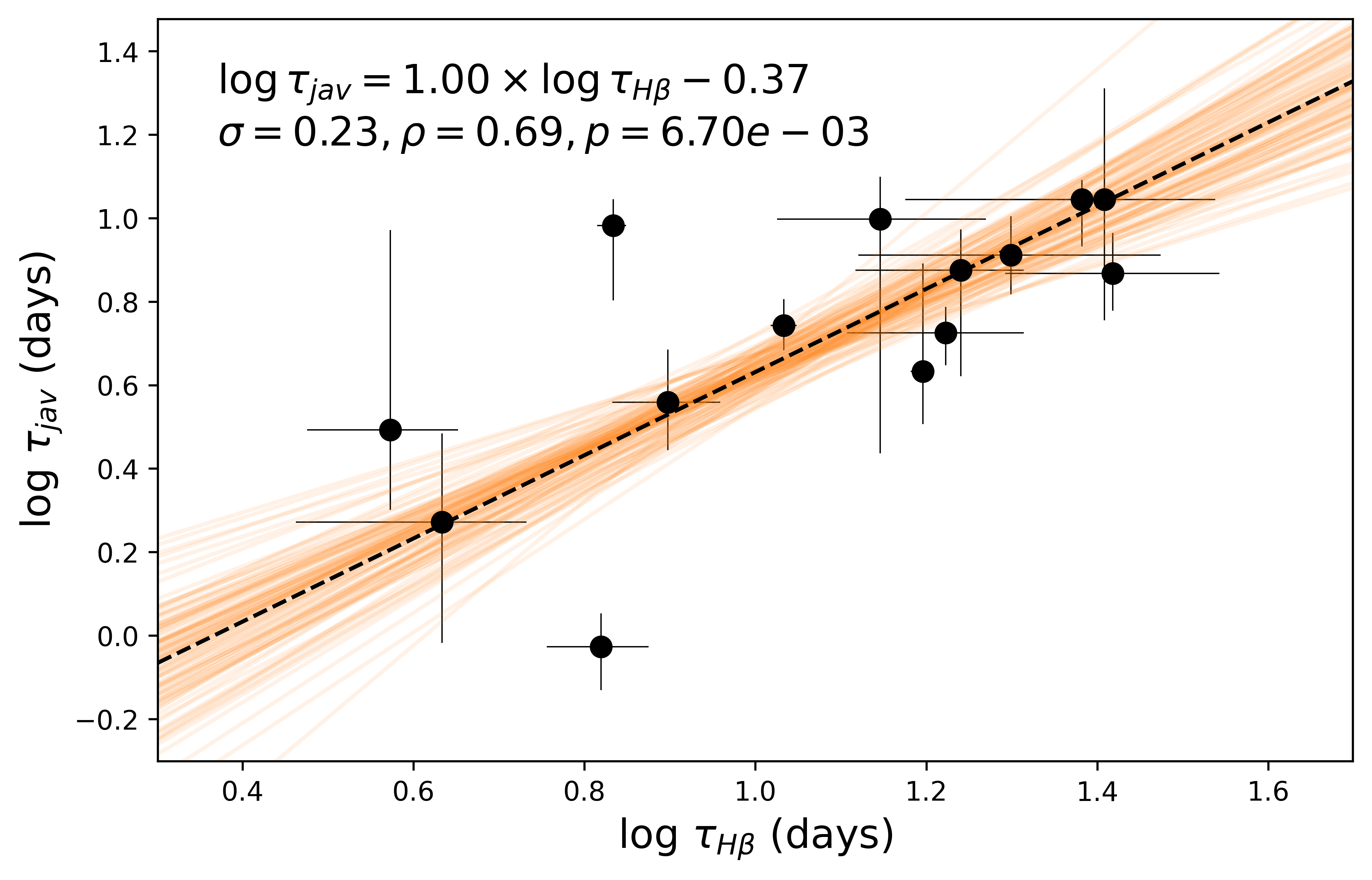}
	\end{minipage}%
 	
	\begin{minipage}[b]{0.45\textwidth}
		\includegraphics[width=\linewidth]{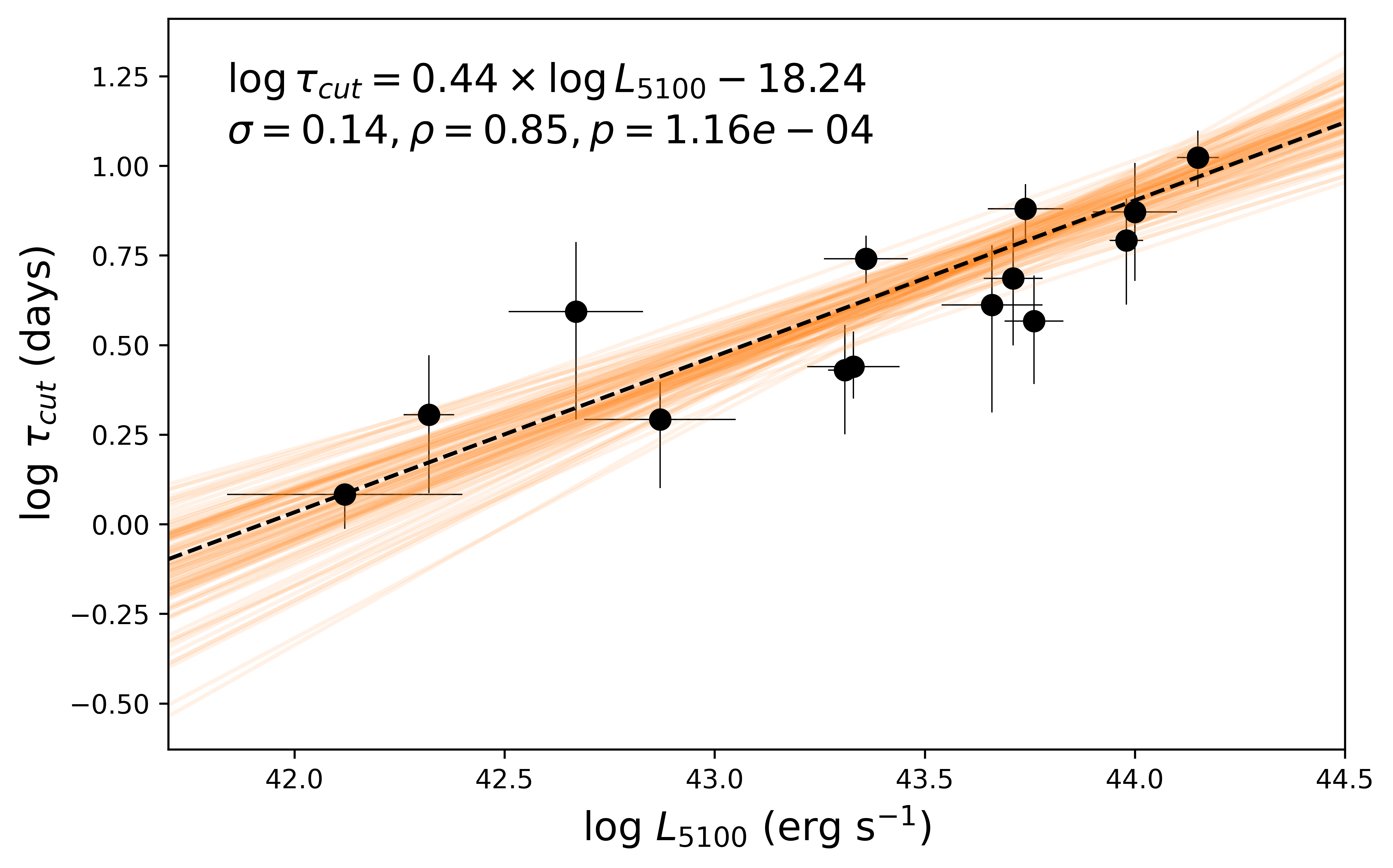}
	\end{minipage}%
        \begin{minipage}[b]{0.45\textwidth}
		\includegraphics[width=\linewidth]{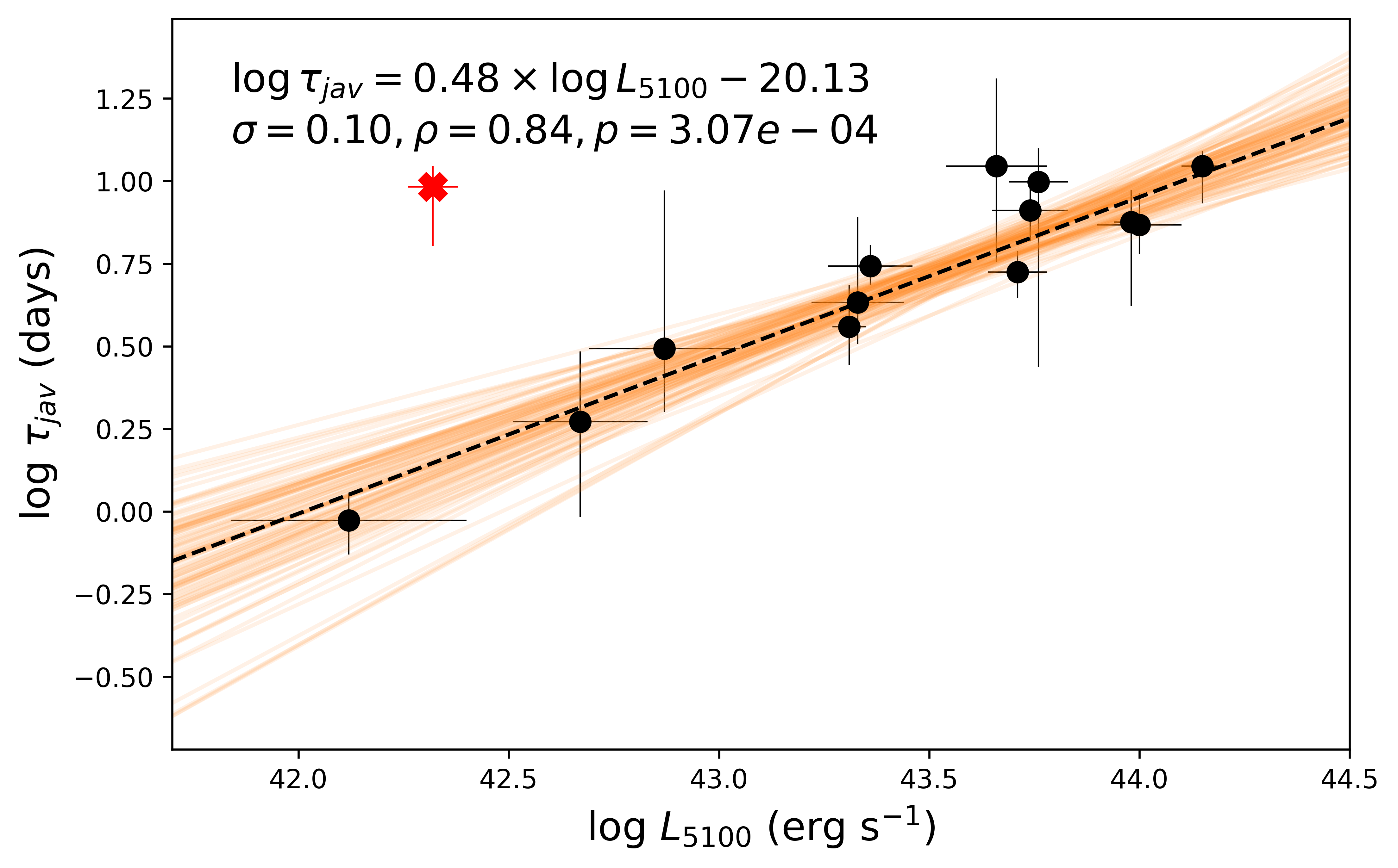}
	\end{minipage}%
    \caption{Top left panel: the relationship between the ICCF-Cut lags and $H_\beta$ lags. Top right panel: the relationship between the Javelin lags and $H_\beta$ lags. Bottom left panel: the relationship between the ICCF-Cut lags and luminosity at 5100 Å. Bottom right panel: the relationship between the Javelin lags and luminosity at 5100 Å. In each panel, the relationships are analyzed using a log-log regression. The fitting results are shown in black solid lines. The specific formulas and scatters are displayed in the upper left corner of each panel. The red dot on the bottom right panel represents NGC 4151, obtained from \cite{2024ApJ...966..149J}. It was removed from the fitting due to the significant deviation.}
    \label{fig:Sample}
\end{figure*}

To quantitatively compare the results derived from the $U$ and $V$ bands, we perform a systematic comparison of five key parameters in Figure \ref{FigUVCompare}: mean signal-to-noise Ratio (SNR; top left), fractional variability amplitude ($F_{var}$; top middle), diffuse continuum ratio ($p_{dc}$; top right), ICCF-Cut lag ($\tau_{cut}$; bottom left), and maximum correlation coefficient ($r_{max}$; bottom right). Here the mean SNR is defined as the average of the epoch-by-epoch flux-to-uncertainty ratios across the light curve. As shown in the bottom left panel of Figure \ref{FigUVCompare}, the lags measured from the $V$ band are generally consistent with those from the $U$ band, lying close to the one-to-one relation (gray dashed line) within the error bars. This further supports that the diffuse continuum emission is not restricted to the $U$ band, but is instead a general feature of the UV/optical continuum.

However, we note that the $V$ band results are typically less robust than those obtained from the $U$ band. In Figure \ref{FigUVCompare}, the bottom left panel shows that the lags derived from the $V$ band light curves generally have larger uncertainties, while the bottom right panel reveals systematically smaller maximum correlation coefficients compared with their $U$ band counterparts. These differences can be attributed to several factors: (1) The $V$ band light curves are of lower quality than those in the $U$ band, characterized by lower mean SNR (see top left panel of Figure \ref{FigUVCompare}), weaker intrinsic variability, and stronger host-galaxy contamination. This is also reflected in the top middle panel of Figure \ref{FigUVCompare}, where the fractional variability amplitudes in the $V$ band are significantly smaller than those in the $U$ band. (2) Although the diffuse continuum component is also present in the $V$ band, its contribution is weaker than in the $U$ band. The top right panel of Figure \ref{FigUVCompare} shows that the diffuse continuum ratios for the $V$ band lie between 0.26 and 0.29, whereas those for the $U$ band extend to $0.36\sim0.39$. (3) The $V$ band emission originates from a larger radius in the accretion disk, where the response to the central variability is intrinsically weaker. Consequently, the ICCF-Cut analysis in this band is more susceptible to larger interband lags. Despite these limitations, we still detect a coherent diffuse continuum lag in the V band. This reinforces our conclusion that the diffuse continuum emission from the BLR makes a significant contribution to the UV/optical continuum lags, and that this contribution is not exclusive to the $U$ band.

\subsection{\texorpdfstring{The $R_{DCR}$-$R_{BLR}$ and $R_{DCR}$-$L_{5100}$ Relations}{The RDCR-RBLR Relations}}\label{subsec:RR-relation}
As mentioned in Section \ref{subsubsec:DCE_BLR}, the outer component, which contributes to the excess lags in the $U$ band, is probably associated with the diffuse continuum emission from BLR. In this section, we build a sample by merging 8 AGNs from this work with 6 AGNs reported in \cite{2024ApJ...966..149J}, aiming to investigate the relationship between the outer component and the BLR. Here, we adopt $\tau_{cut}$ or $\tau_{jav}$ to denote the diffuse continuum region (DCR) size $R_{DCR}$ for the outer component. The BLR size $R_{BLR}$ is denoted by $\tau_{H\beta}$ in Table \ref{tab:Properties}. For those targets with multiple lag measurements, we adopt the most reliable result (flagged as 1 in Table \ref{Table:ICCF-Cut}) for analysis, defined as the case where the ICCF-Cut and JAVELIN results are most consistent. For Fairall 9 and NGC 6814, we specifically adopt the detrended measurements, as these two AGNs exhibit long-term variability unrelated to the disk reprocessing scenario \citep{2020MNRAS.498.5399H,2024MNRAS.527.5569G}. 

First, we plot the direct comparison between $R_{DCR}$ and $R_{BLR}$ in the top panels of Figure \ref{fig:Sample}. The left top panel exhibits the $\log \tau_{cut}-\log\tau_{H\beta}$ relationship, while the right top panel exhibits the $\log\tau_{jav}-\log\tau_{H\beta}$ relationship. To quantify these relationships, we performed a linear regression analysis using \texttt{emcee}\footnote{\url{https://github.com/dfm/emcee}} package \citep{2013PASP..125..306F}. Here, we adopt flat prior distributions within reasonable parameter space. The best-fitting values are determined as the median of the posterior distribution, with uncertainties represented by the range between the 16th and 84th percentiles. In addition, we determined the scatter ($\sigma$) of the residuals to assess the goodness of fit, the Spearman correlation coefficient ($\rho$) to evaluate the strength and direction of the correlation, and the $p$-value to test its statistical significance. Finally, the best-fitting relations are described as
\begin{gather}
    \log \tau_{cut}=1.03^{+0.15}_{-0.13}\times \log \tau_{H\beta}-0.54^{+0.14}_{-0.15}, \label{eq:RR1} \\
    \log \tau_{jav}=1.00^{+0.19}_{-0.17}\times \log \tau_{H\beta}-0.37^{+0.18}_{-0.20}. \label{eq:RR2}
\end{gather}
According to the BLR cloud model in \cite{2020MNRAS.494.1611N}, the predicted diffuse continuum lag is $\tau_{dc}\simeq0.5\tau_{H\beta}$. In logarithmic form, this corresponds to $\log \tau_{dc}\simeq 1\times \tau_{H\beta}-0.3$. We notice that the best-fitting slopes for the $\log\tau_{cut}-\log\tau_{H\beta}$ and $\log\tau_{jav}-\log\tau_{H\beta}$ relationships are both nearly equal to 1, which is highly consistent with the predicted slope. Although the fitted intercept is slightly smaller than the predicted value, this discrepancy is understandable given the approximate nature of the prediction and the non-negligible scatter in the fit. Moreover, \cite{2023ApJ...948L..23W} found a tight relation between the BLR size and the continuum emission size, using a sample of 21 AGNs. This relation is extended to higher-redshift, higher-luminosity gravitationally lensed quasar in \cite{2025A&A...695A..10H}. These findings are similar to the $R_{DCR}-R_{BLR}$ relation in our work, indicating a strong physical connection between the DCR for the outer component and the BLR.

Analogous to the well-known radius–luminosity relation of the broad H$\beta$ line, the DCR size for the outer component is expected to follow a similar relationship with the continuum luminosity if the diffuse continuum emission from the BLR dominates the observed continuum lags. Therefore, we employ the same MCMC linear regression method for the $\log\tau_{cut}-\log L_{5100}$ (bottom left panel) and $\log\tau_{jav}-\log L_{5100}$ (bottom right panel) relationships in Figure \ref{fig:Sample}. We obtain the best-fitting relations as
\begin{gather}
    \log \tau_{cut}=0.44^{+0.06}_{-0.06}\times \log L_{5100}-18.24^{+2.46}_{-2.70}, \label{eq:RL1} \\
    \log \tau_{jav}=0.48^{+0.07}_{-0.07}\times \log L_{5100}-20.13^{+2.92}_{-3.35}. \label{eq:RL2}
\end{gather}
When fitting the $\log\tau_{jav}-\log L_{5100}$ relationship, we first excluded an outlier with an inflated $\tau_{jav}$, which corresponds to NGC 4151 obtained from \cite{2024ApJ...966..149J}. Our fitting results yield $\sigma=0.14, \rho=0.85, p=1.16\times10^{-4}$ for the $\log\tau_{cut}-\log L_{5100}$ relationship, and $\sigma=0.10,\rho=0.84,p=3.07\times10^{-4}$ for the $\log\tau_{jav}-\log L_{5100}$ relationship. It suggests a tight $R_{DCR}-L_{5100}$ relation with a strong positive correlation and high statistical significance. The best-fitting slope is $0.44^{+0.06}_{-0.06}$ for the $\log\tau_{cut}-\log L_{5100}$ relationship, and $0.48^{+0.07}_{-0.07}$ for the $\log\tau_{jav}-\log L_{5100}$ relationship. Although the slopes are slightly smaller than 0.5, it generally follows the continuum lag-luminosity relation ($R\sim L^{0.5}$) predicted in \cite{2022MNRAS.509.2637N}. \cite{2022ApJ...940...20G} further confirmed this relation using a sample including 49 AGNs. Their fitting result yielded a slope of $0.48^{+0.04}_{-0.04}$, which is fully consistent with the slope of the $\log\tau_{jav}-\log L_{5100}$ relationship in our work. In addition, \cite{2020ApJ...903..112D} derived a nearly identical slope of $0.48^{+0.03}_{-0.03}$ for the H$\beta$ $R_{BLR}-L_{5100}$ relation, using a sample of AGNs with luminosity well constrained using Hubble Space Telescope images. Recently, \cite{2024ApJS..275...13W} revisited the H$\beta$ size-luminosity relation using 157 AGNs with the best-quality lag measurements. They found a shallower slope of $0.43^{+0.02}_{-0.02}$, which is also consistent with the slope of the $\log\tau_{cut}-\log L_{5100}$ relationship in our work. In summary, the $R_{DCR}$-$L_{5100}$ is likely a parallel version of the H$\beta$ $R_{BLR}-L_{5100}$ relation. This further supports that the dominant contribution to the UV/optical continuum may come from the BLR.

%\subsection{\texorpdfstring{The $R_{DCR}$-$L_{5100}$ Relation}{The RDCR-L5100 Relation}}\label{subsec:RL-relation}
\section{Summary} \label{sec:conclusion}
In this paper, we perform a comprehensive CRM study for 8 AGNs based on the Swift archive. The main results are summarized as follows:
\begin{enumerate}
\item We not only reproduce the UVOT light curves for AGNs studied in previous CRM campaigns but also provide several sets of unpublished light curves for additional AGNs. Using these high-quality, high-cadence, and multiband light curves, we perform the time series analysis and revisit the lag-wavelength relation. All targets show strongly correlated variation throughout the UV/optical bands. Our results are largely consistent with a disk reprocessing model, with larger lags at longer wavelengths, following the $\tau \approx \lambda^{4/3} $ relation. However, the observed interband lags are significantly larger than that predicted by the standard thin disk model. In particular, the excess lags are observed in the $U$ band for most targets in our sample, with values on average $\sim2.2$ times larger than that predicted from the surrounding band data and fits. These findings suggest that the UV/optical interband lag structure is strongly affected by an outer component, which is most likely the diffuse continuum emission from the BLR.
\item We apply the ICCF-Cut method to extract the outer component embedded in the Swift $U$ band light curves. The outer component exhibits a strong correlation with the central disk emission, as the maximum correlation coefficient is larger than 0.6 for all targets, and above 0.8 for 7 out of 8 targets. The cut light curves yield larger lags ($\tau_{cut}$) than the original lags ($\tau_{u}$) derived from the $U$ band light curves for all targets. In addition, the other parallel method, the JAVELIN Pmap Model, also obtains similar lag measurements. The strong correlation and consistent measurements with large lags further indicate the presence of the outer component caused by the diffuse continuum emission from the BLR. A supplementary analysis using the Swift $V$ band light curves yields similar results, further confirming that this diffuse continuum contribution is not unique to the $U$ band.
\item Based on the lag measurements for 14 AGNs in this work and \cite{2024ApJ...966..149J}, both the $\log\tau_{cut}-\log\tau_{H\beta}$ and $\log\tau_{jav}-\log\tau_{H\beta}$ relationships exhibit a positive correlation with slope nearly equal to 1. It suggests an $R_{DCR}-R_{BLR}$ relation, in broad agreement with the predictions by \cite{2020MNRAS.494.1611N}. Similar to the $R-L$ relation of the broad H$\beta$ line, we also found a tight $R_{DCR}-L_{5100}$ relation according to the $\log\tau_{cut}-\log L_{5100}$ and $\log\tau_{cut}-\log L_{5100}$ relationship. Although the slopes are slightly smaller than 0.5 predicted by \cite{2022MNRAS.509.2637N}, they align well with the slopes of H$\beta$ $R-L$ relation given in recent studies \citep{2020ApJ...903..112D,2024ApJS..275...13W}.

\end{enumerate}

In summary, our work provides further evidence for a significant contribution of diffuse continuum emission from the BLR to the AGN continuum lags.

\section*{ACKNOWLEDGEMENTS}
We are thankful for the support of the National Science Foundation of China (12133001) and the National Key R \& D Program of China (2022YFF0503401). This work made use of data supplied by the UK Swift Science Data Centre at the University of Leicester. We acknowledge the Swift teams for their dedication to designing the UVOT filters and ensuring the flawless execution of the monitoring program. We are deeply grateful to the investigators who proposed the relevant Swift observation proposals, which made this study possible. This study would not have been feasible without the UVOT dataset.

\section*{DATA AVAILABILITY}
The raw data can be downloaded from the
Swift archive at 
\href{https://www.swift.ac.uk/}{https://www.swift.ac.uk/.} The processed light curves are available in electronic format via Zenodo
\href{https://zenodo.org/records/14930797}{https://zenodo.org/records/14930797.}

\bibliography{sample701}{}
\bibliographystyle{aasjournalv7}

\appendix
Here we provide the multiband light curves and interband lag measurements (see Figure Set \ref{FigSetA}), as well as the ICCF-Cut and JAVELIN Pmap Model results (see Figure Set \ref{FigSetB}) for the remaining 7 targets in our sample. In addition, we also present the ICCF-Cut results derived from the $V$ band light curves (see Figure \ref{FigVbandICCF}), which further demonstrate that the diffuse continuum contribution is not unique to the $U$ band.

%=============================
% Appendix A Figure Set
%=============================
\figsetstart
\figsetnum{A}
\figsettitle{Multiband light curves and interband lag measurements for the remaining seven targets, with the same details as those in Figure \ref{fig:Fairall9_lc}}

\figsetgrpstart
\figsetgrpnum{A1}
\figsetgrptitle{3C 120}
\figsetplot{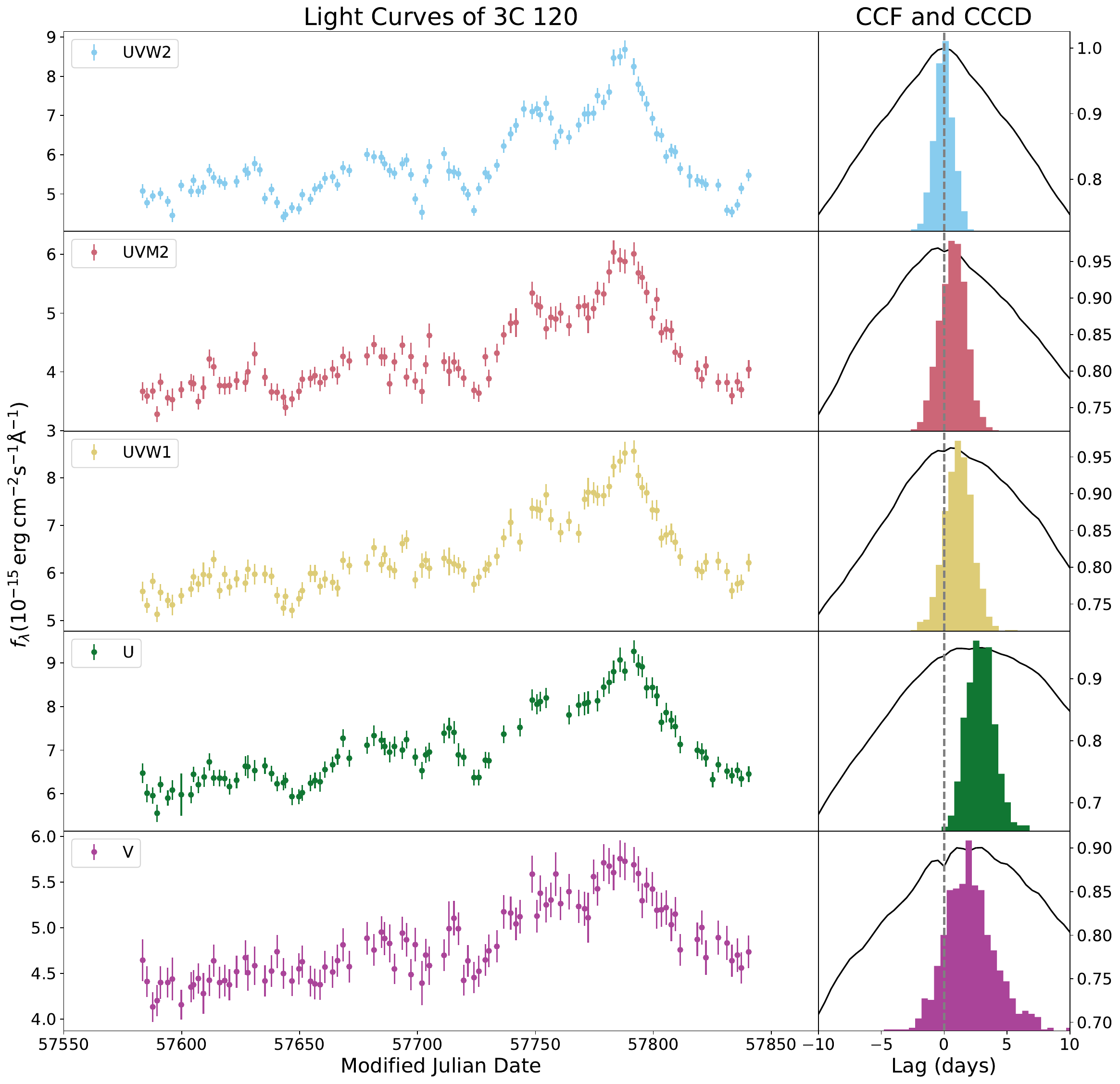}
\figsetgrpnote{The same as Figure \ref{fig:Fairall9_lc}, but for 3C 120. Note that the $B$ band data is unavailable in the Swift archive.}
\figsetgrpend

\figsetgrpstart
\figsetgrpnum{A2}
\figsetgrptitle{MCG+08-11-011}
\figsetplot{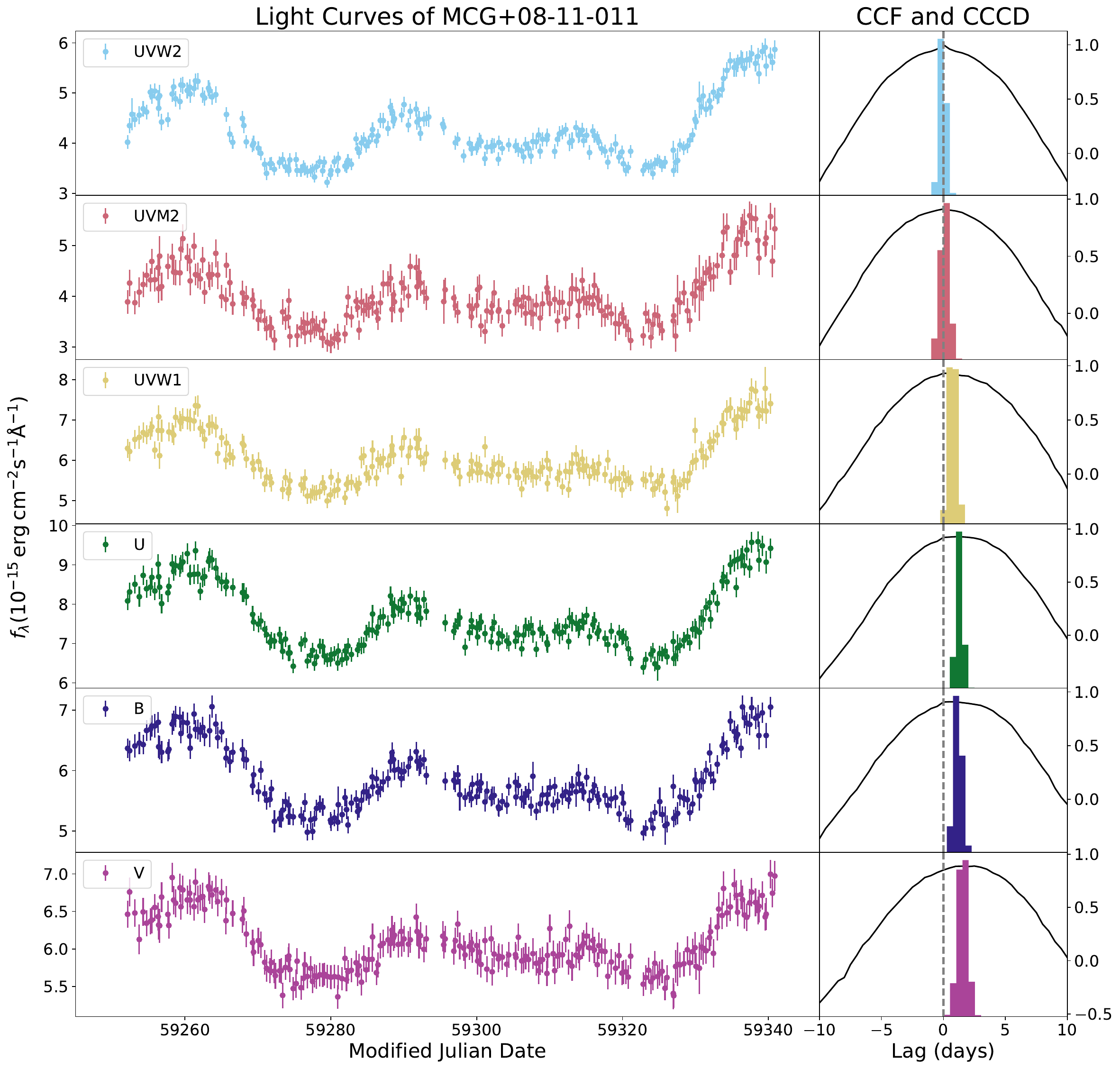}
\figsetgrpnote{The same as Figure \ref{fig:Fairall9_lc}, but for MCG+08-11-011.}
\figsetgrpend

\figsetgrpstart
\figsetgrpnum{A3}
\figsetgrptitle{Mrk 110}
\figsetplot{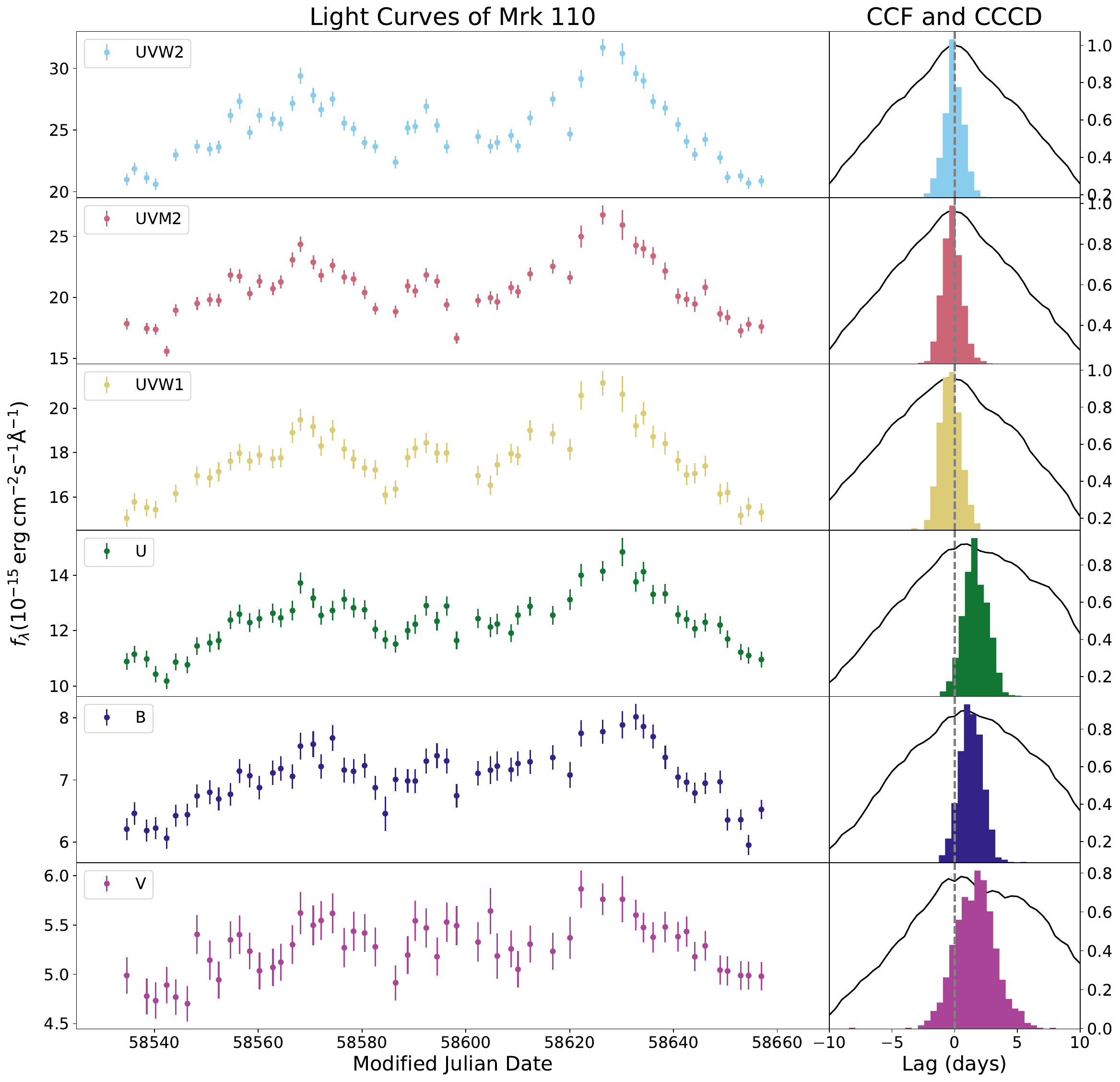}
\figsetgrpnote{The same as Figure \ref{fig:Fairall9_lc}, but for Mrk 110.}
\figsetgrpend

\figsetgrpstart
\figsetgrpnum{A4}
\figsetgrptitle{Mrk 279 (1)}
\figsetplot{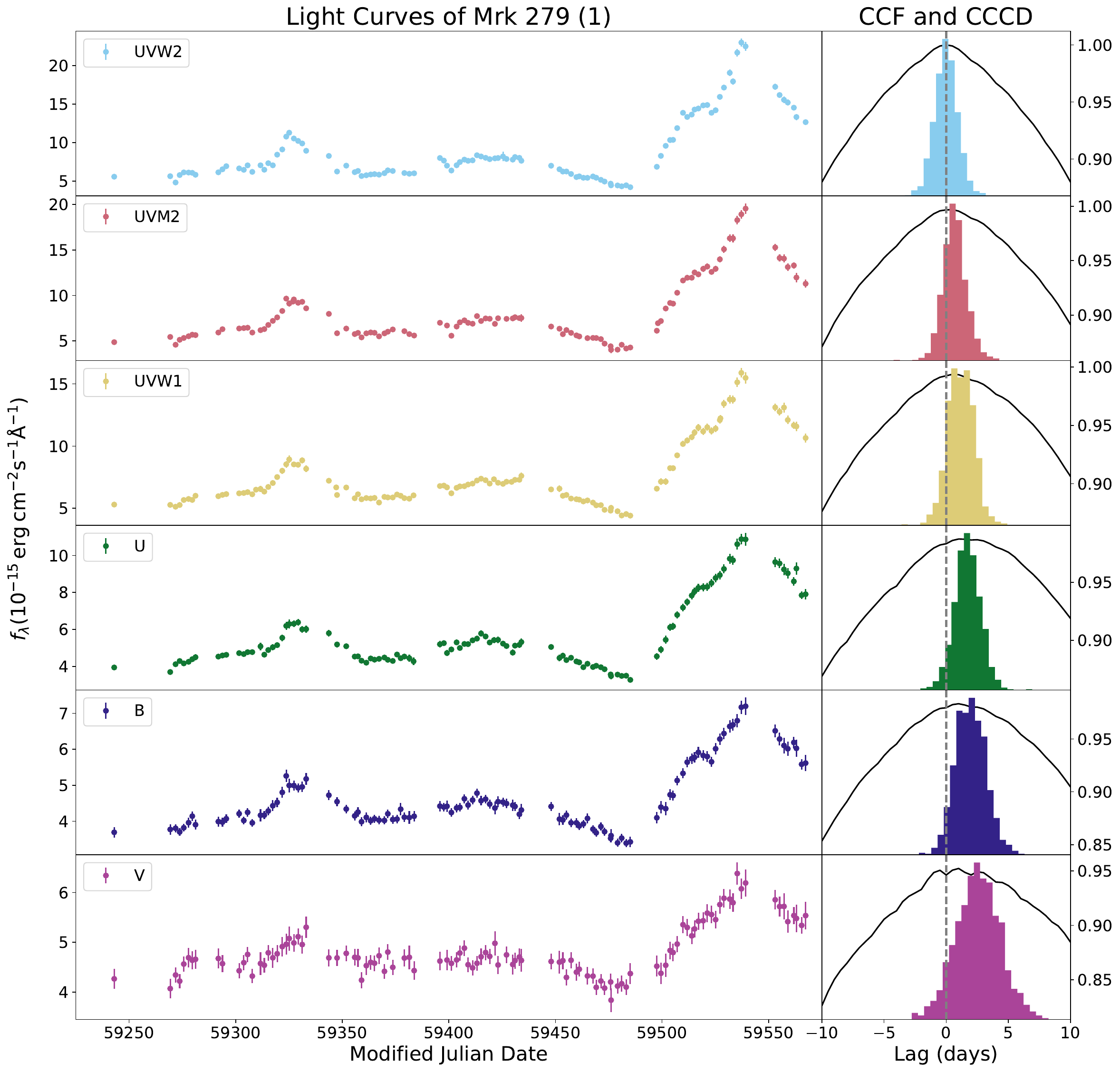}
\figsetgrpnote{The same as Figure \ref{fig:Fairall9_lc}, but for the first part light curves of Mrk 279.}
\figsetgrpend

\figsetgrpstart
\figsetgrpnum{A5}
\figsetgrptitle{Mrk 279 (2)}
\figsetplot{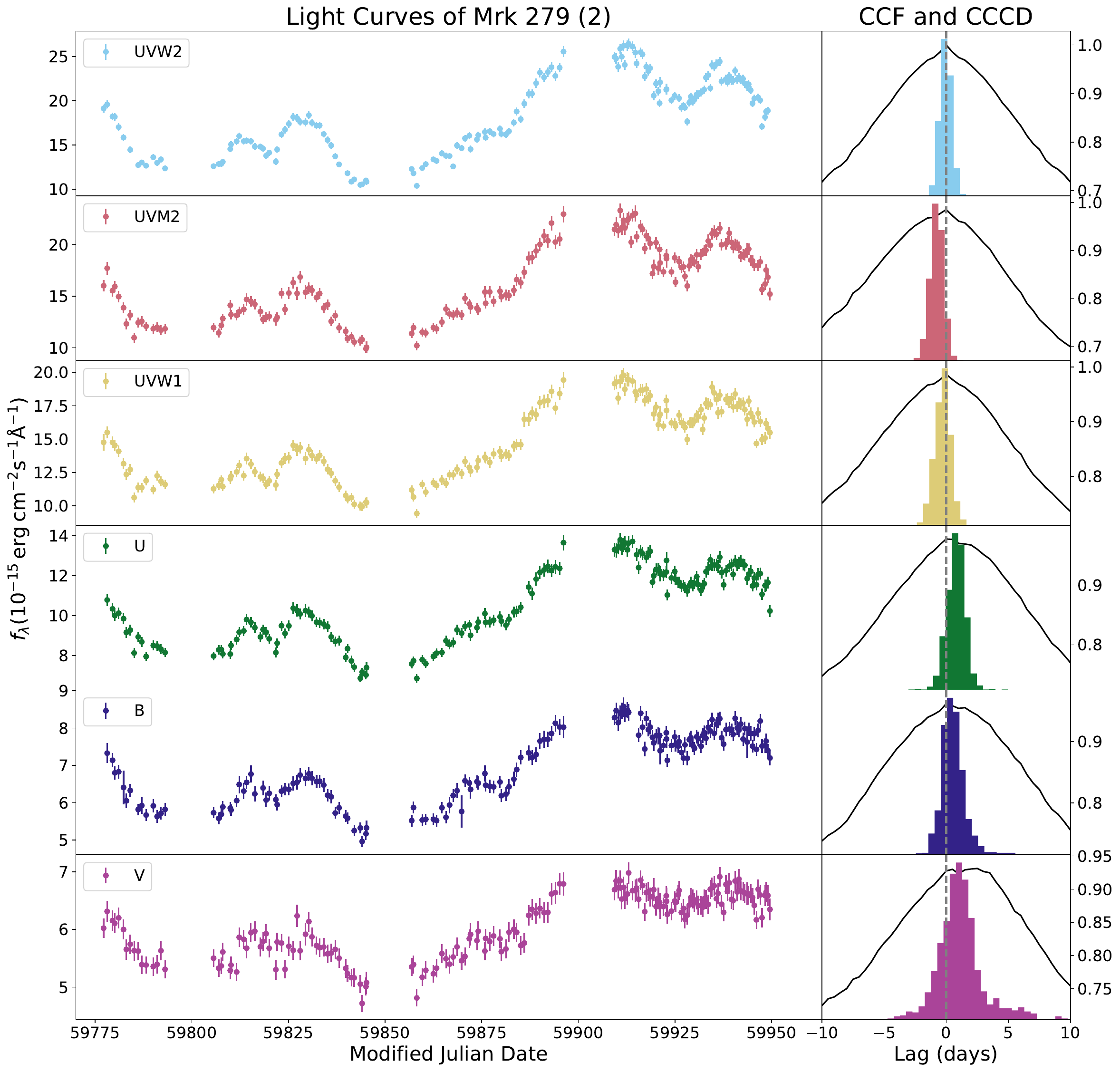}
\figsetgrpnote{The same as Figure \ref{fig:Fairall9_lc}, but for the second part light curves of Mrk 279.}
\figsetgrpend

\figsetgrpstart
\figsetgrpnum{A6}
\figsetgrptitle{Mrk 279 (3)}
\figsetplot{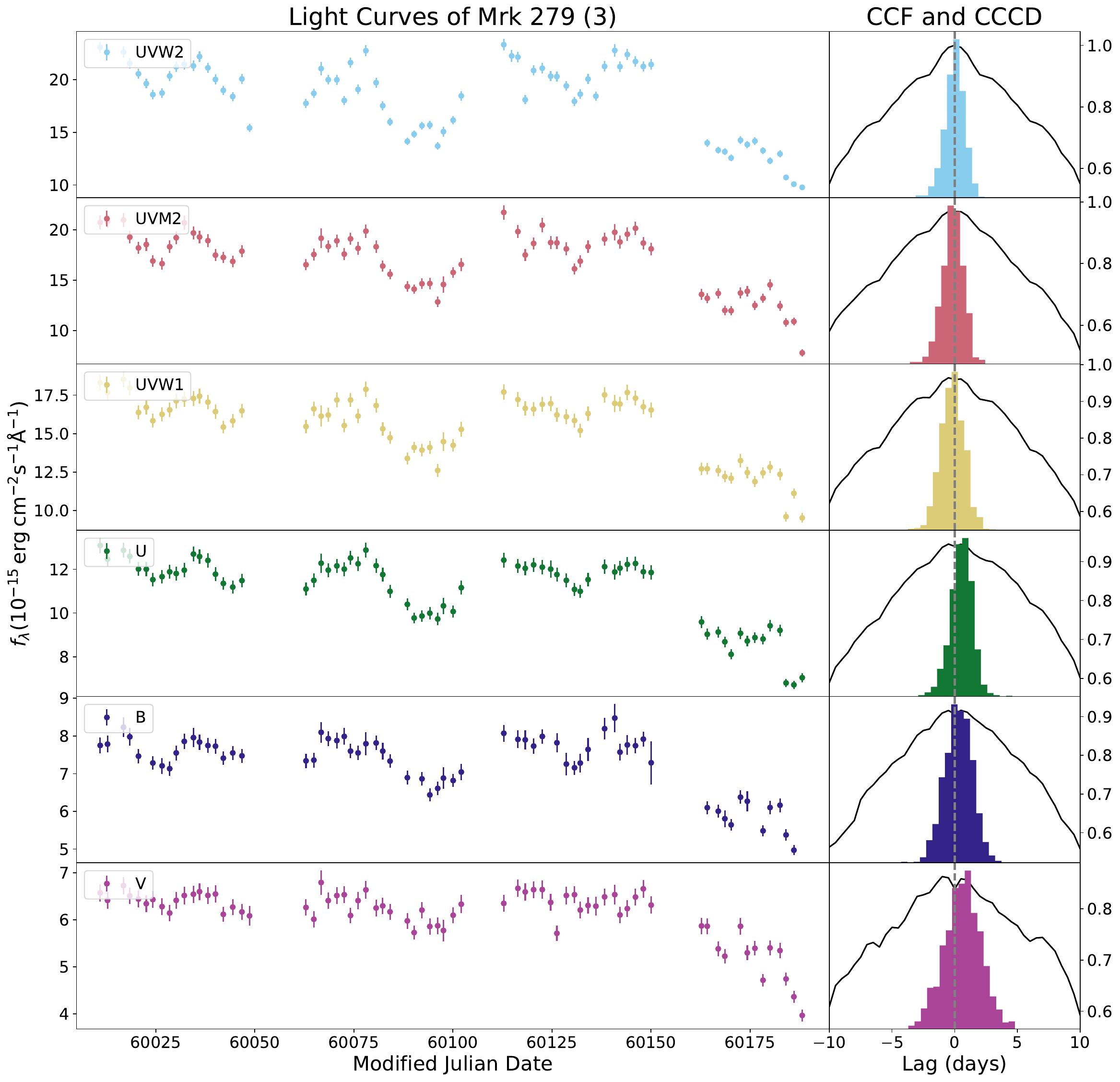}
\figsetgrpnote{The same as Figure \ref{fig:Fairall9_lc}, but for the third part light curves of Mrk 279.}
\figsetgrpend

\figsetgrpstart
\figsetgrpnum{A7}
\figsetgrptitle{Mrk 335}
\figsetplot{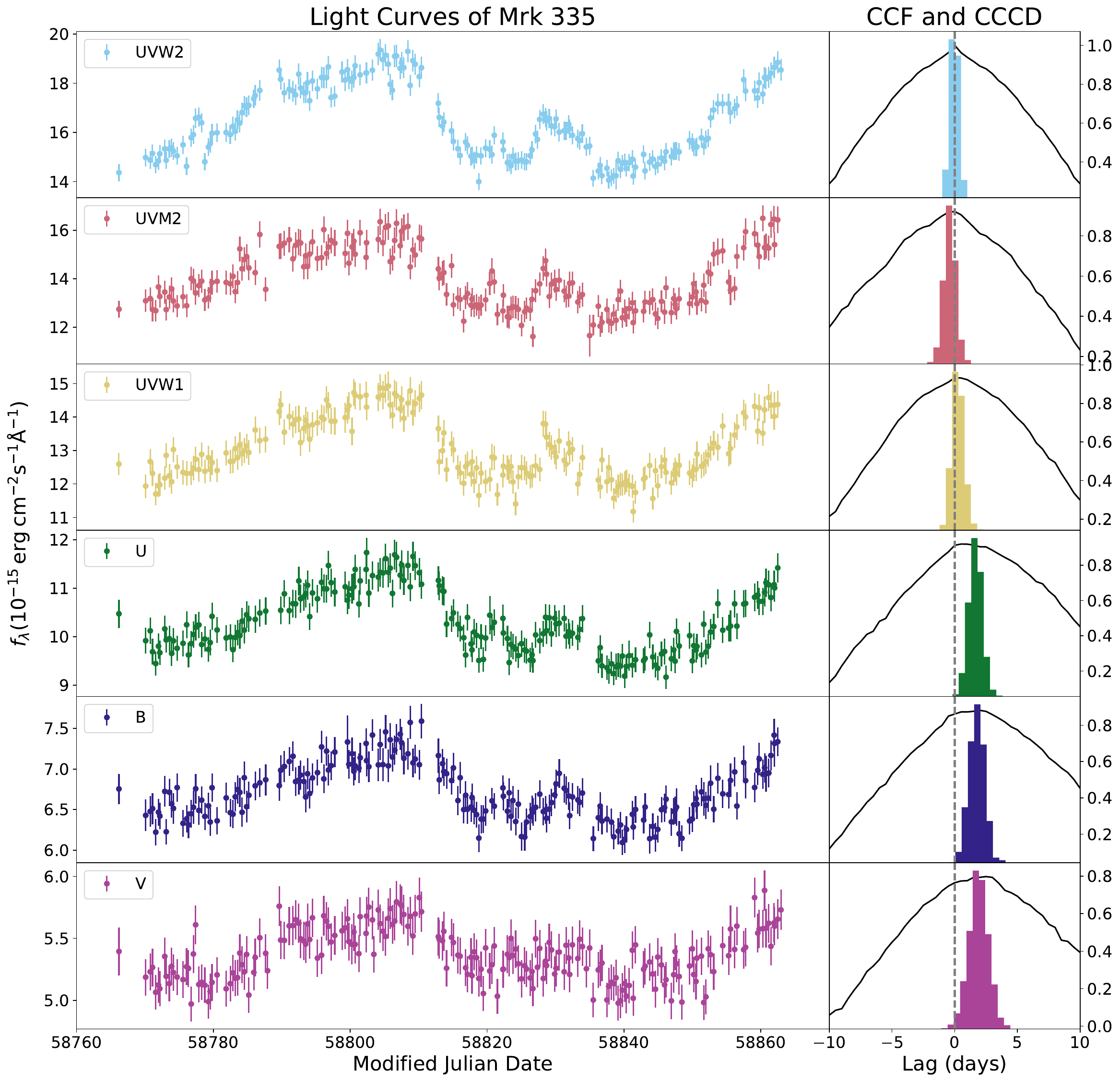}
\figsetgrpnote{The same as Figure \ref{fig:Fairall9_lc}, but for Mrk 335.}
\figsetgrpend

\figsetgrpstart
\figsetgrpnum{A8}
\figsetgrptitle{Mrk 817}
\figsetplot{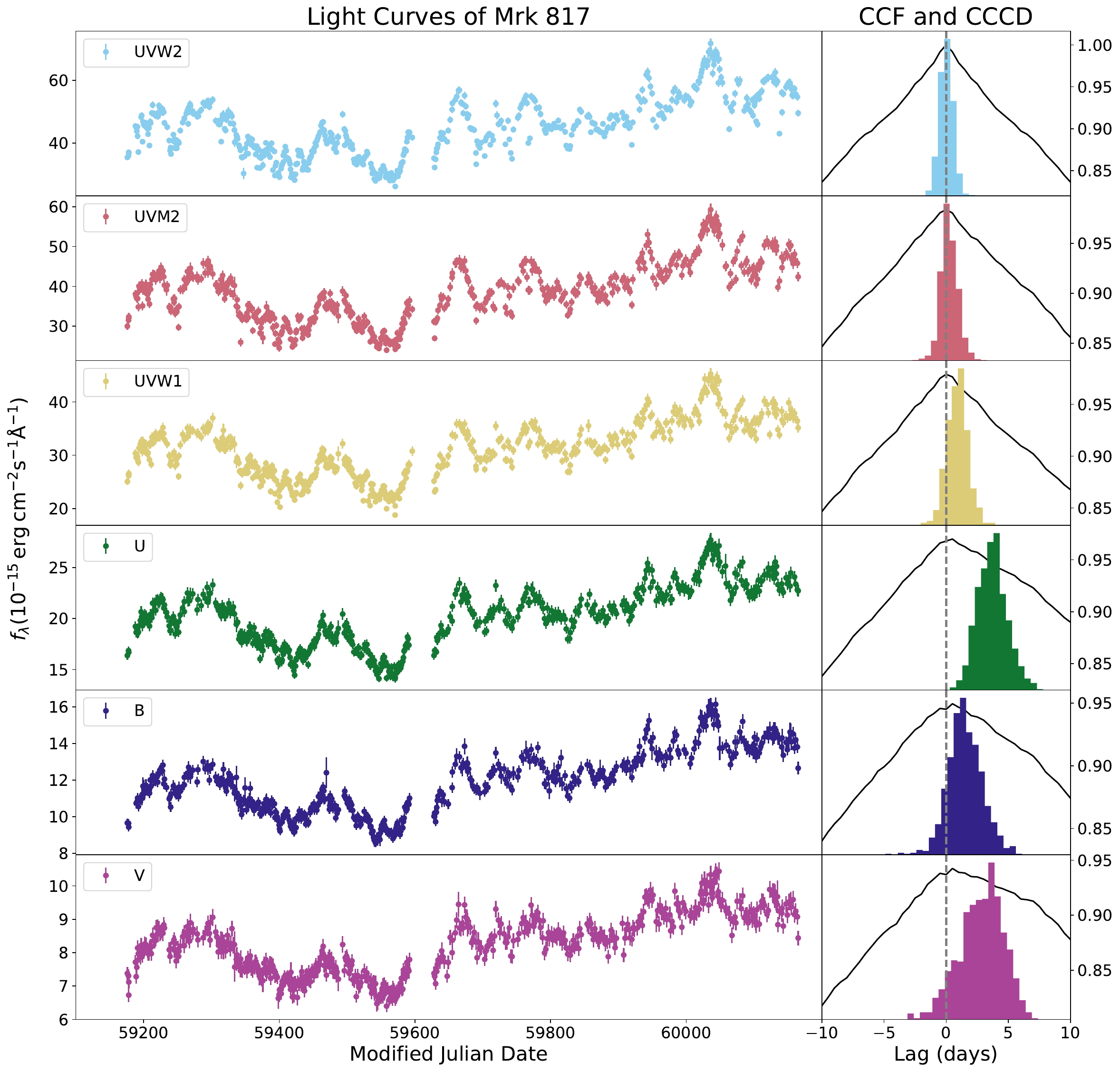}
\figsetgrpnote{The same as Figure \ref{fig:Fairall9_lc}, but for the whole light curves of Mrk 817.}
\figsetgrpend

\figsetgrpstart
\figsetgrpnum{A9}
\figsetgrptitle{Mrk 817 (1)}
\figsetplot{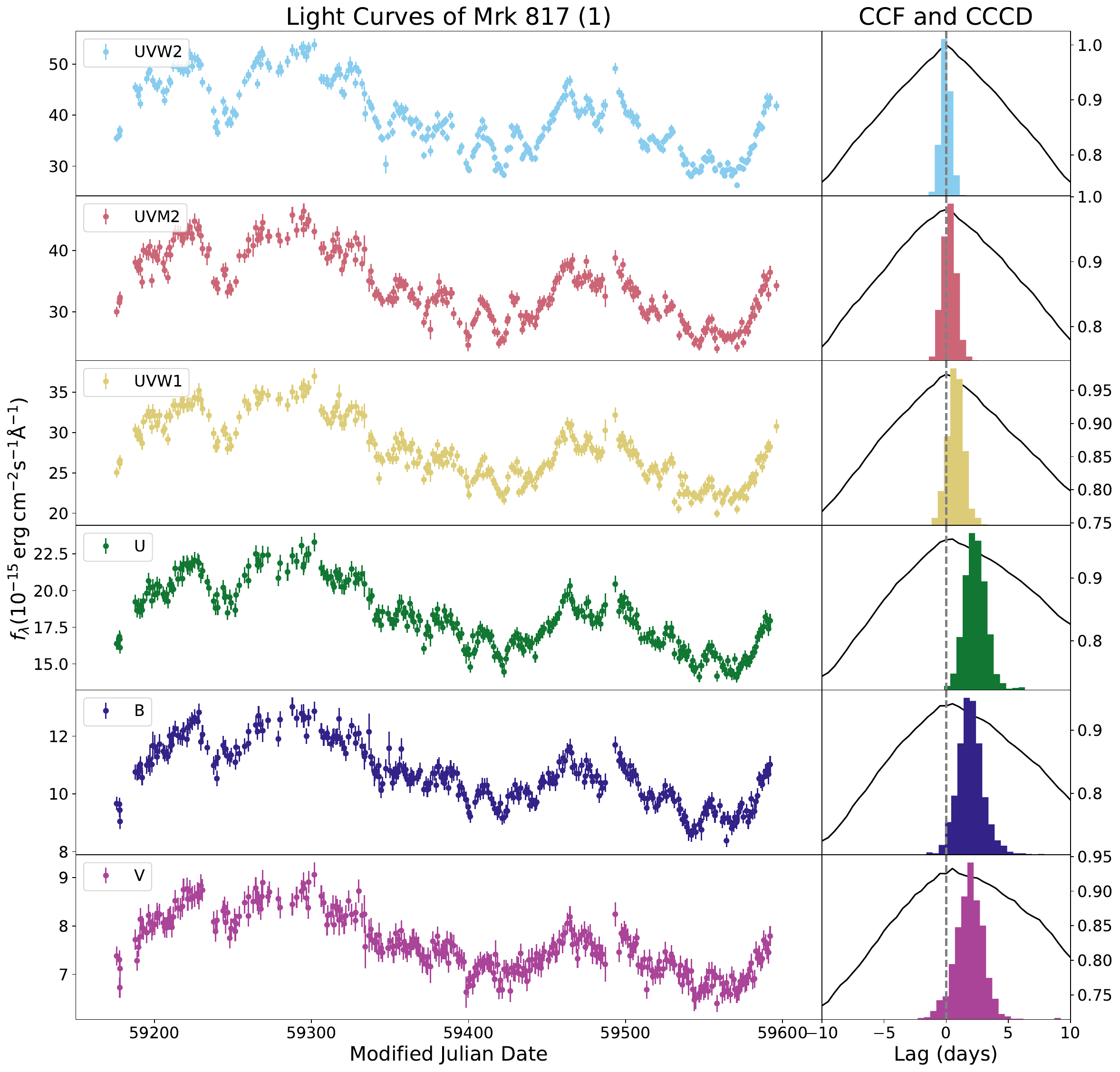}
\figsetgrpnote{The same as Figure \ref{fig:Fairall9_lc}, but for the first part light curves of Mrk 817.}
\figsetgrpend

\figsetgrpstart
\figsetgrpnum{A10}
\figsetgrptitle{Mrk 817 (2)}
\figsetplot{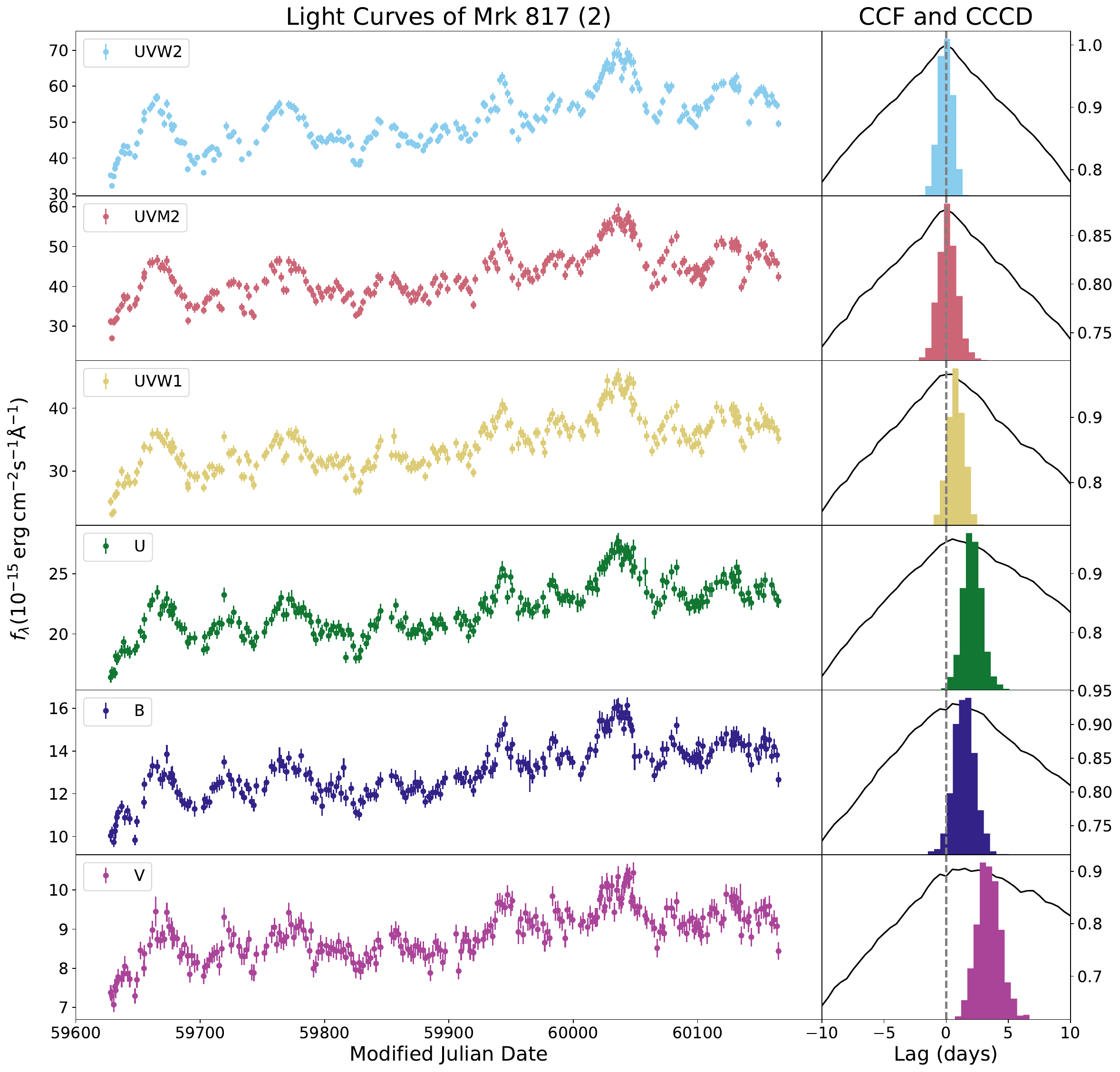}
\figsetgrpnote{The same as Figure \ref{fig:Fairall9_lc}, but for the second part light curves of Mrk 817.}
\figsetgrpend

\figsetgrpstart
\figsetgrpnum{A11}
\figsetgrptitle{NGC 6814}
\figsetplot{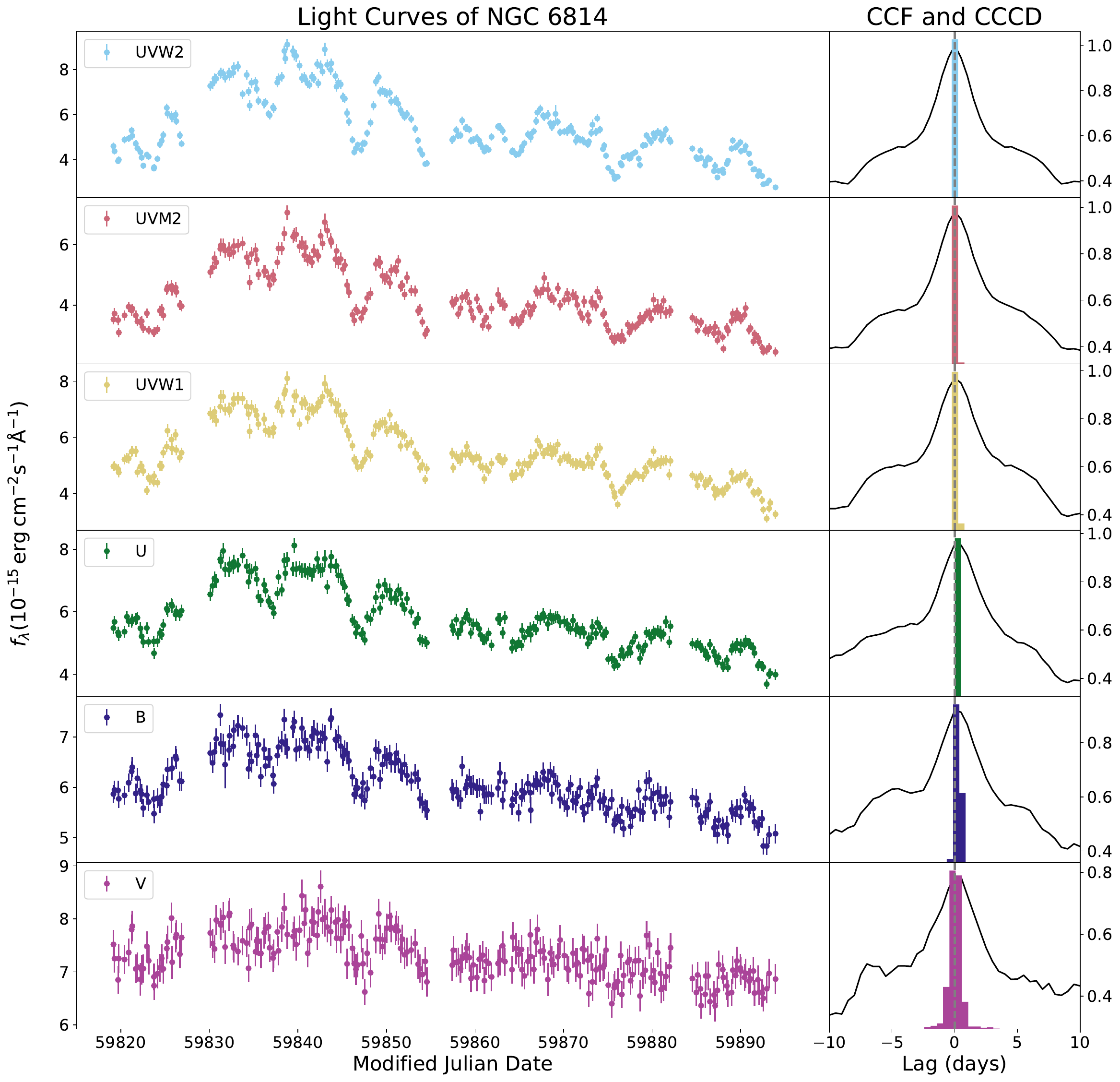}
\figsetgrpnote{The same as Figure \ref{fig:Fairall9_lc}, but for NGC 6814.}
\figsetgrpend

\figsetgrpstart
\figsetgrpnum{A12}
\figsetgrptitle{NGC 6814 (D)}
\figsetplot{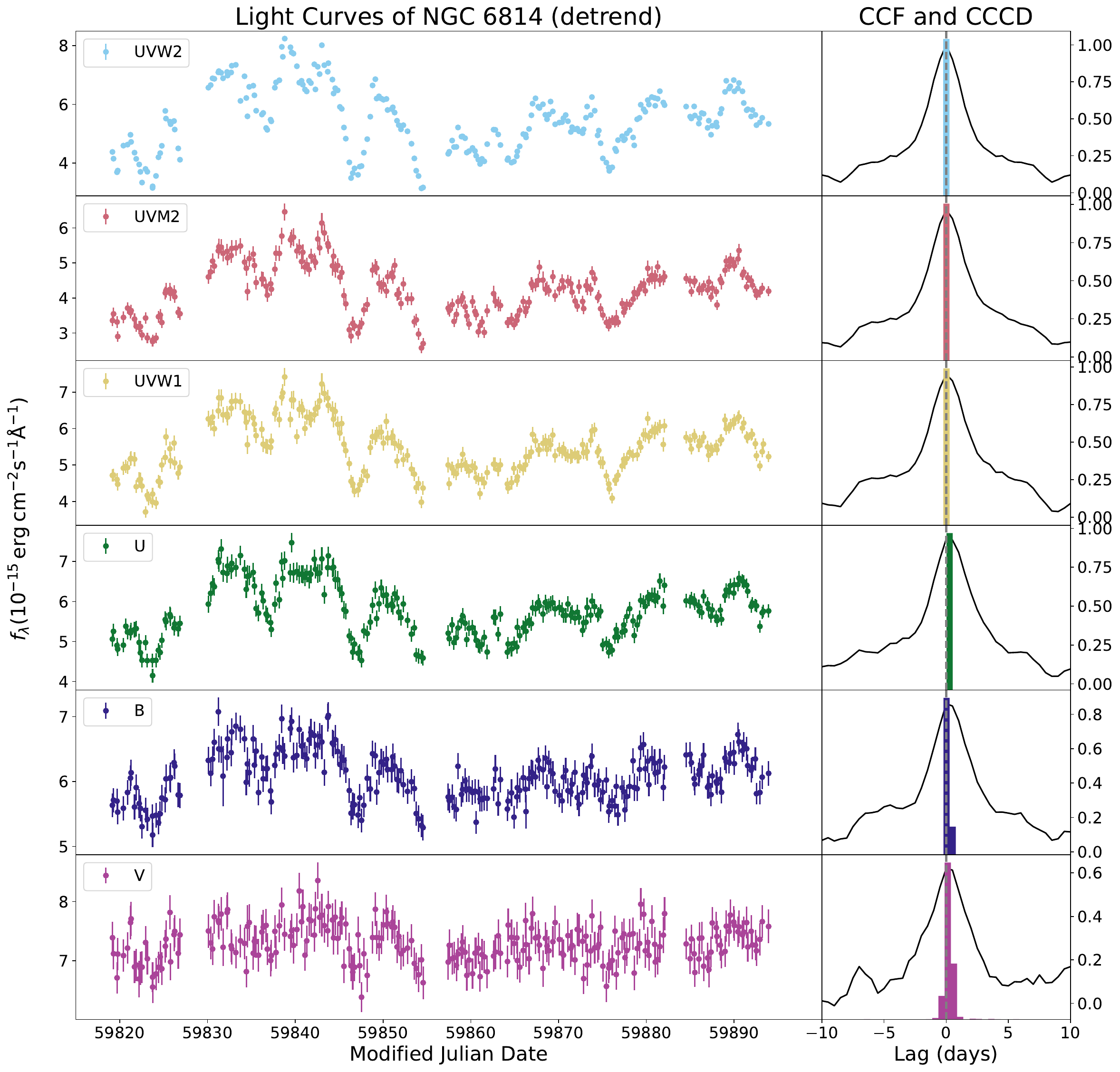}
\figsetgrpnote{The same as Figure \ref{fig:Fairall9_lc}, but for the detrended light curves of NGC 6814.}
\figsetgrpend

\figsetend

\renewcommand{\thefigure}{A\arabic{figure}}
\setcounter{figure}{0}

\begin{figure*}[h]
\centering
\includegraphics[width=0.85\textwidth]{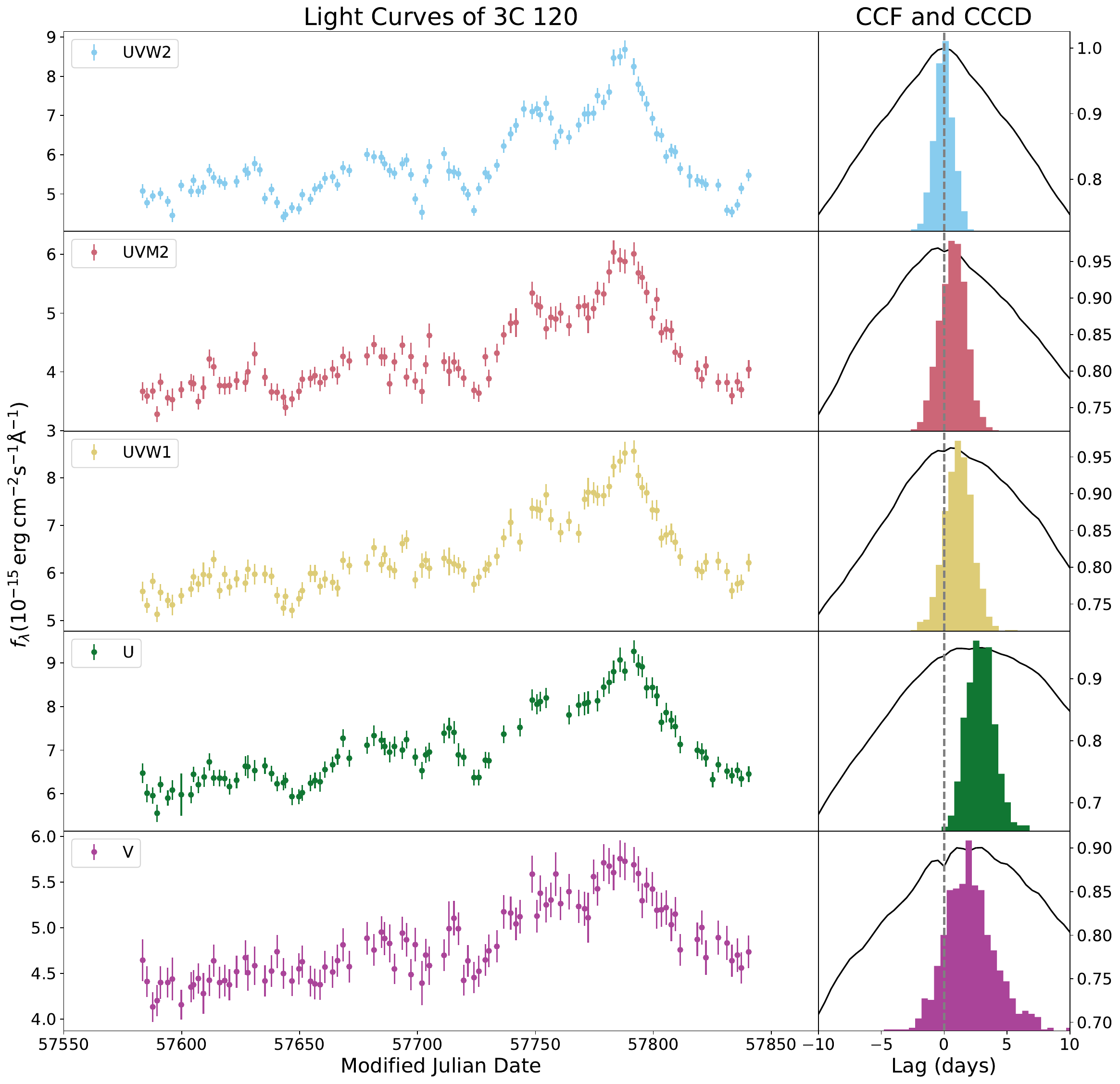}
\digitalasset
\caption{The same as Figure \ref{fig:Fairall9_lc}, but for 3C 120. Note that the $B$ band data for 3C 120 is unavailable in the Swift archive. The complete figure set (12 images) is available in the online journal.}
\label{FigSetA}
\end{figure*}

\newpage

%=============================
% Appendix B Figure Set
%=============================
\figsetstart
\figsetnum{B}
\figsettitle{ICCF-Cut and JAVELIN Pmap Model results for the remaining seven targets, with the same details as those in Figure \ref{fig:Fairall9_result}}

\figsetgrpstart
\figsetgrpnum{B1}
\figsetgrptitle{3C 120}
\figsetplot{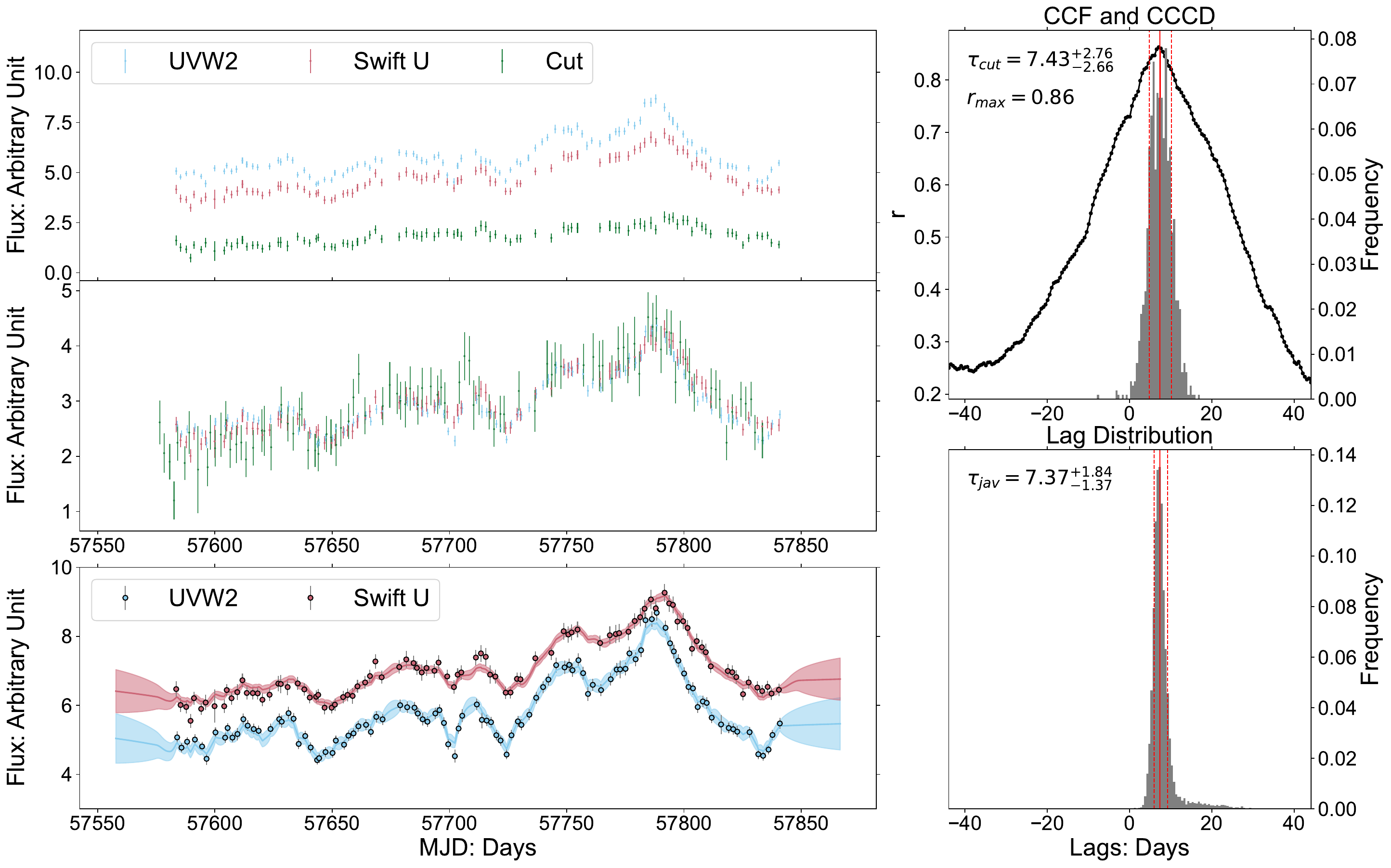}
\figsetgrpnote{The same as Figure \ref{fig:Fairall9_result}, but for 3C 120.}
\figsetgrpend

\figsetgrpstart
\figsetgrpnum{B2}
\figsetgrptitle{MCG+08-11-011}
\figsetplot{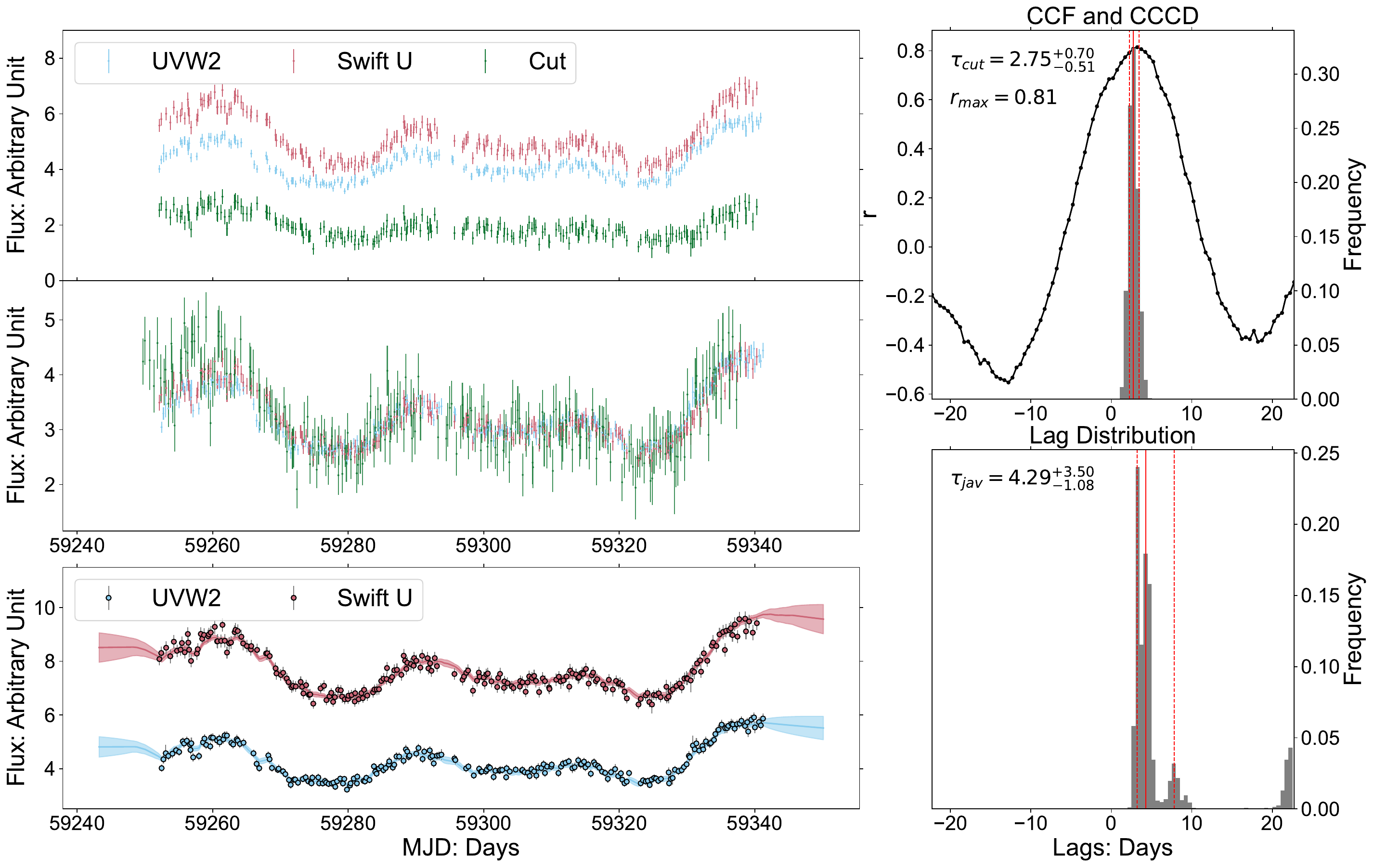}
\figsetgrpnote{The same as Figure \ref{fig:Fairall9_result}, but for MCG+08-11-011.}
\figsetgrpend

\figsetgrpstart
\figsetgrpnum{B3}
\figsetgrptitle{Mrk 110}
\figsetplot{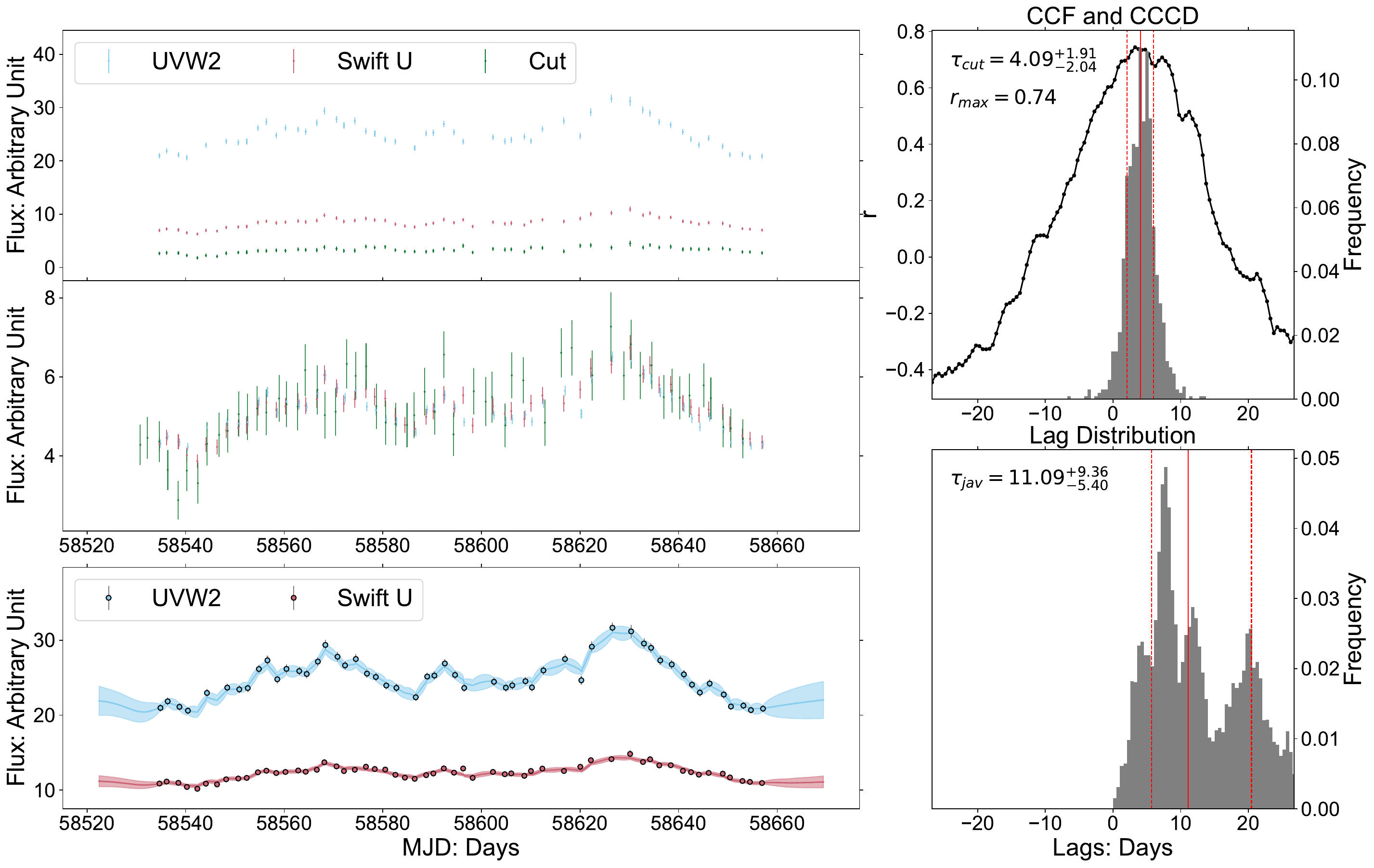}
\figsetgrpnote{The same as Figure \ref{fig:Fairall9_result}, but for Mrk 110.}
\figsetgrpend

\figsetgrpstart
\figsetgrpnum{B4}
\figsetgrptitle{Mrk 279 (1)}
\figsetplot{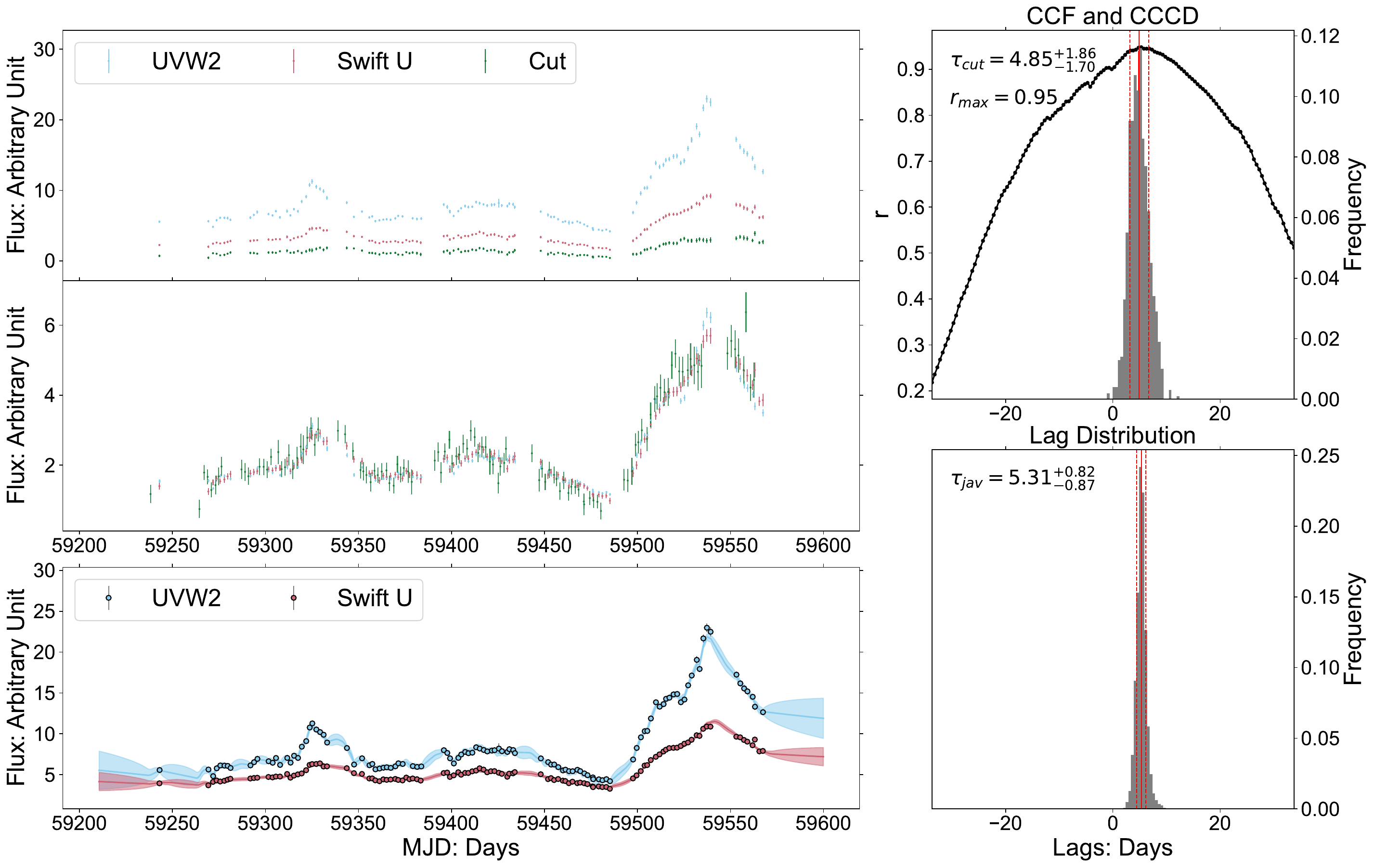}
\figsetgrpnote{The same as Figure \ref{fig:Fairall9_result}, but for the first part light curves of Mrk 279.}
\figsetgrpend

\figsetgrpstart
\figsetgrpnum{B5}
\figsetgrptitle{Mrk 279 (2)}
\figsetplot{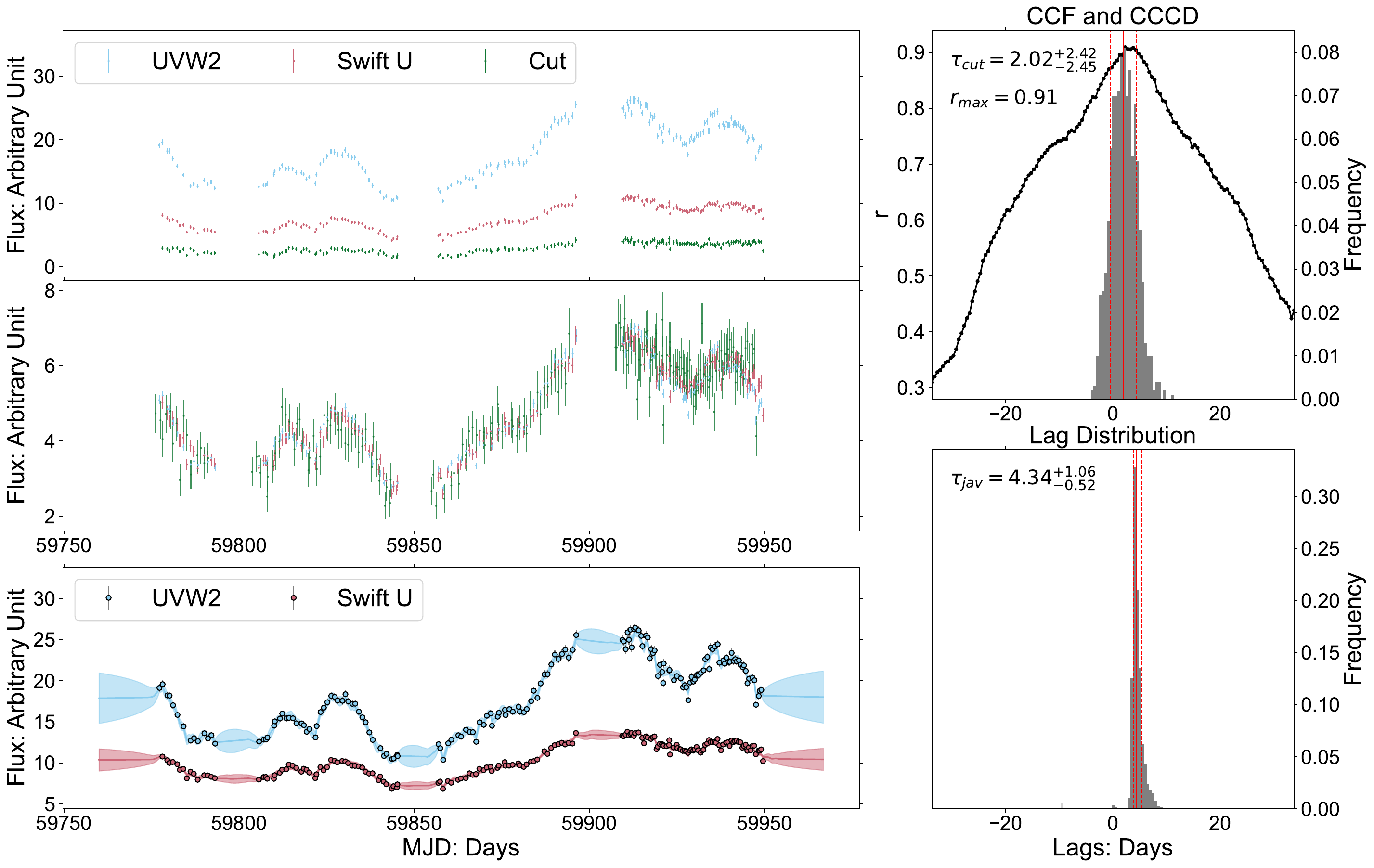}
\figsetgrpnote{The same as Figure \ref{fig:Fairall9_result}, but for the second part light curves of Mrk 279.}
\figsetgrpend

\figsetgrpstart
\figsetgrpnum{B6}
\figsetgrptitle{Mrk 279 (3)}
\figsetplot{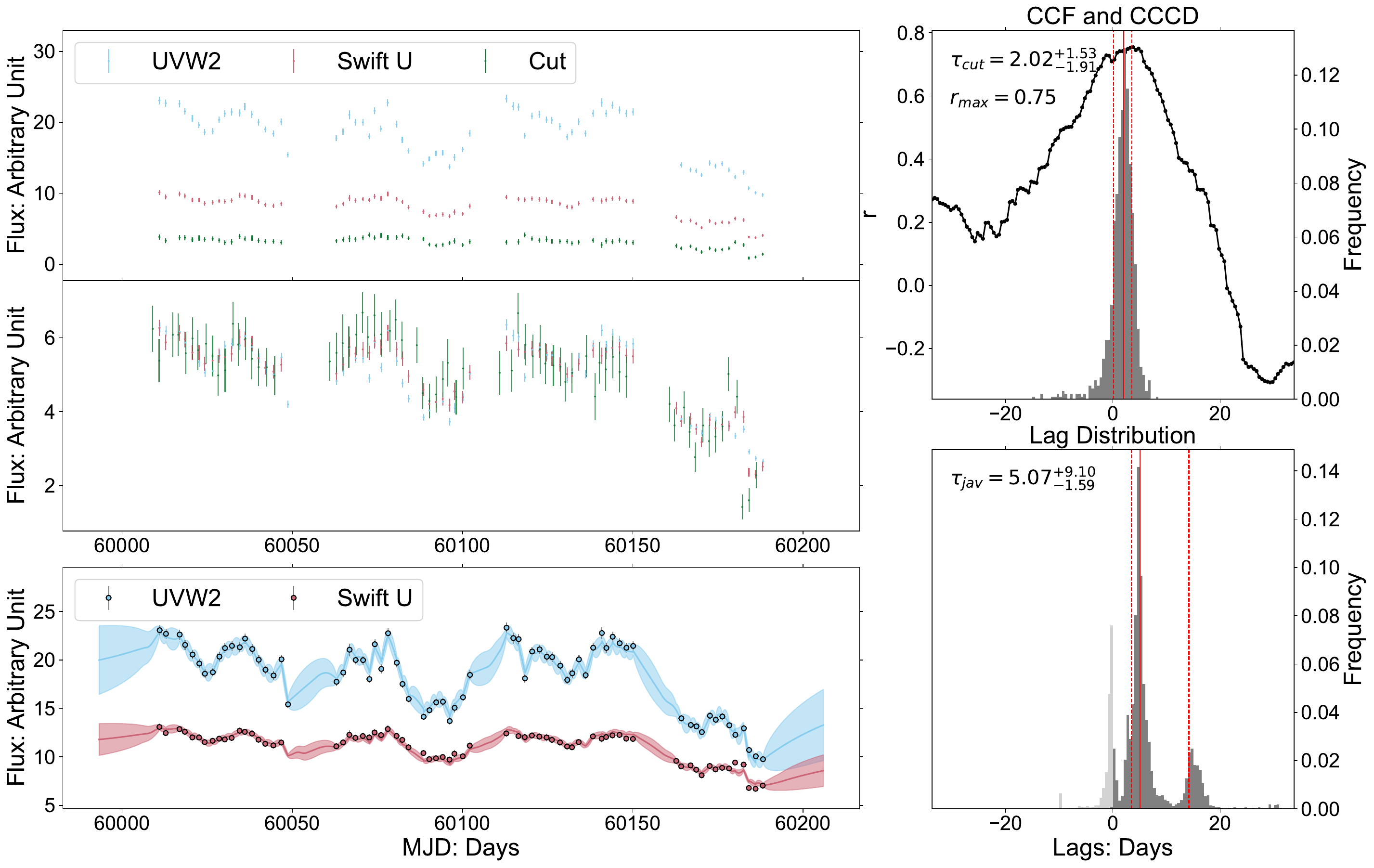}
\figsetgrpnote{The same as Figure \ref{fig:Fairall9_result}, but for the third part light curves of Mrk 279.}
\figsetgrpend

\figsetgrpstart
\figsetgrpnum{B7}
\figsetgrptitle{Mrk 335}
\figsetplot{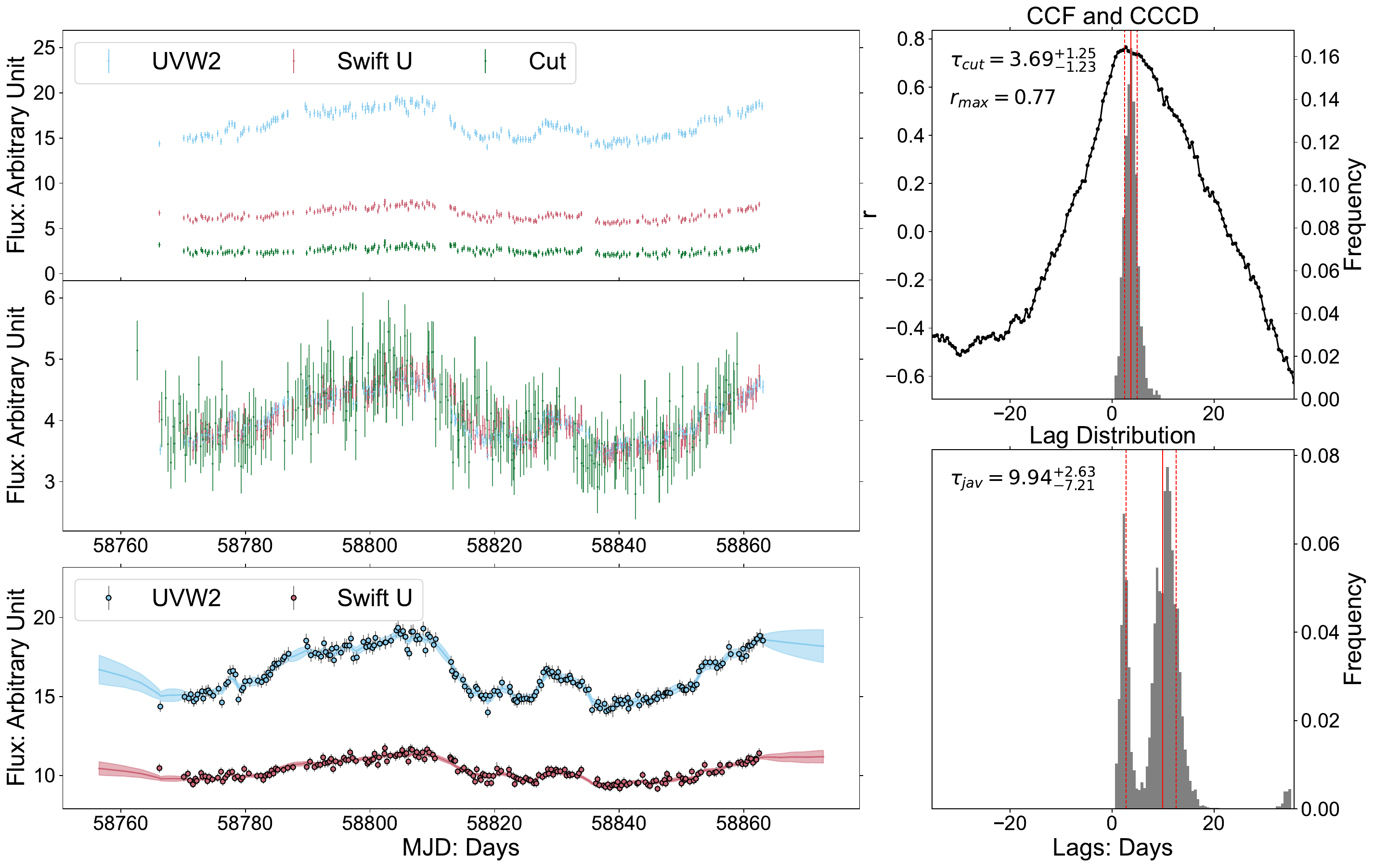}
\figsetgrpnote{The same as Figure \ref{fig:Fairall9_result}, but for Mrk 335.}
\figsetgrpend

\figsetgrpstart
\figsetgrpnum{B8}
\figsetgrptitle{Mrk 817}
\figsetplot{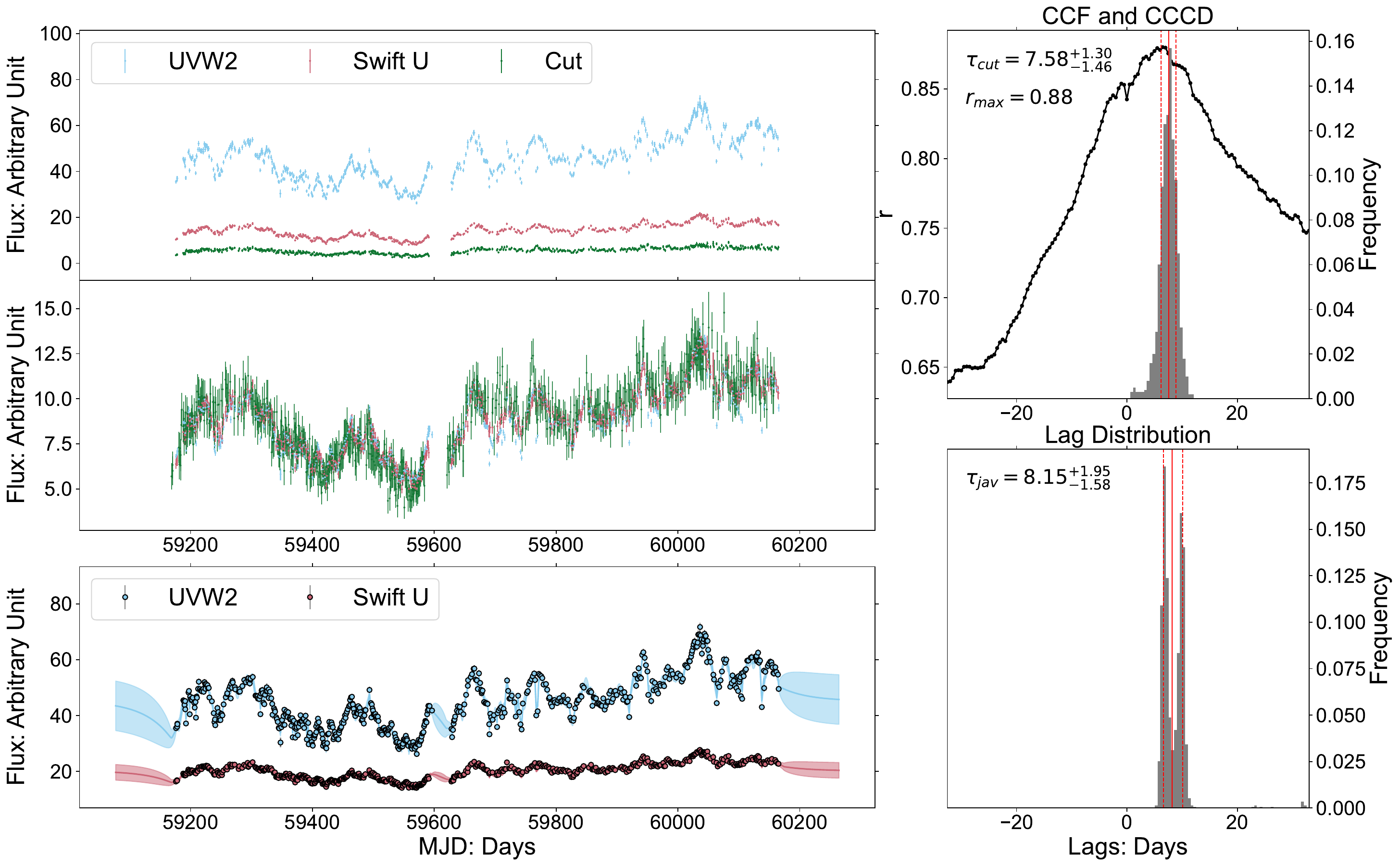}
\figsetgrpnote{The same as Figure \ref{fig:Fairall9_result}, but for the whole light curves of Mrk 817.}
\figsetgrpend

\figsetgrpstart
\figsetgrpnum{B9}
\figsetgrptitle{Mrk 817 (1)}
\figsetplot{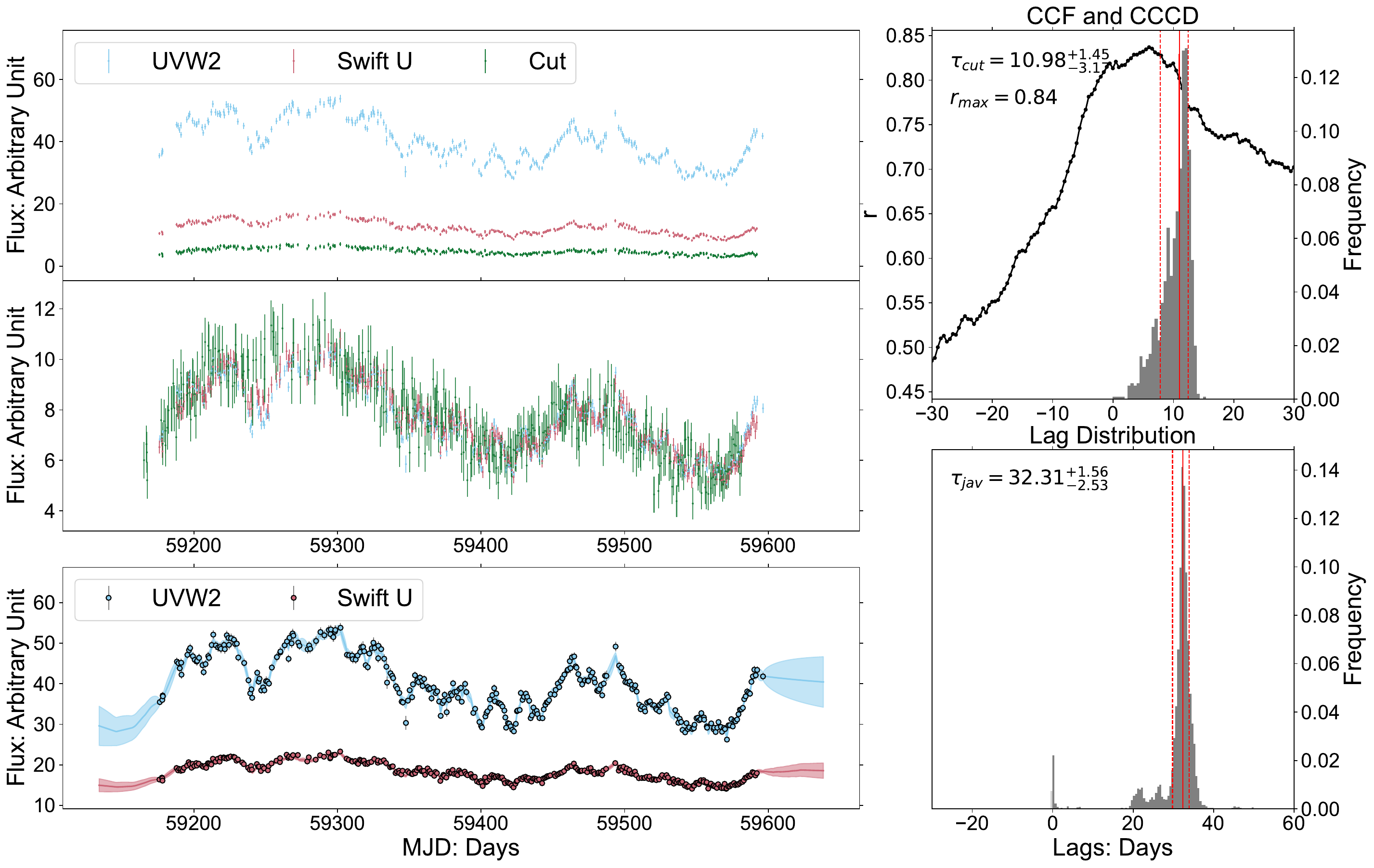}
\figsetgrpnote{The same as Figure \ref{fig:Fairall9_result}, but for the first part light curves of Mrk 817.}
\figsetgrpend

\figsetgrpstart
\figsetgrpnum{B10}
\figsetgrptitle{Mrk 817 (2)}
\figsetplot{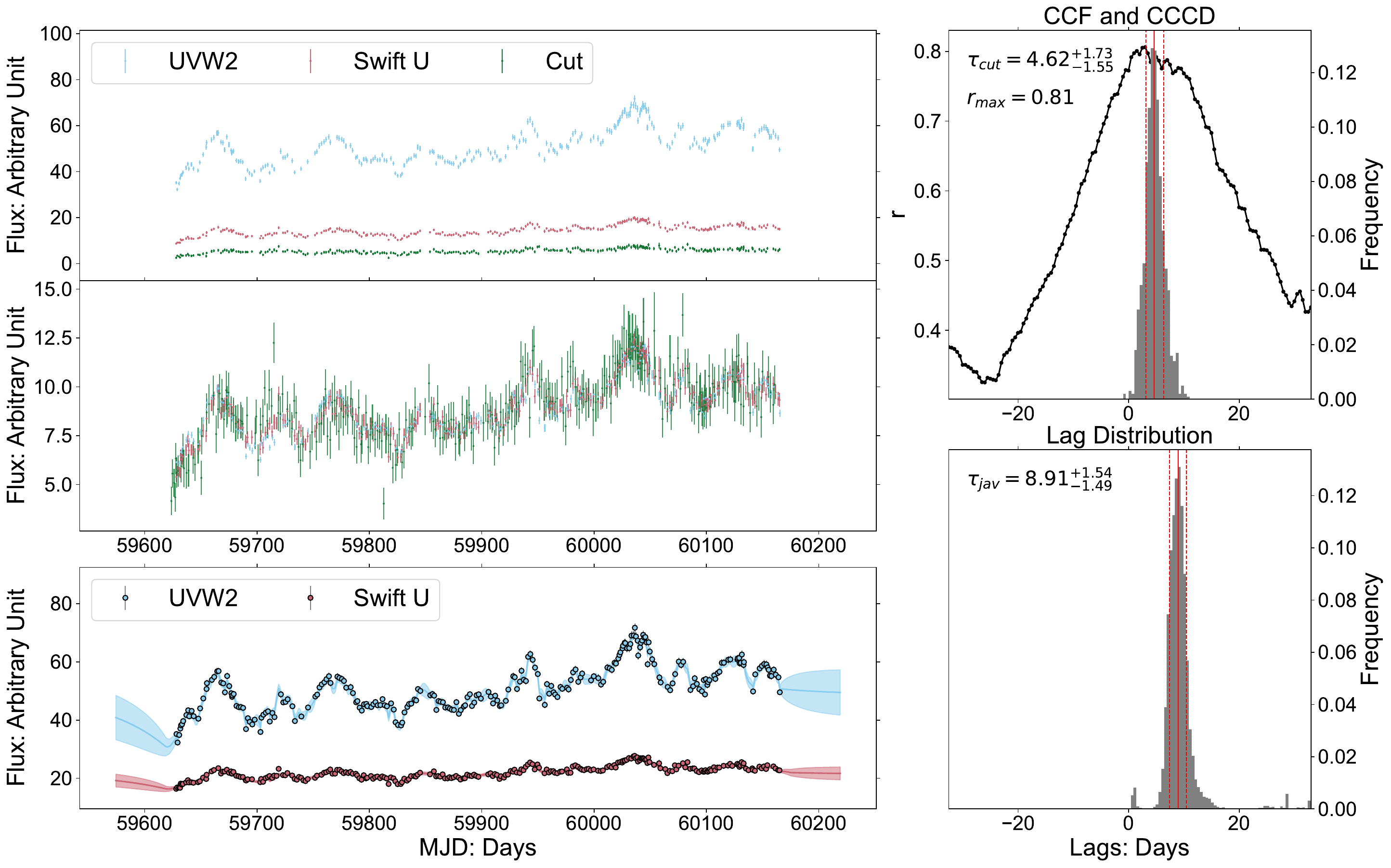}
\figsetgrpnote{The same as Figure \ref{fig:Fairall9_result}, but for the second part light curves of Mrk 817.}
\figsetgrpend

\figsetgrpstart
\figsetgrpnum{B11}
\figsetgrptitle{NGC 6814}
\figsetplot{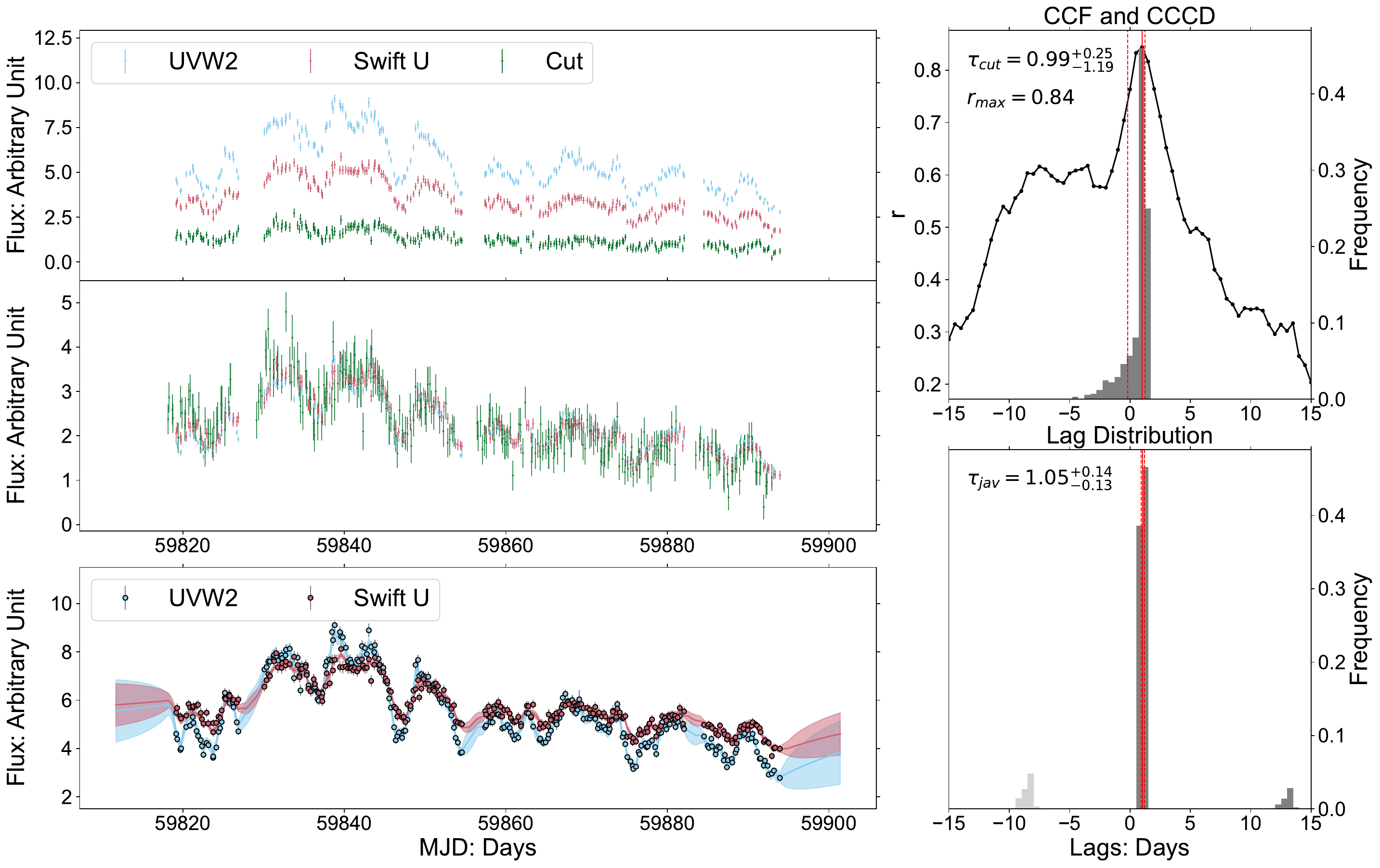}
\figsetgrpnote{The same as Figure \ref{fig:Fairall9_result}, but for light curves of NGC 6814.}
\figsetgrpend

\figsetgrpstart
\figsetgrpnum{B12}
\figsetgrptitle{NGC 6814 (D)}
\figsetplot{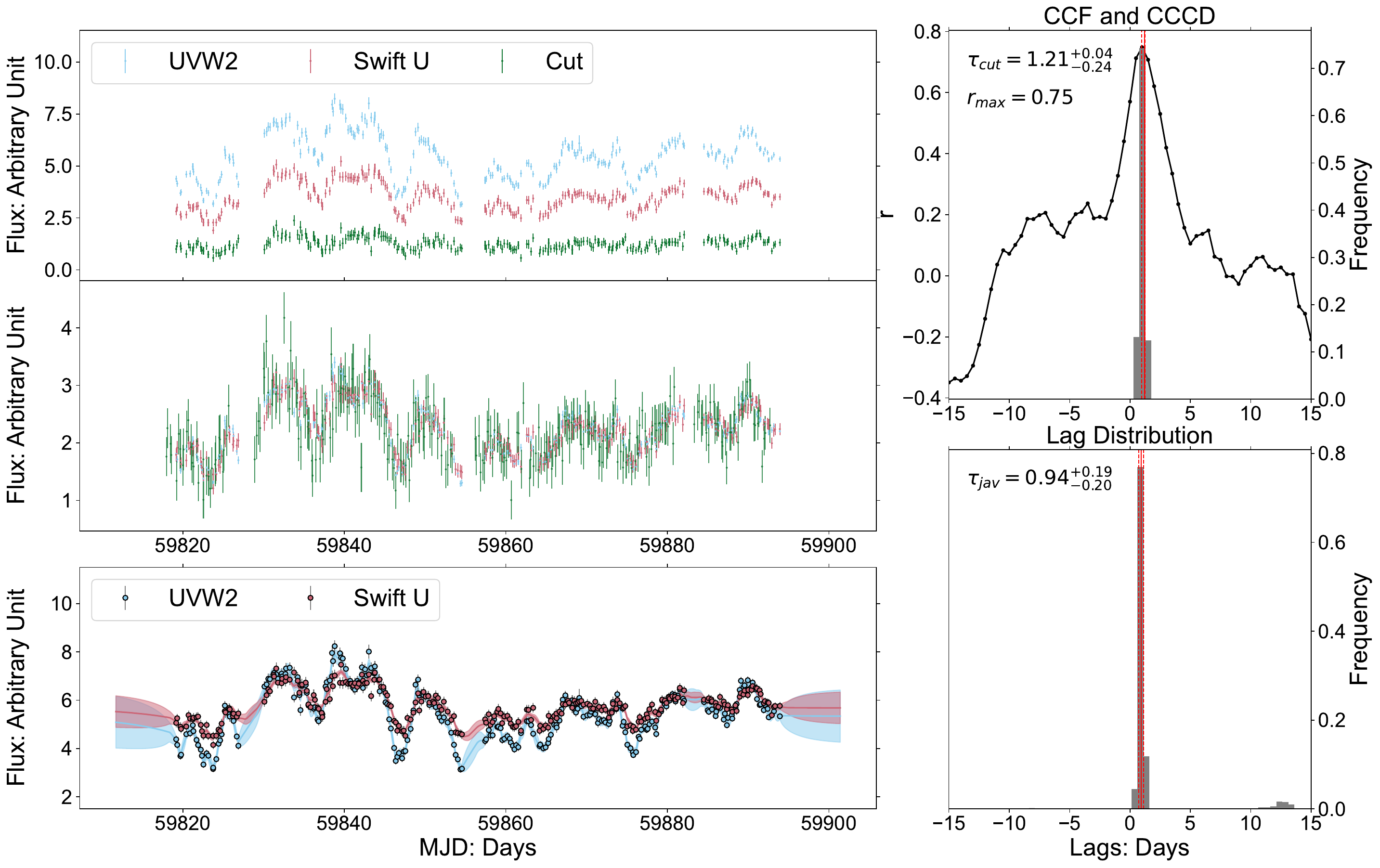}
\figsetgrpnote{The same as Figure \ref{fig:Fairall9_result}, but for the detrended light curves of NGC 6814.}
\figsetgrpend

\figsetend

\renewcommand{\thefigure}{B\arabic{figure}}
\setcounter{figure}{0}

\begin{figure*}[h]
\centering
\includegraphics[width=0.85\textwidth]{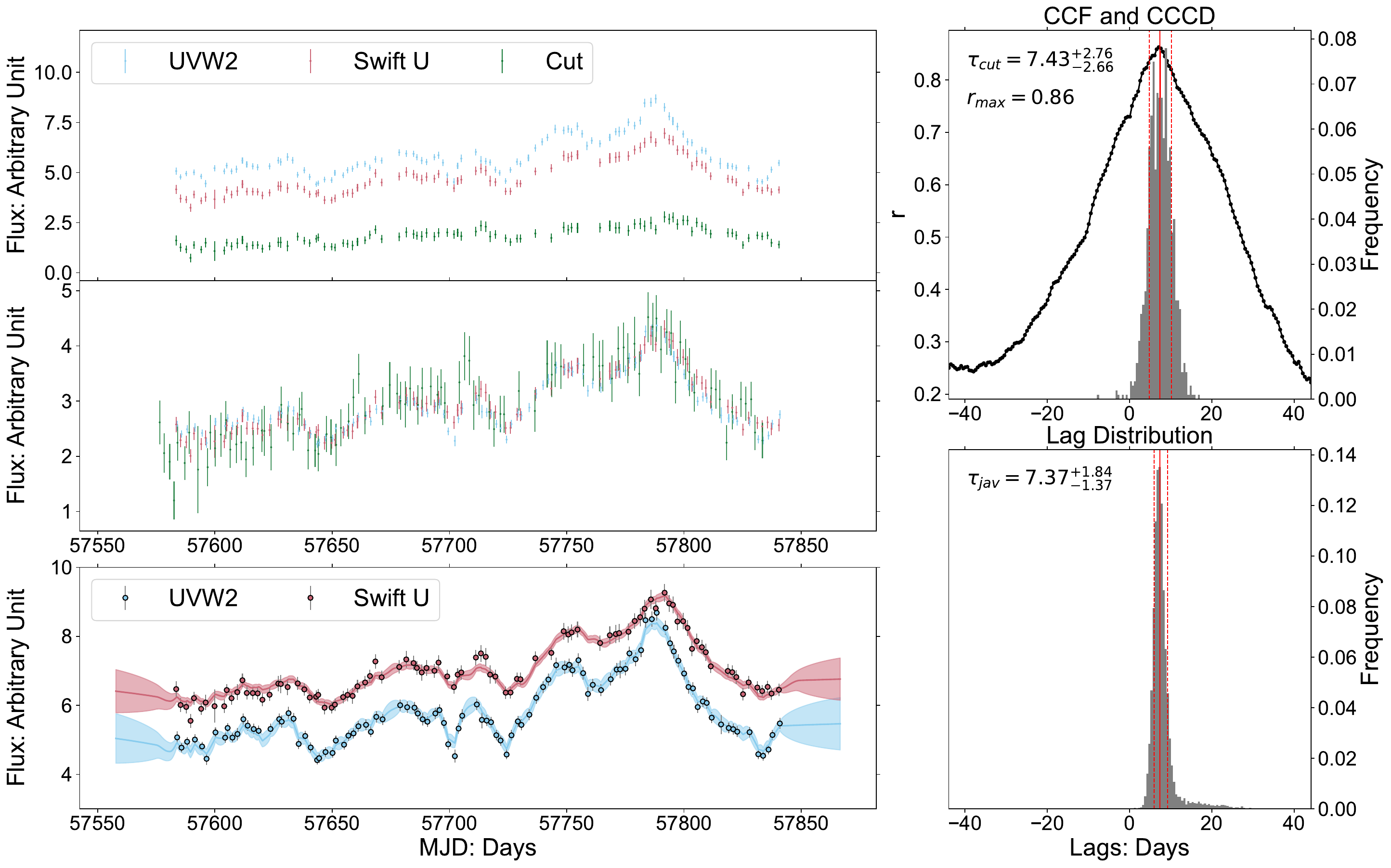}
\digitalasset
\caption{The same as Figure \ref{fig:Fairall9_result}, but for 3C 120. The complete figure set (12 images) is available in the online journal.}
\label{FigSetB}
\end{figure*}

\renewcommand{\thefigure}{C\arabic{figure}}
\setcounter{figure}{0}

\begin{figure*}[h]
\centering
\includegraphics[width=0.9\textwidth]{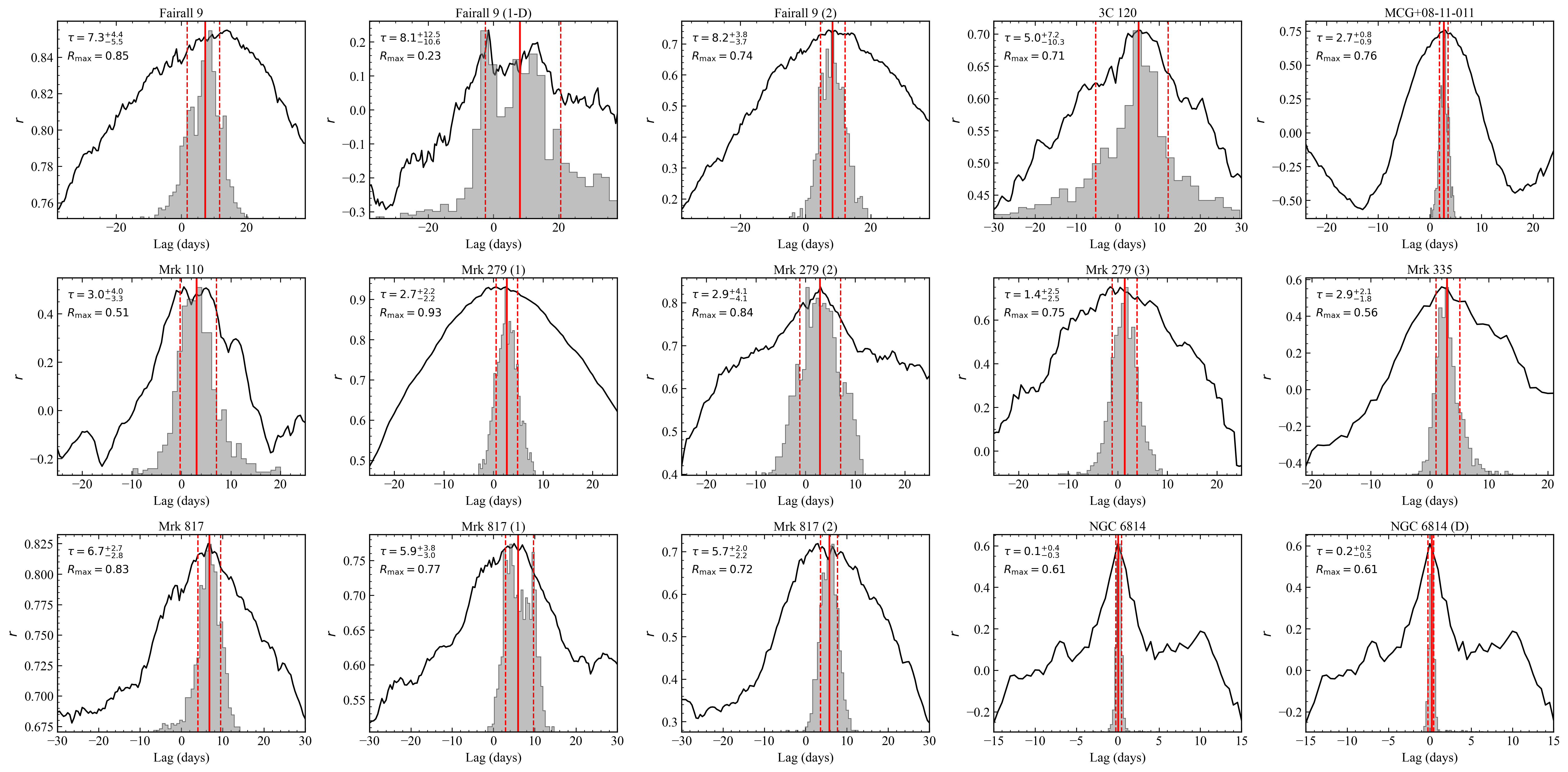}
\caption{ICCF-Cut results derived from the $V$ band light curves. Each panel shows the CCF (black solid line) and the CCCD (gray shaded region) between the $UVW2$ band and cut light curves. The maximum of the CCF and the corresponding lag estimation are indicated in the upper-left corner of each panel.}
\label{FigVbandICCF}
\end{figure*}

\end{document}